\documentclass[12pt]{article}
\usepackage{jheppub}
\usepackage{tikz}
\usetikzlibrary{decorations.markings, shapes.geometric,calc}
\newcommand\cH{\mathcal H}

\newcommand\cA{\mathcal A}

\newcommand\cL{\mathcal L}
\newcommand\cM{\mathcal M}
\newcommand\cS{\mathcal S}
\newcommand\cT{\mathcal T}
\newcommand\Bun{\mathrm{Bun}}
\newcommand\fg{\mathfrak{g}}
\newcommand\fq{\mathfrak{q}}
\newcommand\fA{\mathfrak{A}}
\newcommand\fD{\mathfrak{D}}
\newcommand\fU{\mathfrak{U}}
\newcommand\R{\mathbb{R}}
\newcommand\bR{\mathbb{R}}
\newcommand\C{\mathbb{C}}
\newcommand\bC{\mathbb{bC}}
\newcommand\bZ{\mathbb{Z}}
\newcommand\Tr{\mathrm{Tr}}

\newcommand\op{\mathrm{op}}

\newcommand\wt{\widetilde}

\newcommand{\rf}[1]{(\ref{#1})}

\newcommand{\fr}[2]{{\textstyle \frac{#1}{#2} }}

\newcommand{\cW}{{\mathcal W}}

\newcommand{\sP}{{\mathsf P}}  
\newcommand{\sR}{{\mathsf R}}

\newcommand{\ii}{\mathrm{i}}

\title{Schur Quantization and Complex Chern-Simons theory}
\author[1]{Davide Gaiotto,}
\author[2]{J\"org Teschner}
\affiliation[1]{Perimeter Institute for Theoretical Physics, 
31 Caroline Street North, Waterloo, ON N2L 2Y5, Canada}
\affiliation[2]{Department of Mathematics, 
University of Hamburg, 
Bundesstrasse 55,
20146 Hamburg, Germany,  and Deutsches Elektronen-Synchrotron DESY, Notkestr. 85, 22607 Hamburg, Germany}

\abstract{Any four-dimensional Supersymmetric Quantum Field Theory with eight supercharges can be associated to a certain complex symplectic manifold 
called the ``K-theoretic Coulomb branch'' of the theory. The collection of K-theoretic Coulomb branches include many complex phase spaces of great interest, 
including in particular the ``character varieties'' of complex flat connections on a Riemann surface. The SQFT definition endows K-theoretic Coulomb branches with a variety of canonical structures, including a deformation quantization. In this paper we introduce  a canonical ``Schur'' quantization of K-theoretic Coulomb branches. It is defined by a variant of the Gelfand-Naimark-Segal construction, applied to protected Schur correlation functions of half-BPS line defects. Schur quantization produces an actual quantization of the complex phase space. As a concrete application, we apply this construction to character varieties in order to quantize Chern-Simons gauge theory with a complex gauge group. 
Other applications include the definition of a new quantum deformation of the Lorentz group, 
and the solution of certain spectral problems via dualities.}
\begin{document}
\maketitle

\section{Introduction}

Schur correlation functions are a special class of protected quantities in four-dimensional ${\cal N}=2$ Supersymmetric Quantum Field Theories which have attracted considerable 
attention in the last few years \cite{Kinney:2005ej,Gadde:2011ik,Gadde:2011uv,Dimofte:2011py,Fukuda:2012jr,Gang:2012yr,Beem:2013sza,Lemos:2014lua,Beem:2014rza,Drukker:2015spa,Tachikawa:2015iba,Cecotti:2015lab,Cordova:2015nma,Cordova:2016uwk,Watanabe:2016bwr,Watanabe:2017bmi,Cordova:2017ohl,Neitzke:2017cxz,Fluder:2019dpf,Pan:2021mrw,Hatsuda:2023iwi}. A main goal of this paper is to employ Schur correlation functions to define an interesting collection of quantum mechanical systems whose properties are determined by the parent 4d SQFTs. The procedure is closely related to 
a previous construction of quantum mechanical systems whose properties are determined by 3d ${\cal N}=4$ SCFTs \cite{Gaiotto:2023hda}.

The operator algebra for the quantum mechanical system associated to a four-dimensional ${\cal N}=2$ supersymmetric quantum field theory $\cT$ is the $*$-algebra double
\begin{equation}
	\fD[\cT] \equiv A[\cT] \times A[\cT]^\op \, ,
\end{equation} 
with a $*$-structure defined below. Here $A^\op$ denotes the algebra with the same elements and addition as $A$ but opposite multiplication and $A[\cT]$ is the ``quantum K-theoretic Coulomb branch algebra''\footnote{We caution the reader that some mathematical papers incorrectly refer to the ``K-theoretic Coulomb branch algebra of a 3d ${\cal N}=4$ SQFT'' when describing the algebra associated to a 4d theory with the same field content.} \cite{Kapustin:2006hi,Kapustin:2007wm,Pestun:2007rz,Drukker:2009tz,Alday:2009fs,Drukker:2009id,Gaiotto:2010be,Gomis:2011pf,Ito:2011ea,Saulina:2011qr,Dimofte:2011py,2015arXiv150303676N,Braverman:2016wma,finkelberg2019multiplicative,schrader2019k}, which describes the fusion of (K-theory classes of) half-BPS line defects in $\cT$.
This algebra inherits many remarkable properties from the webs of dualities typical of these SQFTs. Our goal is to identify a natural choice of Hilbert spaces $\cH[\cT]$ on which $\fD[\cT]$ is represented unitarily. The quantum mechanical system defined in this way will reflect important properties of the parent SQFT.
 
Schur correlation functions can be defined as Witten indices of certain spaces of local operators, as reviewed below, or equivalently as 
supersymmetric partition functions on an $S^3 \times S^1$ geometry, with line defects 
wrapping the $S^1$ factor. Intuitively, the relation with a quantum mechanical system arises from an unusual factorization of the 4d
geometry along a supersymmetric $S^2 \times S^1$ slice  including line defect insertions \cite{Drukker:2010jp,Dimofte:2011py,Cordova:2016uwk,Dedushenko:2018tgx}, leading to 
a representation of  correlation functions as expectation values of  elements of $\fD$ between states produced by the path integral over half of the geometry. 

This intuition is motivated by some formal properties enjoyed both by the explicit localization formulae which compute the Schur index of Lagrangian gauge theories 
and by the conjectural IR localization formulae which compute the Schur index of any SQFT with a Seiberg-Witten effective description. 
In both cases, the correlation function takes the form of an expectation value of certain operators acting on an auxiliary Hilbert space $\cH_{\mathrm{aux}}$.
The relations between the auxiliary Hilbert spaces associated to dual presentations of the same theory can be
far from obvious in general, although one may identify unitary operators which intertwine different presentations  in a few specific cases. One would like to argue that these auxiliary Hilbert spaces carry different presentations of a structure which is canonically associated to the SQFT itself.\footnote{It is a bit challenging to make this intuition precise. The natural way to give a physical construction of an Hilbert space equipped with a positive-definite inner product
is to consider a unitary supersymmetric quantum mechanical system and project to its ground states. The original $S^3 \times S^1$ geometry is not equipped with an isometry 
which could play the role of an Hamiltonian for the $S^2 \times S^1$ slice. Presumably, one may seek a family of rigid supergravity backgrounds which 
interpolates between the original $S^3 \times S^1$ geometry and a situation where the required SQM setup can be defined at least locally around the $S^2 \times S^1$ slice.
To the best of our knowledge, the required tools have not yet been developed. We will not attempt to do so.}

We will produce a candidate $\cH$ via a GNS-like construction which only employs the Schur correlation functions and an expected positivity property which has been verified 
in great generality. The candidate $\cH$ is defined as the closure of $A$ under a certain positive-definite inner product, equipped with the natural left- and right- action of $A$ on itself. 
In particular, it is equipped with a dense collection of states $|a\rangle$ labelled by elements $a\in{A}$ and generated from a ``spherical'' vector $|1\rangle$ associated to the identity element.
Intuitively, $|1\rangle$ represents the path integral over half of the $S^3 \times S^1$ geometry and $|a\rangle$ a path integral with an extra line defect insertion. The $|a\rangle$ states are not orthonormal, but have inner products explicitly given by the Schur correlation functions. 

The pair $(\fD,\cH)$ reflects various important properties of the parent SQFT. %It is thus well suited to our purposes. 
The formal factorization of localization formulae can be 
recast as the existence of an isometry mapping $\cH$ to the corresponding auxiliary Hilbert spaces $\cH_{\mathrm{aux}}$. One may then investigate if the isometry may actually define an isomorphism, intertwining conjectural equivalences between different auxiliary descriptions.\footnote{Analogously, if an alternative physical Hilbert space $\cH_{\mathrm{phys}}$ with the desired properties can be defined as in the previous footnote, it will necessarily include analogous vectors $|1\rangle_{\mathrm{phys}}$ and $|a\rangle_{\mathrm{phys}}$ with the same inner products as $|a\rangle$ and thus will include an isometric image of $\cH$.}

An important application this approach is to give an uniform characterization of the quantum mechanical systems associated to 
theories of class $\cS$ \cite{Witten:1997sc,Gaiotto:2009hg,Gaiotto:2009we}. Recall that theories of class $\cS$ are labelled by the data of an ADE Lie algebra $\fg$ and a Riemann surface $C$, possibly decorated in a manner we will not review here \cite{Chacaltana:2012zy}. This data is used to define a supersymmetric compactification on $C$ of the six-dimensional $(2,0)$-SCFT labelled by $\fg$, leading to a 4d ${\cal N}=2$ theory $T[\fg,C]$. Remarkably, the corresponding K-theoretic Coulomb branch algebra $A[\fg,C]$ has a geometric description in terms of skeins on $C$ labelled by finite-dimensional $\fg$ representations \cite{Gaiotto:2010be}. 

We will derive a dual description of the Schur correlation functions as $C \times D^2$ correlation functions in the four-dimensional Kapustin-Witten theory \cite{Kapustin:2006pk}. Factorization 
along a diameter allows us to identify $\cH$ as the Hilbert space of a Chern-Simons theory with complex gauge group \cite{Witten:2010cx,Gaiotto:2021tsq,Gaiotto:2024tpl}. The algebra
$\fD$ maps to the algebra of space-like skeins of Wilson line operators in Chern-Simons theory and the spherical vector to the boundary state for a very special topological boundary condition. The construction is somewhat analogous to the quantum double construction of conventional 3d TFTs \cite{Turaev:1992hq,Levin:2004mi}. When $\fg=\mathfrak{sl}_2$, we expect the construction to be related to a Lorentzian de Sitter variant of the Ponzano-Regge model \cite{Ponzano:461451}. There are strong similarities with 3d loop quantum gravity constructions \cite{Buffenoir:2002tx,Bonzom:2014bua} but the unitary structure appears to be novel.

We will verify this proposal for several four-dimensional ${\cal N}=2$ supersymmetric quantum field theories which have a simple class $\cS$ description with $\fg = \mathfrak{sl}_2$, 
and compare it with a more conventional approach to the  quantization of complex Chern-Simons theory
in a typical example. For reason of space, we will focus on UV localization formulae in this paper and 
make connections to IR formulae in a companion paper \cite{next}. We will also give a general comparison between our 
``Schur quantization'' approach to complex Chern-Simons theory and previous approaches \cite{Witten:1989ip,Dimofte:2016pua}. 

Some of the existing approaches to complex Chern-Simons theory are topological in nature. Constructions based on the 3d-3d correspondence \cite{Dimofte:2011py,Dimofte:2016pua}
also implicitly or explicitly employ the relation to the 6d $(2,0)$-SCFTs and are obviously closely related to this work. Comparison with these approaches will be mostly be postponed to 
our companion paper \cite{next}, as IR formulae play a crucial role. 
%Constructions based on the representation theory of a quantum deformation of the Lorentz group $SL(2,\bC)$
%or more general complex quantum groups are also an important antecedent to this work {\bf citations}. 
Another approach is based  on a quantum deformation of the Lorentz group $SL(2,\bC)$ \cite{Buffenoir:2002tx}.
Remarkably, we will find that the quantum theory defined by Schur correlation functions is related to a quantum deformation of $SL(2,\bC)$ that is different from the one used in \cite{Buffenoir:2002tx}. Both appear to fit into a larger family associated to Schur correlation functions decorated by surface defects \cite{Cordova:2017ohl}, but we expect the construction we propose to be special within this larger class of options: the surface defects will generically not be canonical nor invariant under dualities.  

Another approach to the quantization of complex Chern-Simons theory uses the splitting of flat connections in to $(1,0)$ and $(0,1)$-parts defined by a complex structure on $C$ \cite{Witten:1989ip}. As discussed in the companion paper \cite{Gaiotto:2024tpl}, one is thereby led to a quantization scheme related to the non-compact WZW model with target $G_{\bC}/G_ c$, with $G_c$ the compact real form of $G_{\bC}$, and to a one-parameter deformation of the analytic Langlands correspondence. 
The relation to complex Chern-Simons theory suggests that the complex-structure dependent quantization is equivalent to the topological quantization. The relations with 
class $\cS$ theories furthermore predict an equivalence with the quantum theories defined by the Schur correlation functions. 

The rest of the introduction will draw a somewhat more detailed picture. 

\subsection{Schur indices}

The Schur indices \cite{Gadde:2011uv} of four-dimensional ${\cal N}=2$ supersymmetric quantum field theories decorated by half-BPS line defects \cite{Gaiotto:2010be,Dimofte:2011py,Cordova:2016uwk} represent the physical basis of our proposal. 
References \cite{Kapustin:2005py,Gaiotto:2010be} review of some of the properties of half-BPS line defects, and \cite{Kapustin:2006hi} introduces the holomorphic-topological twist as a tool to study them. 
A mathematical definition of a monoidal, $\bC^*$-equivariant category 
expected to capture the properties of half-BPS line defects in Lagrangian gauge theories has more recently been given in \cite{cautis2023canonical}. We expect that an analogous category $\mathsf{Lines}[\cT]$ exists for any 4d ${\cal N}=2$ SQFT $T$. 
Decorated Schur indices only depend on $\bC^*$-equivariant  K-theory classes of line defects, which 
define the algebra   
\begin{equation}
	A_\fq[\cT] \equiv K_{\bC^*}(\mathsf{Lines}[\cT])
\end{equation}
over $\bZ[\fq, \fq^{-1}]$, where $\fq$ is the $\bC^*$-equivariant parameter but also plays the role of the spin fugacity in the Schur index context. From now on, whenever we mention a line defect, we usually refer to its K-theory class.

Given two half-BPS line defects $L_a$ and $L_b$, one may consider the space of local operators which may appear at a junction between $L_a$ and $L_b$, i.e. the space of line defect-changing local operators. The line defect Schur index
$I_{a,b}(\fq)$ can be defined as the Witten index of this space of local operators, graded by $\mathrm{Spin(2)}$ rotation quantum numbers with fugacity $\fq$ \cite{Gadde:2011uv, Dimofte:2011py}.\footnote{If thus gives the equivariant character of morphisms in $\mathsf{Lines}[\cT]$, with $\fq$ being the equivariant parameter for the $\bC^*$ action on the category.}
The Schur indices often admit an interpretation as a partition function 
of superconformal ${\cal N}=2$ supersymmetric quantum field theories 
on the euclidean four-manifold $S^1 \times S^3$. 
Schur indices can also be defined for theories which are not super-conformal and are expected to still admit an $S^1 \times S^3$ interpretation for some rigid supergravity background. To the best of our knowledge, such a background has not yet been described in detail yet, though its existence follows from general considerations about the holomorphic-topological twist \cite{Kapustin:2006hi} of the theory.\footnote{Indeed, the Schur index can be computed in the HT twist of the theory placed on a quotient of $\bR^2\times \bC$ by a dilatation which acts on $\bC$ by a factor of $\fq$.} 

The Schur indices $I_{a,b}(\fq)$ give a pairing on $A_\fq$. Our main conjecture is that 
\begin{equation}\label{positivity}
I_{a,a}(\fq)>0\quad \text{for all}\;\,a\in A_\fq, \;\,a\neq 0. \qquad \text{(Positivity)}
\end{equation}
for $0<\fq^2<1$.
This conjecture will be checked in many examples later in this paper.  Conjecture \rf{positivity} implies that the
hermitian form on the complexification of $A_\fq$
defined by $\langle a|b\rangle=I_{a,b}(\fq)$ is positive definite, and therefore defines a scalar product on $A_\fq$.

The $L^2$ closure of $A_\fq$ under such pairing defines the Hilbert space $\cH_\fq$ of interest here: $L^2$-normalizable linear combinations of the vectors $|a\rangle$ associated to the line defects $L_a$.

The representation of $A_\fq$ on $\cH_\fq$ has remarkable properties. The space $\cH_\fq$ contains a distinguished 
vector $|1\rangle\in\cH_\fq$ associated to the unit element of $A_\fq$. There are two natural 
actions of $A_\fq$ on $\cH_\fq$, associated to left- and right multiplication in $A_\fq$,
\begin{equation}\label{Wa-tildeWa}
W_a|b\rangle=|ab\rangle,\qquad 
 \wt{W}_a|b\rangle=|ba\rangle,
\end{equation}
respectively.  It is clear that $|1\rangle$ is cyclic with respect to these actions, in the sense that the space
spanned by the vectors $W_a |1\rangle$ is dense in $\cH_\fq$. From \rf{Wa-tildeWa} it follows that
\begin{equation}\label{sphdef} 
W_a|1\rangle=\wt{W}_a|1\rangle \,.
\end{equation}
Vectors $|1\rangle$  satisfying \rf{sphdef} will be called {\it spherical}.

General properties of the Schur index also predict the Hermiticity properties of the inner products: 
there exist an automorphism $\rho:A_\fq\rightarrow A_{\fq}$, defined over $\bZ[\fq, \fq^{-1}]$ and naturally extended to be anti-linear over $\bC$,
such that 
\begin{equation}
	\wt{W}_{a}^{\dagger}=W_{\rho(a)} \, ,
\end{equation}
and thus $W_{a}^{\dagger}=\wt{W}_{\rho^{-1}(a)}$. We will discuss the physical interpretation of $\rho$ in the main text.
This makes 
the representation of 
$\fD_\fq \equiv A_\fq \times A_\fq^\op$ on $\cH_\fq$ unitary with respect to the $*$-algebra structure 
defined by $\wt{a}^\ast=\rho(a)$, 
using the notation $\wt{a}$ for the element of $A^\op$ corresponding to $a\in A$. 
%By considering algebras of the form $\mathfrak{A}=A \times A^\op$, and  
%automorphisms $\eta$ acting trivially on $A^\op$, 
%one sees that $\wt{a}=a^{\ast}$
%is equivalent to $\wt{a}=\rho(a)^{\star}$.}
%\begin{equation}
%	(a,b)^* = \left(\rho(b), \rho^{-1}(a)\right)\, ,
%\end{equation}
%and makes the action on $\cH_\fq$ unitary. 
%\footnote{The $*$-algebra structure is obviously isomorphic to $(a,b)^* = (b,a)$,
%but we prefer to use the spherical condition as a reference.}

The spherical condition implies that the expectation values
\begin{equation}
	\Tr \, a \equiv I_{1,a}(\fq) = \langle 1|W_a|1\rangle \, ,
\end{equation}
define a {\it twisted trace} 
\begin{equation}
	\Tr\, a b = \Tr \rho^2(b) a \, .
\end{equation}
The positivity condition can be written as 
\begin{equation}
	\Tr\, \rho(a) a > 0 \, .
\end{equation}
We will later argue that there is a one-to-one correspondence between positive traces on algebras $A_\fq$ 
and unitary representations of $\fD_\fq$ containing a spherical vector $|1\rangle$. Both 
descriptions involve the automorphism $\rho$ as a characteristic piece of data.\footnote{We will see in the main text that the construction of $\cH_\fq$ can be modified by the insertion of surface defects in the Schur index. This can lead to positive traces on $A_\fq$ twisted by automorphisms $\rho'$ distinct from $\rho$. They lead to spherical unitary representations of the corresponding $*$-algebra doubles $\fD'_\fq$. We will discuss in the main text the relation between the $*$-algebras $\fD_\fq$ and $\fD'_\fq$ and their unitary representations. }

Mathematically, one can identify a linear space of possible twisted traces on the algebra $A_\fq$ for any given
automorphisms $\rho$. Characterizing the convex cone of positive traces is an 
interesting mathematical problem.  %This is essentially a ``bootstrap'' problem. 
The mathematical problem to classify positive 
traces of potential relevance for 
 Abelian gauge theories has been studied in \cite{Klyuev_2022}. 
 The choice of $\rho$ from Schur quantization appears to be distinguished by two properties: a positive $\rho^2$-twisted trace exists and is unique.
 It would  be very interesting to find generalizations of this result. 

We expect that the supergravity backgrounds representing the Schur indices as partition functions on $S^1 \times S^3$ are reflection positive, implying \rf{positivity} on general grounds.
However, as this has not been demonstrated yet, we will later verify \rf{positivity} 
in many examples by direct computations based on Lagrangian descriptions
of the theories $\cT$. We should also observe that positivity is built into the conjectural IR formulae for the 
Schur indices \cite{Cordova:2016uwk}.

It should be noted that the theories $\cT$ may admit several Lagrangian 
descriptions, leading to different formulae for the Schur indices of one and the same theory
$\cT$. The fact that the Schur indices do not depend on the couplings suggests that 
all these different formulae represent  the same function of $\fq$. This is a highly 
non-trivial property which is challenging to prove even in simple examples.

\subsection{Schur quantization of K-theoretic Coulomb branches}

The quantum system abstractly defined by the above construction has an intimate connection with the K-theoretic Coulomb branch $\cM[T]$, i.e. the
moduli space of Coulomb vacua of the four-dimensional ${\cal N}=2$ supersymmetric quantum field theories compactified on a circle while preserving all supercharges. 
The moduli space $\cM[T]$ is a hyper-K\"ahler manifold which is a complex integrable system in one of the complex structures \cite{Seiberg:1996nz}.

Half-BPS line defects wrapping the circle provide a basis of the commutative algebra 
$A_{\mathrm{cl}}$ of holomorphic functions on $\cM[T]$ \cite{Kapustin:2006hi,Kapustin:2007wm,Gaiotto:2010be} in a different (generic) complex structure. 
The algebra $A_{\mathrm{cl}}$ is isomorphic to the classical limit $\fq \to 1$ of $A_{\fq}$.\footnote{There are actually two classical limits $\fq \to \pm 1$ and two 
closely related versions $\cM_\pm[T]$ of the K-theoretic Coulomb branch \cite{Gaiotto:2010be}, depending on the circle-compactification being twisted by 
the fermion number or by the center of the $SU(2)_R$ symmetry of the theory.} 
A precise mathematical definition of the K-theoretic Coulomb branches of quiver gauge theories 
has been given in \cite{Braverman:2016wma},
leading to powerful techniques for the computation of difference operator realisations of $A_\fq$
\cite{finkelberg2019multiplicative} compatible with localization formulae for the Schur indices.

The quantum system $(\fD_\fq,\cH_\fq)$ defined from Schur indices defines a quantization of the complex symplectic space space $\cM[\cT]$ as a real phase space, 
with $\fq = e^{- \hbar}$ for real $\hbar$, henceforth called Schur quantisation. The $*$-algebra $\fD$ quantizes the classical Poisson algebra generated by holomorphic and anti-holomorphic functions on $\cM[T]$.\footnote{The classical definition of the automorphism $\rho$ which appears in the $*$-structure is subtle and interesting. The moduli space $\cM[T]$ is hyper-K\"ahler, with a circle worth of complex structures which give essentially the same complex manifold. An holomorphic function $a$ on $\cM[T]$ can be ``hyper-K\"ahler rotated'' along this circle and mapped to an holomorphic function in the opposite complex structure. Complex conjugation maps it back to an holomorphic function $\rho(a)$. }   
%, equipped with a complex conjugation $\rho_{\mathrm{cl}}$. 

Schur quantization inherits extra structures from a larger collection of protected Schur correlation functions. In particular, Schur ``half-indices'' which count protected local operators supported on 
half-BPS boundary conditions or interfaces for $T$ can be interpreted as distributional states or kernels in $\cH_\fq$. The physical interplay between lines and boundaries/interfaces equips these states/kernels with a specific action of $\fD_\fq$. For example, certain interfaces implement unitary equivalences associated to dualities or RG flows of $T$ \cite{Gaiotto:2008ak,Dimofte:2013lba}.

Schur quantization can also be regarded as a four-dimensional uplift of the ``sphere quantization'' introduced in \cite{Gaiotto:2023hda} for the Coulomb branch of three-dimensional ${\cal N}=4$ SCFTs. It is furthermore related to brane quantization \cite{Gukov:2008ve,Kapustin:2001ij,Bressler:2002eu,Kapustin:2005vs,Pestun:2006rj,Gualtieri:2007bq,Aldi:2005hz}.

%UV Formulae for Schur indices $\rightarrow$ Hilbert spaces, explicit descriptions of algebra  $A_\fq$ from K-theoretic Coulomb branches, localisation of line operators $\rightarrow$  
%finite difference operator representations. Verification of crucial positivity conjecture.

\subsection{Class $\cS$ examples}\label{Intro-classS}

Explicit descriptions of the algebras $A_\fq$ are also known
whenever the four-dimensional ${\cal N}=2$ supersymmetric quantum field theories $\cT$ are in class $\cS$ \cite{Witten:1997sc,Gaiotto:2009we,Gaiotto:2009hg}. Such theories
can, by definition, be described as compactifications of the $(2,0)$-supersymmetric six-dimension\-al theory on
Riemann surfaces $C$. 
This description implies a description of the K-theoretic Coulomb branches of the moduli spaces of 
vacua associated to such theories as 
moduli spaces $\cM(G,C)$ of flat complex $G_\bC$-connections on $C$.\footnote{The reader may be confused by the jump from the ADE Lie algebra $\fg$ labelling $T[\fg,C]$ to 
the global form of a group $G_\bC$ in $\cM(G,C)$. There are some subtleties concerning $T[\fg,C]$ being a relative theory \cite{Witten:2009at,Gaiotto:2010be} which we will neglect as much as possible in this paper.}

The Poisson algebra $\mathrm{Sk}(C,G)$ of algebraic functions on $\cM(G,C)$ is generated by the $W_{a,\mathrm{cl}}$ trace functions
\begin{equation}
		W_{a,\mathrm{cl}} \equiv \Tr_R \,\mathrm{Pexp} \oint_\ell \cA
\end{equation}
labelled by pairs $a=(R,\ell)$, with $\ell$ being a simple closed curve $\ell$ on $C$, and $R$ being a finite-dimensional representation $R$ of $G$,
as well as functions labelled by more general networks $a$ of holonomies along open paths on $C$ 
contracted by intertwining maps. The Poisson bracket relations among the functions $W_{a,\mathrm{cl}}$ on $\cM(C,G)$ admit a simple diagrammatical description via skein manipulations.

A lot is known about the quantization of such moduli spaces on the algebraic level.
The quantization of the Poisson algebra  $\mathrm{Sk}(C,G)$ is essentially canonical. It 
yields the skein algebra
$\mathrm{Sk}_\fq(C,G)$, a non-commutative algebra having generators 
 $W_a$,  satisfying explicitly known
diagrammatic relations.\footnote{We will ignore here some interesting subtleties about $\cM(G,C)$ being related to the $\fq \to 1$ or $\fq \to -1$ classical limits.} 

The representation theory of the algebra $\mathrm{Sk}_\fq(C,G)$ is highly non-trivial. 
It depends heavily on the 
allowed range of values of the parameter $\fq$. We are here interested in the case $0<\fq^2<1$ and in
unitary representations of $A_\fq$ where the generators of $\mathrm{Skein}_\fq(C,G)$ will get represented by normal operators
on a Hilbert space $\cH_\fq$. Schur quantization of theories of class $\cS$ gives us precisely such a quantization which is conjecturally canonical, i.e. it only depends on $C$ and $G$.

The representations of interest in the context of  Schur quantisation are
distinguished from previously studied representations by  the existence of a cyclic spherical vector. 
Later in the paper, we will discuss in a typical  example a more conventional approach to the quantisation
of $\cM(C,G)$, and show how a spherical vector can be constructed in this approach. 
 Once a spherical vector is found, 
expectation values $\langle 1|W_a|1\rangle$ give a positive twisted trace. We will show that $\langle 1|W_a|1\rangle$
coincides with Schur indices $I_{1,a}(\fq)$ derived using Lagrangian descriptions of the associated theory 
of class $\cS$.\footnote{The check is relatively straightforward, as the coordinate system traditionally used to quantize $\cM(G,C)$
happens to be compatible with the localization procedure employed in the calculation of the Schur index. It is nevertheless instructive.}

Observe that a mathematical proof of the uniqueness of positive twisted traces on $\mathrm{Sk}_\fq(C,G)$ with the correct $\rho$ 
would allow one to streamline the quantization of $\cM(G,C)$, making many of the properties suggested by the 
connections to  theories of class $\cS$ and their Schur indices manifest.

\subsection{Lift to Kapustin-Witten theory and a dictionary to Schur quantization}
There is a  relation between Schur quantization and complex Chern-Simons theory which can be motivated by a chain of dualities involving six-dimensional maximally-supersymmetric SCFTs, as discussed in more detail in our companion paper  \cite{Gaiotto:2024tpl}. 

The first half of the duality chain maps the Schur index of a class $\cS$ theory to a partition function of the Kapustin-Witten twist \cite{Kapustin:2006pk} of ${\cal N}=4$ Supersymmetric Yang Mills gauge theory with gauge group $G$, which is placed on the product of $C$ with a disk $D^2$ having Neumann boundary conditions. The original half-BPS line defects in the Schur index map to Wilson lines wrapping skeins in $C$, placed at the boundary of the disk\footnote{We are working in the generic KW twist, which does not admit bulk line defects.} in the same order as in the trace.\footnote{A disk geometry is a very natural way to define a trace of boundary local operators in a 2d TFT. In general, there is a whole collection of possible traces labelled by insertions of one bulk operators in the middle of the disk. Here that would necessarily be some 4d bulk local operator placed at points in $C$ or a bulk surface defect wrapping $C$.
Back along the duality chain this would map to the insertion of a surface defect in the Schur index, transverse to the plane supporting the line defects. The insertion of surface defects appear to modify $\rho$. Positivity properties may still hold, see \cite{Klyuev_2022} for some Abelian examples, but a physical explanation is more challenging.} 

The second part of the duality chain cuts the disk along a segment. The space of states which 
the KW theory associates to the segment appears in a  natural embedding of complex Chern-Simons theory into the KW twist \cite{Witten:2010cx,Gaiotto:2024tpl}. This is similar  to the duality chains previously considered in \cite{Mikhaylov:2017ngi}
for the case of partition functions on deformed $S^4$, leading to segment compactifications of KW theory  with suitable choices of boundary conditions. A related approach had previously been discussed in \cite{Cordova:2013cea}.\footnote{The 4d geometry can also be seen as a 4d uplift of a 2d qYM construction \cite{Gadde:2011ik,Fukuda:2012jr} and it would be interesting to formulate Schur quantization (and in particular positivity) directly in that language.} 

The KW path integral on each half disk is then predicted to produce a specific state $|1\rangle$ in complex CS theory, so that a Schur correlation function maps to an expectation value:
\begin{equation}
I_{1,a}(\fq) = \langle 1|a|1\rangle \, .
\end{equation}
The justification for this statement is somewhat non-trivial, involving the deformation of the half-disk to 
a quotient $\bR\times [1,-1]$ by a $\bZ_2$ reflection of both factors.

\subsection{Relation with complex Chern-Simons theory}

In this way one arrives at a conjectural representation of the Schur indices in terms of complex Chern-Simons (CS) theory. One may recall that 
the classical equations of motion of Chern-Simons theory require the complex connection $\cA$ to be flat. On 
a compact two-dimensional surface $C$, the theory has a finite-dimensional phase space,
the moduli space $\cM(C,G)$ of flat $G_\C$ connections $\cA$ on $C$, equipped with a symplectic form proportional to 
\begin{equation}
i \int_C \left[\delta \cA \wedge \delta \cA - \delta \bar \cA \wedge \delta \bar \cA \right] \, .
\end{equation} 
Finite-dimensional descriptions of $\cM(C,G)$ can offer a convenient starting point 
to the quantization using some convenient coordinate systems, but establishing 
independence on the choices of coordinates may require additional work. 

Topological invariance of the Chern-Simons functional suggests 
that the complex CS theory should associate a Hilbert space $\cH_{\rm CS}(C,G)$
to any surface $C$, with $\cH_{\rm CS}(C,G)$
depending only the topological type of $C$. 
The algebra of observables should coincide with $\mathrm{Sk}_\fq(C,G)\times 
\mathrm{Sk}_\fq(C,G)^{\rm op}$, with the first factor generated by the quantized holomorphic trace
functions $W_{a}$ (aka space-like Wilson lines for $\cA$) and the second factor generated by the quantized anti-holomorphic trace
functions $\wt{W}_{a}$ (aka space-like Wilson lines for $\bar \cA$).
%As discussed in the general case, the choice of a twist $\rho_{\mathrm{cl}}$ in the definition of complex conjugation is immaterial until we encounter a specific choice of spherical vector.

The path integral over  three-manifolds $M_3$ having boundary $C$
is expected to define states $|M_3\rangle\in\cH_{\rm CS}(C,G)$.  
One may also consider path integrals over three-manifolds 
of the form $\bR^+ \times C$,
with boundary conditions $B$ imposed at $0\times C$, in order to define distributions
$|B\rangle$. It was argued in \cite{Gaiotto:2024tpl} there should exist a distinguished boundary condition $B_c$ 
characterized by the condition that the holonomy of $\cA$, restricted to the boundary $C$, is unitary.
It should define a state $|1\rangle\in\cH_{\rm CS}(C,G)$ which satisfies $W_a|1\rangle=\wt{W}_{ a}|1\rangle$. Here we assume having chosen labelling  conventions 
in such a way that we have  $W_{a,\mathrm{cl}}=\wt{W}_{a,\mathrm{cl}}$ when the connection is unitary. This corresponds to a specific Hermiticity condition $\wt W_a = W_{\rho(a)}^\dagger$.\footnote{In the absence of irregular singularities, we have $\rho^2=1$. For $\fg=\mathfrak{sl}_2$, $\rho=1$. Irregular singularities on $C$ will complicate the story. Based on the properties of Schur indices and of class $\cS$ theories, we expect $\rho$ to act on line defects ending on irregular singularities by shifting the endpoint from one Stokes sector to the next one around the puncture. i.e. a ``pop'' in the notation of \cite{Gaiotto:2009hg}. In the class $\cS$ theory, this corresponds to an anomalous $U(1)_r$ rotation by $\pi$, which leads to $\theta$-angle shifts and Witten effect on dyonic lines \cite{Gaiotto:2010be}. It would be nice to have a clearer understanding of this point. We will continue the discussion in \cite{next}.}

Furthermore, it was argued that $B_c$ arises in the chain of dualities mentioned above as the path integral of KW theory on an half-disk. 
As a consequence, expectation values $\langle 1|W_a|1\rangle$ are predicted to match Schur indices $I_{1,a}(\fq)$, giving 
an isometry $\cH_\fq \to \cH_{\rm CS}(C,G)$ which is compatible with the action of 
$\mathrm{Sk}_\fq(C,G)\times \mathrm{Sk}_\fq(C,G)_{\rm op}$. Analogous arguments predict that the isometry should be compatible with 
\begin{itemize}
\item The action of the mapping class group of $C$. Indeed, the mapping class group is simply the duality group of $T[\fg,C]$ \cite{Gaiotto:2009we}. 
\item The collection of states $|M_3\rangle$ labelled by three-manifolds \cite{Dimofte:2011py,Dimofte:2013lba}.
\item A richer collection of TFT structure based on the factorization properties of quantum group representations (see \cite{jordan2023quantum} for a brief review and further references), which can be expressed in terms of physical operations on theories of class $\cS$ \cite{Gaiotto:2009hg,Gaiotto:2010be,Gaiotto:2012rg,Dimofte:2014ria,Neitzke:2020jik,Freed:2022yae}.
\end{itemize}

We conjecture that the isometry is an isomorphism and thus Schur quantization of theories of class $\cS$ provides a consistent quantization of 
complex Chern-Simons theory. A crucial aspect of this conjecture is that it requires the  states $W_a|1\rangle$ created from $B_c$ decorated by boundary skeins
to be dense in the Hilbert space of the theory.

One should, of course,  compare this approach to previous approaches to the quantisation of complex Chern-Simons theory. We will briefly review the comparison to 2d CFT-based methods, previously discussed in \cite{Gaiotto:2024tpl}, later in this paper. We also 
refer to \cite{Dimofte:2016pua} for a review of cluster algebra-based quantization strategies and to our upcoming work \cite{next} for a comparison based on the IR description of Schur indices. In both cases, the comparison proceeds by identifying canonical analogues of the spherical vector $|1\rangle$ to build an isometry from $\cH_\fq$. 
 
\subsection{Relations to quantum groups}

Relations to quantum group theory have played an important, in many cases a basic role in most of the previous
studies of quantum CS theories associated to compact or non-compact groups. Quantum 
group representation theory in particular represents the foundation of the approach to quantum CS theory
pioneered by Reshetikhin-Turaev \cite{Reshetikhin:1990pr}. 
Quantum groups furthermore represent the quantisation of the residual gauge
symmetries in the 
Hamiltonian quantisation of Alekseev-Grosse-Schomerus 
\cite{Alekseev:1994pa,Alekseev:1994au}.  

Deeply related connections
to quantum group representation theory have been observed in quantum Teichm\"uller theory
in \cite{kashaev2001spectrum,Nidaiev:2013bda}. Quantum Teichm\"uller theory is
related to a sector of the  $\mathrm{PSL}(2,\bR)$ CS-theory.
The modular double of
$U_q(\mathfrak{sl}_2)$ can serve as  a crucial link 
between quantum cluster variables associated to triangulations, and the modular functor structure
associated to pants decompositions in this context  \cite{Nidaiev:2013bda,Teschner:2013tqy}. 
A generalisation to higher Teichm\"uller theory has been 
developed in \cite{schrader2019cluster,schrader2017continuous}. 

The factorization algebra approach reviewed in \cite{jordan2023quantum} 
unifies and streamlines many of these conceptual threads and connects them directly to KW theory along the lines of \cite{Witten:2010cx}: 
the category of representations of quantum groups can be used to describe the theory algebraically, as a generalized Crane-Yetter theory \cite{kinnear2024nonsemisimple}. 

Quantisation of complex Chern-Simons theory has previously been studied in the regime $\fq\in\bR$
of our interest in particular in \cite{Buffenoir:2002tx}. The approach taken in \cite{Buffenoir:2002tx} follows the strategy of 
Alekseev-Grosse-Schomerus, using  the  quantum group  $U_q(\mathrm{SL}(2,\bC)_\bR)$ 
constructed in \cite{PodWoron} and further studied in \cite{Buffenoir:1997ih}
instead to $U_q(\mathrm{SU}(2))$. 

We are here going to present  evidence  that Schur quantisation defines  a quantisation 
of complex Chern-Simons theory related to 
a quantum deformation of the group $\mathrm{SL}(2,\bC)$. However, we will see that 
the quantum group relevant in this context is different  from the quantum group
used  in \cite{Buffenoir:2002tx} to construct a quantisation of complex Chern-Simons theory.
The variant of $U_q(\mathrm{SL}(2,\bC)_\bR)$
coming from Schur quantisation deserves 
further study. It should, in particular, help  to develop the quantisation of complex Chern-Simons theory
in close analogy to the quantum Teichm\"uller  theory.

\subsection{Relation with conformal field theory}
An alternative strategy to quantize complex Chern-Simons theory is to pick a complex structure on $C$ and 
use it to polarize the phase space, treating the $(0,1)$ part of the connection as coordinates and 
the $(1,0)$ part as momenta \cite{Witten:1989ip}. 

Essentially, one focusses on a family of distributional states $\langle\mathbf{x}|$ associated to certain 
boundary conditions $B_{\mathrm{WZW}}$ for the 3d theory, which fix the gauge 
equivalence class of $\bar{\cA}_{\bar{z}}$, or equivalently a holomorphic bundle on $C$. 
We are using $\mathbf{x}$ as the notation for a collection of parameters labelling a family
of holomorphic bundles on $C$. States $|\psi\rangle\in\cH_{\rm CS}$ can thereby be represented
by wave-functions $\psi(\mathbf{x})=\langle\mathbf{x}|\psi\rangle$. 

One may naturally consider the space of $L^2$-normalizable twisted half-densities, 
\begin{equation}\label{dR-Hilbert}
	\cH^{\mathrm{dR}}_s(C,G):= L^2(\Bun_G, |\Omega|^{1 + i \frac{s}{2\kappa_c}}),
\end{equation}
on the space/stack $\Bun_G$ of $G$-bundles on $C$ \cite{BraKazh}.\footnote{The Hilbert space itself can be defined in terms of twisted half-densities on some convenient non-singular open patch in $\Bun_G$. The intricacies of $\Bun_G$, though, can affect the definition of a rigged Hilbert space and of distributional states.}
In order to see that this is a natural scalar product one may first note that
variations in the complex structure of $C$ are represented by the projectively flat 
KZB connection \cite{Witten:1989ip}. One may furthermore check that
the parallel transport defined by the KZB connection is formally unitary in $L^2(\Bun_G, |\Omega|^{1 + i \frac{s}{2\kappa_c}})$.
This suggests, in particular, that  the KZ connection can be integrated to a unitary representation of the mapping class group. 

% The definition of $\cH^{\mathrm{Hol}}_s(C,G)$ and of the associated unitary KZ connection can be considered as a quantum version \cite{Gaiotto:2024tpl} of the analytic Geometric Langlands program \cite{Etingof:2019pni}. 

As discussed in more detail in \cite{Gaiotto:2024tpl}, 
one may then consider the wave-functions $\mathcal{Z}(\mathbf{x})=\langle\mathbf{x}|1\rangle$, 
or,  more generally $\mathcal{Z}_a(\mathbf{x})=\langle\mathbf{x}|W_a|1\rangle$. 
One of the main objectives of \cite{Gaiotto:2024tpl} is to propose a definition of the wave-functions 
$\mathcal{Z}_a(\mathbf{x})$ based on conformal field theory. We conjecture, in particular, that
the wave-functions $\mathcal{Z}(\mathbf{x})$ can be identified with the partition functions of 
the  WZW models with target $G_\C/G$ \cite{Gawedzki:1988hq}.
This  CFT has a partition function $\mathcal{Z}_{\rm WZW}$ 
which can be represented by a twisted half-density on $\Bun_G$ satisfying the KZB equations \cite{Gawedzki:1991yu}.  
$\mathcal{Z}_{\rm WZW}$ should in particular 
be invariant under the mapping class group of $C$. If  the WZW level $\kappa$ satisfies $\kappa-\kappa_c \in i \bR$,
we expect that  the partition functions $\mathcal{Z}_{\rm WZW}$ 
represent  elements  of $\cH^{\mathrm{dR}}_s(C,G)$, though normalizability is not  obvious. 

As furthermore discussed in \cite{Gaiotto:2024tpl}, it is natural to modify the partition functions 
$\mathcal{Z}_{\rm WZW}(\mathbf{x})$ by the insertion of Verlinde line operators. Representing 
$\mathcal{Z}_{\rm WZW}(\mathbf{x})$ as an integral over products of holomorphic and anti-holomorphic contributions
allows us to define two types of Verlinde line
operators, labelled by the same data $a$ as used to label trace functions, 
defining modified partition functions $(\cW_a\mathcal{Z}_{\rm WZW})(\mathbf{x})$
and $(\wt{\cW}_a\mathcal{Z}_{\rm WZW})(\mathbf{x})$, respectively. 
The main proposal made in \cite{Gaiotto:2024tpl} is the correspondence
\begin{equation}
(\cW_a\mathcal{Z}_{\rm WZW})(\mathbf{x})=\langle\mathbf{x}|W_a|1\rangle_{\rm CS}, 
\qquad 
(\wt{\cW}_a\mathcal{Z}_{\rm WZW})(\mathbf{x})=\langle\mathbf{x}|\wt{W}_a|1\rangle_{\rm CS}.
\end{equation}
The crucial consistency condition 
$(\cW_a\mathcal{Z}_{\rm WZW})(\mathbf{x})=(\wt{\cW}_a\mathcal{Z}_{\rm WZW})(\mathbf{x})$
can be verified with the help of CFT technology.

In order to round off the discussion let us note that the physics background outlined above 
predicts  that 
\begin{equation}
	\langle 1|W_a |1\rangle_{\mathrm{Schur}} = 
	\big\langle \mathcal{Z}_{\rm WZW}, {\cW}_a\mathcal{Z}_{\rm WZW}\big\rangle_{\rm dR},
\end{equation}
using the notation $\langle .,.\rangle_{\rm dR}$ for the scalar product in $\cH^{\mathrm{dR}}_s(C,G)$.
This is a rather non-trivial prediction. It would be nice to check it directly. 

\subsection{Structure of the paper}
In Section \ref{sec:schur} we discuss Schur quantization in greater detail. In Section \ref{sec:examples} we present a series of examples of increasing complexity where the rank of the gauge group is $1$. In Section \ref{sec:qg} we discuss in greater detail the occurrence of complex quantum groups in Schur quantization. In Section \ref{sec:charq} we discuss a relevant example of quantization of complex character varieties based on Fenchel-Nielsen coordinates. Section \ref{sec:hol} discusses the relation to complex Chern-Simons theory. 
Section \ref{sec:real} presents a tentative ``real'' generalization of Schur quantization, with algebra of observable $\fA = A_\fq$ equipped with some $*$-structure $\tau$. It should be applicable to a quantization of complex Chern-Simons theory on surfaces with boundaries or cross-caps. We conclude with two Appendices containing some useful formulae for gauge theories with $U(N)$ gauge groups.

\section{Schur quantization of K-theoretic Coulomb branches} \label{sec:schur}

For the sake of clarity, we begin by briefly reviewing a crucial relation between two mathematical structures which can be associated to an algebra $A$ defined\footnote{We could relax the condition to $A$ being defined over $\bC$ and $\rho$ being anti-linear with respect to scalar multiplication. The definitions below can be adjusted accordingly.} over $\bR$
and equipped with an invertible automorphism $\rho: A \to A$:
\begin{itemize}
	\item {\it Positive twisted traces}, i.e. linear maps $\Tr: A \to \bR$ which satisfy
	\begin{align}
		\Tr \, a b = \Tr \, \rho^2(b)a \cr
		\Tr \, \rho(a) a >0 \, .
	\end{align}
	\item {\it Spherical unitary representations} $\cH$ of the $*$-algebra\footnote{A $*$-algebra  $\mathfrak{D}$ is an algebra equipped with a star-structure. A star-structure is an involutive antilinear map $\ast:\mathfrak{D}\rightarrow \mathfrak{D}$, $\ast(a)=:a^\ast$, satisfying,   $(ab)^{\ast}=b^\ast a^\ast$. Unitary representations 
of a star-algebra $\mathfrak{D}$ are representations of $\mathfrak{D}$ on an Hilbert space $\mathcal{H}$ by operators $W_a$ such that ${W}_a{}^{\dagger}=W_{a^\ast}$. } ``double'' defined as $\mathfrak{D} = A \otimes A^\op$ with star structure $\wt{a}^\ast_{}=\rho(a)$, 
using the notation $\wt{a}$ for the element of $A^\op$ corresponding to $a\in A$ and
with $\rho$ being an automorphism of $A$.\footnote{In defining the $*$-algebra double $\fD$, we take the underlying vector space of $A$ and $A^\op$ to be literally the same. With this choice, $\rho$ is intrinsic to the definition and the spherical condition below is natural. If one forgets the choice of isomorphism of the underlying vector spaces, the $*$-algebras associated to the same $A$ and different $\rho$'s are equivalent and the choice of $\rho$ only affects the definition of spherical vectors.} 

Denoting  the normal %, possibly unbounded 
	operators representing $a\in A$ and $\wt{a}\in A^\op$ by  $W_a$ and $\wt{W}_{{a}}$, respectively, 
	unitarity requires $\wt{W}_{a}{}^\dagger=W_{\rho(a)}$. The term ``spherical'' refers to the existence of a spherical vector,  a cyclic\footnote{I.e. a vector $|1\rangle$ such that $\fD|1\rangle$ is dense in $\cH$.} vector $|1\rangle \in \cH$ 
	satisfying 
	\begin{equation}\label{spherical-def}
		W_a |1\rangle = \wt{W}_{{a}} |1\rangle.
	\end{equation} 
\end{itemize}
We will use the notation $|a\rangle=W_a|1\rangle$, $a\in A$.
It is useful to observe that  \rf{spherical-def} relates the representation of $A^\op$ on $\cH$ to the right action of $A$ on itself,
\[
{\wt W}_{b}\,|a\rangle={\wt W}_{b}W_a\,|1\rangle=W_a{\wt W}_{b}\,|1\rangle=W_aW_b|1\rangle=W_{ab}|1\rangle=|ab\rangle.
\]
It is straightforward to see how spherical unitary representations define positive traces: 
\begin{equation}
	\mathrm{Tr}\, a= \langle 1|W_a|1\rangle
\end{equation}
defines a positive twisted trace. Positivity follows immediately from 
\begin{equation}
	\Tr \, \rho(a) b = \langle 1|W_{\rho(a)} W_b|1\rangle=
	\langle 1|\wt{W}_{a}{}^\dagger W_b|1\rangle =\langle a|b\rangle,
\end{equation}
and the  twisted trace condition is also straightforward:
\begin{equation}
 \mathrm{Tr}\, \rho^2(b) a=\langle \rho(b)|a\rangle=\langle 1|W_{\rho(b)}{}^\dagger|a\rangle=\langle 1|\,{\wt W}_{b}\,|a\rangle
=\langle 1|ab\rangle=\mathrm{Tr}\,ab.	
\end{equation}

We would also like to argue that positive traces canonically define 
spherical unitary representations. 
The first step is to make the underlying vector space of $A$ into a 
module for $A \otimes A^{\op}$. In order to avoid confusion, we denote as $|a\rangle$ the element of the module corresponding to the element $a \in A$ and thus as $|1\rangle$ the element corresponding to the identity. We will use the canonical left and 
right actions of $A$ on itself in order to introduce the structure as a $A\otimes A^\op$-module, using the notations
\begin{equation}\label{WwtW-def}
	W_a \wt{W}_{{c}} \,|b\rangle := |a b c\rangle.
\end{equation}
Obviously, the vector $|1\rangle$ is cyclic for the module and satisfies \rf{spherical-def}.
%\begin{equation}
%	W_a |1\rangle = \wt W_a |1\rangle.
%\end{equation} 
The key step is to define the positive-definite inner product
\begin{equation}\label{def-scalar}
	\langle a|b\rangle \equiv \Tr \, \rho(a) b.
\end{equation}
We may then define an Hilbert space $\cH$ as the $L^2$ closure of $A$ under the inner product. 
The algebra $A \otimes A^{\op}$ acts on $\cH$ by densely-defined operators. We may observe that
\begin{equation}
\langle a|W_{\rho(b)}|c\rangle = \Tr \, \rho(a)\rho(b) c = \Tr \, \rho(ab) c = \langle ab|c\rangle =\langle\,  a\,|\,{\wt W}_{b}{}^\dagger |c\rangle,
%\left( \wt W_b |a\rangle\right)^\dagger |c\rangle
\end{equation} 
indicating that the hermitian conjugation defined by the scalar product \rf{def-scalar} makes the representation
of $A \otimes A^{\op}$ on $\cH$ into a spherical unitary representation of $\fD$.

One should note, however,  that the operators $W_a$ and ${\wt W}_{a}$ defined in \rf{WwtW-def} will be unbounded, in general. 
We will not attempt to determine under which conditions $W_a$ and ${\wt W}_{a}$ admit  extensions
defining normal operators on $\cH$.  

A classical example of this construction is the definition of spherical principal series representations of complex reductive Lie algebras $\fg_\bC$ starting from 
the unique traces on the central quotients of $U(\fg)$. This example and many more occur in the context of sphere quantization \cite{Gaiotto:2023hda}: the positive twisted traces are provided by 
protected correlation functions of 3d ${\cal N}=4$ SCFTs and are studied mathematically in the context of ``short star products'' \cite{Etingof:2019guc}.

Schur quantization similarly produce candidate positive twisted traces on many algebras of interest, including central quotients of $U_q(\fg)$ with $q=\fq^2$. It includes trigonometric deformations of the classical representation theory results found in 3d ${\cal N}=4$ SCFTs and much more. 

We will sometimes use the notation $\fD[A,\rho]$ to denote the $*$-algebra double of a given algebra $A$ with automorphism $\rho$.
\subsection{Schur correlation functions as a twisted trace}

The Schur index $I(\fq)$ was originally introduced as a specialization of the superconformal index of four-dimensional ${\cal N}=2$ SCFTs \cite{Gadde:2011uv}. It can either be interpreted as a supersymmetric partition function on a ``$S^1 \times_{\fq^2} S^3$'' geometry,\footnote{For real $0<\fq<1$, this denotes a geometry where the radius of $S^1$ is $-\log |\fq|^2$ times the radius of the sphere, decorated by some extra complexified R-symmetry backgrounds to preserve a specific amount of supersymmetry.} or as a graded Witten index of the space of local operators. Compared with the reference \cite{Cordova:2016uwk}, we define $\fq = q^{\frac12}$ to avoid square roots in our formulae. 

The Schur index can be generalized to a family of line defect Schur indices $I_{a,b}(\fq)$, graded Witten indices of the space of local operators intertwining between supersymmetric line defects $L_a$ and $L_b$. In terms of partition functions, this matches a correlation function with two line defect insertions in $S^1 \times_{\fq^2} S^3$: the defect $L_b$ is inserted at a specific point in the sphere and wraps $S^1$, while $L_a$ is inserted at an antipodal point on the sphere and wraps $S^1$ in the opposite direction. 

\begin{figure}[h]
\centering
\begin{tikzpicture}

% First part: a shorter vertical line with a single arrow and L_a label

% Define the origin and draw a black dot at the center
\coordinate (O) at (0, 0);
\fill[black] (O) circle (1.5pt);

% Draw a shorter vertical line with a single arrow pointing up at 2/3 height
\draw[thick, decoration={markings, mark=at position 0.67 with {\arrow{>}}, mark=at position 0.33 with {\arrow{>}}}, postaction={decorate}] (0, -2) -- (0, 2);
\node at (-0.3, 2.2) {$L_{b}$};
\node at (-0.3, -2.2) {$L_{a}$};

% Draw a thin circle around the origin
\draw[thin] (0,0) circle (1.5);

% Move second part to the right
\begin{scope}[xshift=7cm]

% Draw the sphere for S^3
\draw[thick] (0,0) circle (2);
\node at (0,2.3) {$S^3$};

% Draw the equator ellipse
\draw plot[domain=pi:2*pi] ({2*cos(\x r)},{.2*sin(\x r)});
\draw[dashed] plot[domain=0:pi] ({2*cos(\x r)},{.2*sin(\x r)});

% Draw two points on the equator
\foreach \i in {0, 180} {
    \fill[black] ({2*cos(\i)}, {0.2*sin(\i)}) circle (1.5pt);
}

% Draw the circle for S^1 closer to the sphere with a double-arrow
\draw[thick, decoration={markings, mark=at position 0.25 with {\arrow{>}}, mark=at position 0.74 with {\arrow{<}}, mark=at position 0.76 with {\arrow{<}}}, postaction={decorate}] (3.7,0) circle (1);
\node at (3.7,1.3) {$S^1$};

% Place the \times_{\mathbb{F}_q} symbol between the sphere and the circle
\node at (2.4, 0) {$\times_{\fq^2}$};

\end{scope}

\end{tikzpicture}
\caption{Left: The line defect Schur index $I_{a,b}(\fq)$ counts protected local operators interpolating between line defects $L_a$ and $L_b$. Right: the state-operator map relates it to a partition function on a twisted $S^3 \times_{\fq^2} S^1$ geometry with antipodal insertions of $L_a$ and $L_b$ wrapping the $S^1$ factor in opposite directions.}
\end{figure}
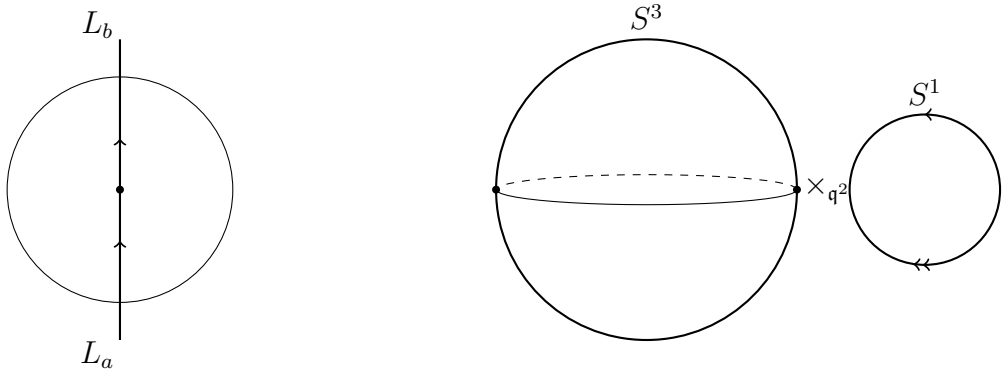

The line defect Schur indices can be generalized further to a collection of ``Schur correlation functions'' $I_{a_1 \cdots a_n}(\fq)$, with insertions of $L_{a_i}$ line defects wrapping $S^1$ at a cyclic sequence of points along a great circle of $S^3$ \cite{Dimofte:2011py}. These can also be understood as graded Witten indices for spaces of local operators sitting at the junction of multiple line defects. The notation reflects the fact that Schur correlation functions only depend on the relative order of the insertion points along the great circle, up to an important subtlety we discuss next.

\begin{figure}[h]
\centering
\begin{tikzpicture}

% First part: half-lines on the plane

% Define the origin and draw a black dot at the center (smaller scale for the first part)
\coordinate (O) at (0, 0);
\fill[black] (O) circle (1.5pt);

% Define angles for the half-lines, rotated by 18 degrees counterclockwise
\def\angles{{18, 90, 162, 234, 306}}

% Draw half-lines with arrows in the middle and labels
\foreach \i in {1,...,5} {
    \draw[thick, scale=0.6, decoration={markings, mark=at position 0.5 with {\arrow{>}}}, postaction={decorate}] (O) -- ({\angles[\i-1]}:4);
    \node at ({0.6*4.5*cos(\angles[\i-1])}, {0.6*4.5*sin(\angles[\i-1])}) {$L^{\vartheta_{\i}}_{a_{\i}}$};
}

% Draw a thin circle around the origin
\draw[thin] (0,0) circle (1.5);

% Move second part to the right
\begin{scope}[xshift=7cm]

% Draw the sphere for S^3
\draw[thick] (0,0) circle (2);
\node at (0,2.3) {$S^3$};
% Draw the equator ellipse
\draw plot[domain=pi:2*pi] ({2*cos(\x r)},{.2*sin(\x r)});
\draw[dashed] plot[domain=0:pi] ({2*cos(\x r)},{.2*sin(\x r)});

% Draw dots on the equator
\foreach \i in {1,...,5} {
    \fill[black] ({2*cos(\angles[\i-1])}, {0.2*sin(\angles[\i-1])}) circle (1.5pt);
}

% Draw the circle for S^1 closer to the sphere
\draw[thick, decoration={markings, mark=at position 0.5 with {\arrow{>}}}, postaction={decorate}] (3.7,0) circle (1);
\node at (3.7,1.3) {$S^1$};
% Place the \times_{\mathbb{F}_q} symbol between the sphere and the circle
\node at (2.4, 0) {$\times_{\fq^2}$};

\end{scope}

\end{tikzpicture}
\caption{Left: Schur correlators such as $I_{a_1 a_2 a_3 a_4 a_5}(\fq)$ count local operators at junctions of multiple line defects. Right: A state-operator map relates this to correlation functions on a twisted $S^3 \times_{\fq^2} S^1$ geometry with line defects inserted along a great circle of $S^3$ and wrapping $S^1$.}
\end{figure}
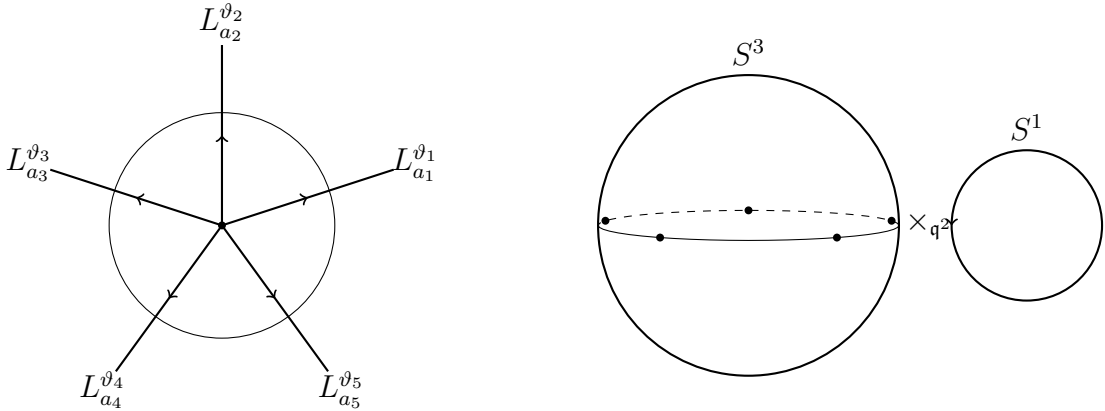

Supersymmetric line defects break the the $U(1)_r$ R-symmetry of the SCFT and thus occur in one-dimensional families $L^\vartheta_a$ rotated into each other by $U(1)_r$ rotations. Different members of the same family preserve different linear combinations of the bulk super-charges. The Schur correlation functions are defined by placing $L^{\vartheta_i}_{a_i}$ at the locations  $\vartheta_i$ on the great circle, so that a line defect will move along the family as the location of the line defect insertion is transported along the great circle of $S^3$. A full circuit along the great circle implements a $U(1)_r$ rotation by $2 \pi$. In Lagrangian SCFTs, the $U(1)_r$ charges which occur in the theory are integral, so that a $2\pi$ rotation is trivial. Accordingly, the $2 \pi$ rotation brings the line defect back to itself. In other SCFTs, such as Argyres-Douglas theories, a $2 \pi$ $U(1)_r$ rotation is non-trivial and gives a defect which preserves the same SUSY as the original one but may be different. 

We denote the effect of the $2\pi$ rotation as a map $a \to \rho^2(a)$, so that $L^{\vartheta}_a \equiv L^{\vartheta+ 2 \pi}_{\rho^2(a)}$. Then cyclic invariance is twisted as 
\begin{equation}
I_{a_1 \cdots a_n}(\fq) = I_{\rho^2(a_n) a_1 \cdots a_{n-1}}(\fq) \, .
\end{equation}
The line defect Schur indices are special cases of Schur correlation functions. The precise relation requires accounting for the opposite orientation of a line defect along the $S^1$ factor. 
A $U(1)_r$ rotation by $\pi$ applied to a line defect wrapping $S^1$ in the opposite direction gives a map $a \to \rho(a)$, so that $I_{a,b}(\fq)$ coincides with a Schur correlation function of $L_{\rho(a)}$ and $L_b$:
\begin{equation}
I_{a,b}(\fq) = I_{\rho(a) b}(\fq) =  I_{b\rho^{-1}(a)}(\fq)\, .
\end{equation}

We will now introduce a notation which anticipates another property of the Schur correlation functions: parallel line defects can be fused and 
the correlation functions are compatible with the fusion operation. A proper definition of the notion of fusion of line defects requires some care \cite{Kapustin:2006hi,Kapustin:2007wm, Gaiotto:2010be,Gaiotto:2023ezy}. We will review some salient aspects momentarily. For now, we recall that one can define a ``quantized K-theoretic Coulomb branch'' algebra $A_\fq$ 
with coefficients in $\bZ[\fq,\fq^{-1}]$, i.e. Laurent polynomials in $\fq$ with integral coefficients, and that wrapped supersymmetric line defects $L_a$ map to elements in $A_\fq$, which we will denote with the same symbol $a$ and refer to as the K-theory class of $L_a$. Then all correlation functions are encoded in 1pt functions $I_{a}(\fq)$ via the algebra relations:
\begin{equation}
	I_{a_1 \cdots a_n}(\fq) = I_{(a_1 \cdots a_n)}(\fq) \, .
\end{equation}
and $\rho$ is an algebra automorphism. 

We will thus define a twisted trace $\Tr$ on $A_\fq$ simply as 
\begin{equation}
	\Tr \, a \equiv I_{a}(\fq) \,
\end{equation}
The trace is twisted by $\rho^2$:
\begin{equation}
	\Tr \, a b = \Tr \,\rho^2(b) a \, ,
\end{equation}

 We are now ready to make a non-trivial claim, supported by the known explicit UV and IR formulae for line defect Schur indices: the pairing 
\begin{equation}
	\langle a|b \rangle \equiv \Tr \, \rho(a) b = I_{a,b}(\fq)
\end{equation}
is positive definite if $\fq \in [-1,1]$. This claim should follow from reflection positivity of the associated Schur two-point functions.

With a slight abuse of notation, we will also denote as $A_\fq$  the algebra over the real numbers obtained from $\bR \otimes A_\fq$ by specializing the variable $\fq$ to a real number between $-1$ and $1$. According to our initial discussion, we immediately gain a spherical unitary representation of the algebra double $\fD_\fq  = A_\fq \times A_\fq^\op$ on a real Hilbert space $\cH_\fq$ defined as the closure of $A_\fq$ under this inner product. 

\subsection{Non-conformal examples and holomorphic-topological twist}
The formulae employed to compute the Schur index and correlation functions apply equally well to non-conformal SQFTs and satisfy the properties described above, with an appropriate choice of $\rho$. This may be surprising, as the original superconformal index only makes sense in the conformal case. Intuitively, this happens because the Schur index does not make use of $U(1)_r$, which is broken for general SQFTs, but only of the Cartan subgroup of the $SU(2)_R$ R-symmetry, which is generically unbroken in the vacuum. 

A sharper justification employs the Holomorphic-Topological (HT) twist of 4d ${\cal N}=2$ SQFTs \cite{Kapustin:2006hi}. A reader interested only in algebraic aspects of our construction can safely skip this discussion and simply keep in mind that the Schur correlation functions technology applies to non-conformal theories as well.

The HT twist is a canonical modification of the physical theory which treats a specific nilpotent supercharge as a BRST charge. Accordingly, three out of four translation generators become gauge symmetries and the twisted theory treats two directions as topological and the remaining two as holomorphic. 
The Schur index ``counts'' local operators in the HT-twisted theory and is thus defined for generic SQFTs as long as the Cartan sub-algebra of the $SU(2)_R$ R-symmetry is unbroken. 

Although the HT twist is the natural setting for discussing many properties of the Schur index, a full discussion of this interesting topic goes well beyond the scope of this paper. We will briefly discuss here some expected properties of the HT twist, leaving a full discussion to future work. 

Even if the original physical theory is not conformal, the HT-twisted theory still enjoys a scale symmetry. Indeed, denoting the holomorphic coordinate as $z$, the only non-trivial part of a scale transformation is the re-scaling of $z$, which 
is implemented by the same $z\partial_z$ generator which implements rotations of the holomorphic plane. The rotation generator in the twisted theory is the combination of the physical rotation generator and of the 
Cartan $R$ of the $SU(2)_R$ R-symmetry . It is also useful to employ conventions where the ghost number grading/homological degree/fermion number is shifted by the $R$, so that the role of ``fermion number'' in indices is played by $(-1)^R$.

In these conventions, the Schur index is the Euler character of the complex of local operators in the HT twist of the physical theory \cite{niu2022local}. The partition function interpretation should also be available within the HT twist: the quotient of $\bR^2_t \times \bC_z$ by $(t,z,\bar z) \to (|\fq|^2 t, \fq^2 z , \bar \fq^2 \bar z)$ endows $S^1 \times_{\fq^2} S^3$ with an HT structure.  

Supersymmetric line defects map to topological line defects wrapping lines in the topological plane in the HT twisted theory.
We will always consider line defects supported at the origin in the complex plane and keep track of the $\bC^*$ rotation symmetry of the complex plane. Schur correlation functions can be defined as counting local operators at junctions of such topological defects or as correlation functions of circle-wrapped topological defects in $S^1 \times_{\fq^2} S^3$.

Recall that topological line defects generically form a category, with morphisms consisting of defect-changing local operators. 
Essentially by definition, circle-wrapped line defects only remember the K-theory class of the corresponding objects and so Schur correlation functions take as inputs {\it K-theory classes of line defects in an HT twist of the physical theory}. We identify the K-theory as $A_\fq$ and identify $\fq^2$ as the equivariant parameter for rotations of the complex plane. 

In the presence of a transverse topological direction, as is the case here, the category has a monoidal structure controlling the fusion of parallel line defects. It is also possible to rotate the support of a line in the topological plane and define a dualization functor $\rho$ which 
maps a line to a line rotated by $\pi$. A topological theory may be {\it framed}, in which case a rotation $\rho^2$ of $2 \pi$ fails to be the identity. Accordingly, $A_\fq$ is an algebra over $\bZ[\fq, \fq^{-1}]$ endowed with an endomorphism $\rho$. 

A full mathematical treatment of the category of line defects for the HT twist of Lagrangian gauge theories and of the associated Schur indices can be found in a series of papers \cite{cautis2018cluster,niu2022local,cautis2023canonical}.

This construction accounts nicely for all the expected properties of the Schur index for conformal or non-conformal theories, except for the crucial positivity property: reflection positivity is a property of the physical theory but not necessarily of the twisted theory. A proof for non-conformal theories would thus require one to write an explicit supergravity background defining the supersymmetric $S^1 \times_{\fq^2} S^3$ partition function for the physical theory and verify reflection positivity. We leave this to future work. Experimentally, positivity holds for non-conformal examples as well, with a $\rho$ discussed below. 
 
\subsection{Schur quantization as a quantization of the K-theoretic Coulomb branch(es)}
The algebra $A_\fq$ has classical limits $\fq \to \pm 1$. In these limits, it reproduces the Poisson algebra of holomorphic functions on two versions  $\cM_\pm$ of the moduli space $\cM$ of 3d Coulomb vacua for supersymmetric circle compactifications of the 4d theory. The two versions differ by the choice of spin structure and central $SU(2)_R$ holonomy placed on the circle \cite{Gaiotto:2010be,Gaiotto:2023ezy}. We will usually disregard this subtlety. The moduli space $\cM$ is a complex symplectic manifold. 

Keeping track of the star structure, the classical limit of $\fD_{\fq}$ reproduces the combined Poisson algebra of holomorphic and anti-holomorphic functions on $\cM$. 
Keeping track of the leading non-commutativity, we see that the holomorphic and anti-holomorphic Poisson brackets have opposite signs, i.e. they arise 
from the Poisson bracket defined in terms of the {\it imaginary} part of the complex symplectic form on $\cM$.\footnote{Notice that the real part of the complex symplectic form
is not exact and thus would not give rise to a continuous family of quantizations.}

We conclude that the quantum system $(\fD_\fq, \cH_\fq)$ provides a quantization of $\cM$ as a real phase space equipped with the {\it imaginary} part of the complex symplectic form.

Following considerations similar to these we employ for theories of class $\cS$ later in the paper, one may argue that the Schur correlation functions of a 4d theory can be recast
as disk correlation functions in an A-twist of the 2d supersymmetric sigma model with target $\cM$. Ultimately, this presents Schur quantization as a computable example of brane quantization \cite{Gukov:2008ve}.

\subsection{Lagrangian building blocks}
The Schur index for Lagrangian SQFTs receives contributions from hypermultiplets and vectormultiplets. 
It can be readily understood as counting gauge-invariant local operators built from BPS letters in the physical theory \cite{Gadde:2011uv}, or superfields of the HT theory \cite{niu2022local}. An hypermultiplet contributes holomorphic derivatives $\partial^n X$ and $\partial^n Y$ of the complex scalar fields, with twisted spin $n+\frac12$ and $(-1)^R =-1$, while vectormultiplets contribute two sets of fermionis generators $\partial^n U$ and $\partial^n V$, with with twisted spin $n+1$ and $(-1)^R =-1$ as well. 

Putting it all together, we get
\begin{equation} \label{eq:int}
	I(\fq;\mu) = \frac{1}{|W_G|}\oint_{|\zeta_i|=1} \prod_i \frac{(\fq^2)^2_\infty d\zeta_i}{2 \pi i \zeta_i} \Delta(\zeta) \frac{\prod_\alpha (\fq^{2} \zeta^\alpha;\fq^2)^2}{\prod_{w,w_f} (-\fq \zeta^w \mu^{w_f};\fq^2)} \, ,
\end{equation}
where $\zeta_i$ are valued in the Cartan torus of the gauge group and the products run over roots $\alpha$ and 
gauge and flavour weights $(w,w_f)$ of hypermultiplet scalars. The factor $\Delta(\zeta)$ is the appropriate Vandermonde determinant 
\begin{equation}
	\Delta(\zeta) \equiv \prod_{\alpha>0} \left(\zeta^{\frac{\alpha}{2}} - \zeta^{-\frac{\alpha}{2}} \right)^2
\end{equation}
for projecting on the character of gauge-invariant operators. We included flavour fugacities $\mu$ for completeness. In the following we will assume $|\mu|=1$.\footnote{This condition seems sufficient, but not strictly necessary for positivity of the Schur indices $I_{ab}$. Other reality conditions may also work. We will not explore this phenomenon.}
 
As a first step towards discussing positivity, observe that both roots and non-zero weights occur in 
opposite pairs, so that the integrand is positive definite when $|\zeta_i|=1$ and $|\mu|=1$. 

The simplest class of supersymmetric line defects are Wilson lines, labelled by an unitary representation $R$ of 
$G$. The insertion of a wrapped Wilson line $w_R$ in a Schur index results in the insertion in the integral (\ref{eq:int}) of 
the character $\chi_R(\zeta)$ of the corresponding representation. The Wilson lines satisfy $\rho(w_R) = w_{R^\vee}$ and
\begin{equation}
\chi_{R^\vee}(\zeta) = \overline{\chi_R(\zeta)} 
\end{equation} 
on the $|\zeta_i|=1$ integration locus. As a consequence, 
\begin{equation} \label{eq:wnorm}
	\Tr \rho(w_R) w_R = \oint_{|\zeta_i|=1} \prod_i \frac{(\fq^2)^2_\infty d\zeta_i}{2 \pi i \zeta_i} \Delta(\zeta) \frac{\prod_\alpha (\fq^{2} \zeta^\alpha;\fq^2)^2}{\prod_{w,w_f} (-\fq \zeta^w \mu^{w_f};\fq^2)}|\chi_R(\zeta)|^2 \, ,
\end{equation}
is positive and more generally $\Tr \rho(a) a >0$ manifestly for any linear combination $a$ of wrapped Wilson lines. 

The most general class of supersymmetric line defects in a Lagrangian gauge theory are 't Hooft-Wilson lines $\ell_{\lambda_m, \lambda_e}$. Naively, these are labelled by a labelled by a pair $(\lambda_m, \lambda_e)$ of magnetic and electric weights modulo the action of the Weyl group. In practice, monopole bubbling makes the definition subtle.\footnote{And so does the option to introduce more elaborate couplings between the gauge fields in the neighbourhood of the defect and auxiliary degrees of freedom supported on the defect. The notion of ``Koszul-perverse coherent sheaf'' from \cite{cautis2023canonical} appears to satisfactorily handle these subtleties in the HT twist of the theory. Simple Koszul-perverse coherent sheaves are labelled by the pair $(\lambda_m, \lambda_e)$ modulo Weyl.} 

The calculation of Schur correlation functions with insertions of non-zero magnetic weight is somewhat intricate and requires one to introduce some more formalism. K-theory classes of 't Hooft-Wilson lines and of the resulting K-theoretic quantum Coulomb branch algebra $A_\fq$ can be handled via the BFN formalism \cite{Braverman:2016wma}: elements of $A_\fq$ are represented as equivariant K-theory classes on a variant of the affine Grassmanian and the product is defined through certain correspondences. In practice, the generators of $A_\fq$ can be presented as multiplicative difference operators $D_a$ acting on a collection of formal variables $v_i$ \cite{Alday:2009fs,Drukker:2009id,Drukker:2009tz,Drukker:2010jp,Dimofte:2011py,Gomis:2011pf,Ito:2011ea,Gang:2012yr,Gaiotto:2014lma,Drukker:2015spa}. In the context of class ${\cal S}$ theories with a Lagrangian description, these differential operators match \cite{Allegretti:2024svn} the description of the Skein algebra $A_\fq = \mathrm{Sk}_\fq$ in terms of quantum Darboux coordinates $(u_i, v_i)$
\begin{align}
	u_i u_j &= u_j u_i \cr
	u_i v_j &= \fq^{2 \delta_{ij}} v_j u_i \cr
	v_i v_j &= v_j v_i \, ,
\end{align} 
of Fenchel-Nielsen type. 

Each difference operator $D_a$ is a sum of terms of the form $D^{(n)}_a(v) u^n$, which shift $v_i \to v_i \fq^{2 n_i}$ \cite{2019,schrader2019ktheoretic}. In particular, the $n=0$ term $D^{(0)}_a(v)$ is some rational function of the $v_i$. The prescription to compute a Schur correlation function with a single line defect insertion is then straightforward: insert $D^{(0)}_a(\zeta)$ in the integral (\ref{eq:int}). Correlation functions of multiple lines can be computed by first composing the respective difference operators.

We should remark that the original BFN construction requires the matter representation $M$ for the hypermultiplet scalar fields to be of cotangent form, i.e. to 
be a direct sum $T^*N$ of a representation $N$ and its dual $N^\vee$. The resulting K-theoretic Coulomb branch algebra is independent of this choice, but 
explicit ``Abelianized'' expressions as difference operators do depend on it. We will discuss momentarily how this dependence cancels out in index calculations. Matter of non-cotangent type 
can be handled by more refined means \cite{Braverman:2022zei,teleman2023coulomb}.

Cyclicity of the resulting twisted trace is far from obvious from this prescription. Based on examples, it should follow from contour deformations which are only unobstructed thanks to delicate cancellations between the poles and zeroes  in the integrand and in the $D_a'$. It would be very nice to formulate an abstract proof in the BFN language.\footnote{The proof is perhaps already implicitly given by the combination of dualizability results in \cite{cautis2023canonical} and the relation to the Schur index in \cite{niu2022local}.}
Notice that the integrand in the Schur index is a ratio of $\theta$ functions:
\begin{equation}
	\theta(x;\fq^2) \equiv (-\fq x;\fq^2)_\infty (-\fq x^{-1};\fq^2)_\infty \, ,
\end{equation}
which transform well under shifts
\begin{equation}
	\theta(\fq^2 x;\fq^2) \equiv (-\fq^3 x;\fq^2)_\infty (-\fq^{-1} x^{-1};\fq^2)_\infty = \fq^{-1} x^{-1} \theta(x;\fq^2) \, ,
\end{equation}
up to an overall factor. The overall factors of gauge fugacities accumulated under the shift from the numerators and denominators cancel out in a conformal theory. In a non-conformal theory, they combine to reproduce a non-trival $\rho^2$ twist of the trace expected from the following gauge theory considerations. 

Namely, the conformal symmetry anomaly in a 4d ${\cal N}=2$ gauge theory is closely associated to the anomaly in the $U(1)_r$ conformal symmetry. The sort of $2 \pi$ $U(1)_r$ rotation 
which would control the framing anomaly in the HT theory can be mapped to a shift in the $\theta$ angle of the theory. By the Witten effect, that results in a shift of the electric charge $\lambda_e$ of a 't Hooft-Wilson line by an amount proportional to the magnetic charge $\lambda_m$ and to the anomaly. The $\pi$ rotation functor $\rho$ flips the signs of both $(\lambda_m, \lambda_e)$ and shifts the electric charge by a certain multiple of the magnetic charges, depending on the specific value of the mixed $U(1)_r$-gauge anomaly coefficients and on precise labelling conventions for the line defects. We refer the reader to concrete examples in the next section.

Another feature which we see in concrete examples is that positivity of $\Tr \rho(a) a$ can also be demonstrated by a contour deformation to a contour where the measure is manifestly positive. This suggests that a combinatorial proof of positivity in the BFN language may be possible as well.

\subsection{An useful isometry}
It is often the case that protected partition functions such as the Schur indices can be factored out into pieces which correspond to a decomposition of the underlying geometry \cite{Drukker:2010jp,Cordova:2016uwk,Dedushenko:2018tgx}. In particular, the Schur index can be factored into two ``half-indices'' $I\!\!I_B(\zeta)$ associated to hemi-spheres with Dirichlet boundary conditions
\begin{equation} \label{eq:hint}
	I\!\!I_B(\zeta) = \delta_{B,0} \prod_i (\fq^2)_\infty \frac{\prod_\alpha (\fq^{2} \zeta^\alpha;\fq^2)}{\prod^N_{w,w_f} (-\fq \zeta^w \mu^{w_f};\fq^2)} \, ,
\end{equation}
glued together (as in a 3d superconformal index for a 3d $G$ gauge theory) by a $\zeta$ contour integral and a sum over magnetic charges $B$ on $S^2$:
\begin{equation} \label{eq:inn}
	I(\fq;\mu) = \frac{1}{|W_G|} \sum_{B \in \Lambda} \oint_{|\zeta_i|=1} \prod_i \frac{d\zeta_i}{2 \pi i \zeta_i} \Delta_B(\zeta) I\!\!I_B(\zeta^{-1}) I\!\!I_B(\zeta) \, .
\end{equation}
Here $\Lambda$ is the lattice of magnetic weights of $G$. The product $\prod^N$ in the definition of the half-index indicates that we have assumed of cotangent type $T^*N$, with $N$ being a representation of $G$, and we only include the weights for the $N$ half. In particular, $I\!\!I_B(\zeta)$ is invariant under simultaneous Weyl reflection of $\zeta$ and $B$.\footnote{If matter is not of cotangent type, the gauge theory has a potential anomaly. If the anomaly cancels, the K-theoretic Coulomb branch and Schur indices are well-defined but there are no Dirichlet boundary conditions which preserve the full $G$ symmetry, making the 3d gluing interpretation of the factorized formula unavailable. Nevertheless, the analysis below essentially goes through even if $N$ is not a representation of $G$. The main difference is that Weyl reflections will implemented via non-trivial transformations $R_{N,N'}$ described below. We expect that difference operator realizations of K-theoretic Coulomb branch generators preserving this modifiel Weyl symmetry will be available.} The factor $\Delta_B(\zeta)$ is a modification of the Vandermonde measure:
\begin{equation}
	\Delta_B(\zeta) \equiv \prod_{\alpha>0} \left(v^{\frac{\alpha}{2}} - v^{-\frac{\alpha}{2}} \right)\left(\wt v^{\frac{\alpha}{2}} - \wt v^{-\frac{\alpha}{2}} \right)\, ,
\end{equation}
where $v= \fq^{-B} \zeta$ and $\wt v = \fq^{B} \zeta$. 

This factorization resembles an inner product $\langle I\!\!I|I\!\!I\rangle$ in an auxiliary Hilbert space 
\begin{equation}
	\cH^{\mathrm{aux}}_\fq \equiv L^2\left(T\times \Lambda\right)^{W_G} \, ,
\end{equation}
where $T$ is the Cartan torus, $\Lambda$ the magnetic weight lattice, we use the modified Vandermonde measure $\Delta_B(\zeta)$ and consider Weyl-invariant wavefunctions only.

Remarkably, such a factorization works well with the insertion of line defects. One can formally define multiplication and shift operators
\begin{align}
	(u^m \psi)_B(\zeta) &=\psi_{B-m}(\fq^m \zeta)\cr 
	(v \psi)_B(\zeta) &= \fq^{-B} \zeta \psi_b(\zeta) \cr 
	(\wt u^m \psi)_B(\zeta)  &= \psi_{B-m}(\fq^{-m} \zeta)\rangle \cr
	(\wt v \psi)_B(\zeta)  &= \fq^{B} \zeta\psi_b(\zeta) \, , 
\end{align}
and specific expressions for $D_a$ and $\wt D_a$ in terms of these operators such that 
\begin{equation} \label{eq:innl}
	\Tr \, a = \langle I\!\!I|D_a|I\!\!I\rangle \, .
\end{equation}
The specific expressions for  $D_a$ and $\wt D_a$ depends on the choice of $N$. 
%The representation of $\wt D_a$ in terms of $\wt u$ and $\wt v$ 
%can be obtained from the representation of $D_a$ in terms of $u$ and $v$ for the opposite choice $N^\vee$ and $\fq \to \fq^{-1}$. 

Formally, the expected properties of the trace/Schur correlators should follow from a non-trivial interplay between the functional form of 
the half-index and the $D_a$'s
\begin{equation}
	D_a|I\!\!I\rangle = \wt D_a |I\!\!I\rangle \,
\end{equation}
as well as a formal adjoint-ness property $D(\rho(a))^\dagger = \wt D(a)$, which involves a non-trivial shift of the $\zeta$ integration contour. 

A concise way to express these relations is to say that the map $A_\fq \to \cH^{\mathrm{aux}}_\fq$ 
\begin{equation}
	\pi: a \to D_a|I\!\!I\rangle
\end{equation}
is an {\it isometry} $\pi$ with respect of the inner product $\langle a|b \rangle$. Taking the closure of $A_\fq$, 
this gives an isometry 
\begin{equation}
	\pi: \cH_\fq \to \cH^{\mathrm{aux}}_\fq: \qquad |a\rangle \to D_a|I\!\!I\rangle
\end{equation}

Bubbling phenomena make it hard to give any more detail about the $D_a$ which is theory-independent. The exception is the part of $D_a$ which contains the largest Abelian magnetic charges, which we could denote as 
\begin{equation}
	U_\lambda = F_\lambda(v) u^\lambda 
\end{equation}
for a magnetic charge $\lambda$. Up to an overall monomial, $F_\lambda$ precisely cancels
all the factors in $u^\lambda I\!\!I_B(\zeta)$ which would obstruct a contour deformation. Namely, 
\begin{equation}
	u^\lambda I\!\!I_B(\zeta) = \delta_{B,\lambda} \prod_i (\fq^2)_\infty \frac{\prod_\alpha (\fq^{2+\lambda\cdot \alpha} \zeta^\alpha;\fq^2)}{\prod^N_{w,w_f} (-\fq^{1+\lambda\cdot w} \zeta^w \mu^{w_f};\fq^2)} \,
\end{equation}
and 
\begin{equation}
	F_\lambda(v)= f_\lambda(v) \frac{\prod^N_{w,w_f} \prod_{n=\lambda\cdot w+1}^{-1} (1+\fq^{1+2 n} v^w \mu^{w_f})}{\prod_\alpha \prod_{n=\lambda\cdot \alpha}^{-1}(1-\fq^{2+ 2n} v^\alpha)}
\end{equation}
where we only include factors with $\lambda\cdot w<0$ or $\lambda\cdot \alpha<0$ and $f_\lambda(v)$ is some monomial. Hence 
\begin{equation}
	U_\lambda I\!\!I_B(\zeta) = f_\lambda(v) \delta_{B,\lambda} \prod_i (\fq^2)_\infty \frac{\prod_\alpha (\fq^{2+|\lambda\cdot \alpha|} \zeta^\alpha;\fq^2)}{\prod^N_{w,w_f} (-\fq^{1+|\lambda\cdot w|} \zeta^w \mu^{w_f};\fq^2)} \,
\end{equation}
The monomials $f_\lambda(v)$ satisfy some constraints described below, but express a potential ambiguity in deciding which dyonic line defects  should be considered ''bare'' 't Hooft lines with no electric charge: a change in conventions would redefine $f_\lambda(v)$ by a power of $v$. Powers of $\fq$ would similarly represent an ambiguity in defining rotation generators in the presence of the defect.

The definition of $\wt U_\lambda$ has the same structure, so that the auxiliary condition 
\begin{equation}
	U_\lambda |I\!\!I\rangle = \wt U_\lambda |I\!\!I\rangle 
\end{equation}
reduces to $f_\lambda(\fq^{-\lambda}\zeta) = \wt f_\lambda(\fq^{\lambda}\zeta)$ which can be satisfied by using the same monomials and adjusting the power of $\fq$. 

On the other hand, when we check the adjointness properties we will compare $\wt U_\lambda^\dagger$ and $U_{-\lambda}$. 
The former contains factors involving, say, $(\wt v^\dagger)^w= v^{-w}$ for positive $\lambda \cdot w$, as tilde variables shift fugacities 
in the opposite manner. The latter contains factors involving $v^w$ for negative $-\lambda \cdot w$. This is not a problem,
as $(1+x^a v^b) = x^a v^b(1+x^{-a} v^{-b})$, but the comparison generates an extra monomial for each factor, which ultimately feed into the non-trivial definition of $\rho$. 

The Abelianized formulae for $D_a$ are often described in the literature directly in terms of shift operators analogue to the $U_\lambda$'s,
constrained by 
\begin{equation}
	U_\lambda U_{\lambda'} = \frac{F_\lambda(v) F_{\lambda'}(\fq^{2 \lambda} v)}{F_{\lambda+\lambda'}(v)}U_{\lambda+\lambda'}
\end{equation}
If $\lambda \cdot w$ and $\lambda' \cdot w$ are both positive, the resulting factors does not enter in the ratio. If they are both negative, 
the resulting factors cancel out in the ratio. The ratio thus only contains contributions from $\lambda \cdot w>0$ and $\lambda' \cdot w<0$ 
or viceversa.  

In particular, the auxiliary algebra formed by $U_\lambda$ and $v$ can be opposite to the algebra formed by 
$\wt U_\lambda$ and $\wt v$, even though the $\wt F_\lambda$ factors have a structure analogous to that of 
 $F_{-\lambda}$ rather than $F_\lambda$.
 
 If $\lambda$ and $\lambda'$ are in the same alcove, so that $\lambda \cdot w$ and $\lambda' \cdot w$ have the same sign for all possible $w$, 
 then it is natural to impose $f_\lambda f_{\lambda'} = f_{\lambda + \lambda'}$, compatible with a convention where products of 't Hooft lines with no electric charge give back a 't Hooft line with no electric charge. Another reasonably natural requirement is to have Weyl-invariant expressions. 
 We will see in examples that it may be useful to relax the latter requirement slightly in order to avoid unpleasant square roots of fugacities. 

We should also observe that multiplicative unitary transformations by factors such as $\zeta^{w\cdot B}$ can be readily employed to redefine $u$'s 
by powers of $v$'s and thus $f_\lambda(v)$'s by some monomial to the power of $\lambda$, allowing for some irreducible freedom in 
choosing the $f_\lambda$'s.

\subsection{Changing $N$}
We can briefly discuss the dependence of this construction on the choice of $N$. We need a bit of notation to compare different ways to split the matter contributions in two halves:
\begin{equation} \label{eq:Dint}
	I\!\!I_B(\zeta;N) = \delta_{B,0} \prod_i (\fq^2)_\infty \frac{\prod_\alpha (\fq^{2} \zeta^\alpha;\fq^2)}{\prod^{N}_{w,w_f} (-\fq \zeta^w \mu^{w_f};\fq^2)} \,
\end{equation}
We should also distinguish the representations $D_a(N)$ and $\wt D_a(N)$ suitable for this choice and the corresponding isometry $\pi_N$.

We then define a collection of ``reflection'' unitary transformations on $\cH^{\mathrm{aux}}_\fq$
which acts as multiplication by 
\begin{equation}
	R_{N,N'} \equiv \frac{\prod^{N'}_{w,w_f} (-\fq \zeta^w \mu^{w_f};\fq^2)}{\prod^{N}_{w,w_f} (-\fq \zeta^w \mu^{w_f};\fq^2)}
\end{equation}
As each factor is either shared between numerator and denominator or appears with opposite fugacities in numerator and denominator, this is manifestly a phase. 
It satisfies 
\begin{equation} 
	I\!\!I_B(\zeta;N) = R_{N,N'} I\!\!I_B(\zeta;N')\,
\end{equation}
but also intertwines the corresponding representations of the $D_a$'s and $\wt D_a$'s as difference operators and thus the isometries $\pi_N$ and $\pi_{N'
}$.

\subsection{Wilson line spectral decomposition}
There is another, powerful perspective on this isometry. The Wilson lines $w_R$ and $\wt w_R$ are a collection of  
commuting normal operators acting on $\cH_\fq$. The images $\chi_R(v)$ and $\chi_R(\wt v)$ 
act diagonally on $\cH^{\mathrm{aux}}_\fq$, with one-dimensional distributional eigenspaces labelled by
points in 
\begin{equation}
	\frac{T\times \Lambda}{W_G} \, ,
\end{equation} 
where $T$ is the Cartan torus and $\Lambda$ the lattice of magnetic weights, and common eigenvalues 
\begin{equation}
	\chi_R(\fq^{-B} \zeta) \qquad \qquad \chi_R(\fq^{B} \zeta) 
\end{equation}

We can attempt a direct diagonalization of the action of Wilson lines on $\cH$. This is possible because we have a lot of information on the products of Wilson lines with more general line defects. 
We can start from the ring of Wilson lines, reproducing the representation ring:
\begin{equation}
	w_R w_{R'} = w_{R\otimes R'}
\end{equation}
The spectrum of this ring is the complexified Cartan torus modulo Weyl and we can easily write infinite formal linear combinations $|0;\zeta\rangle$
of $|w_R\rangle$'s such that 
\begin{equation}
	w_R |0;\zeta\rangle = \wt w_R |0;\zeta\rangle = \chi_R(\zeta) |0;\zeta\rangle 
\end{equation}
Hermiticity imposes $|\zeta|=1$. It is easy to see that the $|0;\zeta'\rangle$ states are delta-function normalizable: they literally map to multiples of $\delta$-function distributions in $\cH^{\mathrm{aux},W_G}_\fq$ supported at $B=0$ and $\zeta =\zeta'$ and Weyl images of that. 

More generally, if we label line defects $D_{m,e}$ by a magnetic weight $m$ and an electric weight $e$, we have  
\begin{align}
	w_R D_{m,e} &= \sum^R_{\lambda} \fq^{- m \cdot \lambda} D_{m,e+\lambda} + \cdots \cr
	D_{m,e}w_R  &= \sum^R_{\lambda} \fq^{m \cdot \lambda} D_{m,e+\lambda} + \cdots
\end{align}
where the sum is over weights in $R$ and the ellipsis denote terms with smaller magnetic charge. We can use a triangularity argument to 
recursively build states $|m;\zeta\rangle$ as linear combinations of $|D_{m,e}\rangle$, corrected by terms of lower magnetic charge,
which are formal eigenvectors of $w_R$ with eigenvalues $\chi_R(\fq^{-m} \zeta)$. 

The triangularity of the relation between $|D_{m,e}\rangle$ and $|m;\zeta\rangle$ strongly suggests that these states exhaust the spectrum
and that the isometry $\cH_\fq \to \cH^{\mathrm{aux},W_G}_\fq$ is really an isomorphism and gives the spectral decomposition of $\cH$ into 
one-dimensional distributional eigenspaces of the Wilson lines. 
%There is a simple multiplicative unitary transformation acting on $\cH_\fq^{\mathrm{aux}}$ which maps $|I\!\!I\rangle$ to 
%\begin{equation} \label{eq:hintp}
%	I\!\!I^+_B(\zeta) = \delta_{B,0} \prod_i (\fq^2)_\infty \frac{\prod_{\alpha>0} (\fq^{2} \zeta^\alpha;\fq^2)^2}{\prod_{w,w_f>0} (-\fq \zeta^w \mu^{w_f};\fq^2)} \,
%\end{equation}
%such that only positive powers (in the sense of positive roots) of $\zeta$ appears in a series expansion. Furthermore, after the unitary transformation, elementary 't Hooft-Wilson lines have the structure:
%\begin{equation}
%	D^+_{m,e} = \fq^{-m \cdot e} u^m v^e + \cdots + \mathrm{Weyl\,images}
%\end{equation}
%where $m$ is a principal magnetic weight, the ellipsis indicates terms with smaller magnetic charge in the principal chamber
%and the Weyl averaging is done with the Weyl symmetry inherited from the original $L^2\left(T\times \Lambda\right)^{W_G}$ description.

\subsection{Schur quantization and gauging}
We will now extend and generalize further the spectral decomposition statement. 
 
Consider now a generic theory $\cT$ with global symmetry $G$ and a theory $\cT/G$ obtained by gauging $G$. A general feature of 
Coulomb branches is that line defects of $\cT$ are inherited by $\cT/G$, except that Weyl-invariant combinations of flavour parameters for the $G$ symmetry are promoted to the corresponding $G$ Wilson lines. In order to express this fact, denote as $A_\fq[\cT,G]$ the result of promoting the Weyl-invariant combinations of flavour parameters in $A_\fq[\cT]$ to central elements. Then we have an algebra embedding $A_\fq[\cT,G] \to A_\fq[\cT/G]$. 

We can also promote the Schur trace $\Tr_{\cT}$ on $A_\fq[\cT]$ to a family of traces $\Tr^\mu_{\cT,G}$ on $A_\fq[\cT,G]$ which just maps the central elements back to specific values $\mu$. Then the trace of inherited operators is simply 
\begin{equation}
	\Tr_{\cT/G} a = \oint \frac{d\zeta}{2 \pi i \zeta}  \Delta(\zeta;\fq) \Tr^\zeta_{\cT,G} a \qquad \qquad a \in A_\fq[\cT,G]
\end{equation}
Here we denoted for brevity the full vectormultiplet contribution
\begin{equation}
	\Delta(\zeta;\fq) \equiv (\fq^2;\fq^2)_\infty^{2 \mathrm{rk}_G}\Delta(\zeta) \prod_\alpha (\fq^{2} \zeta^\alpha;\fq^2)^2
\end{equation}
We will now attempt to give a general characterization of the Schur quantization for $\cT/G$ in terms of the Schur quantization of $\cT$. We begin with the case of Abelian $G$.

If $G$ is Abelian, general operators in $A_\fq[\cT/G]$ will carry quantized magnetic charge $m$ so that they lie in $A_\fq[\cT,G]$ if $m=0$ and in some Harish-Chandra-like bimodules $M^{(m)}_\fq[\cT,G]$ otherwise. The $G$ Wilson lines of charge $e$ are multiplied by appropriate powers $\fq^{-2 m \cdot e}$ when brought across an operator of given magnetic charge. 

The trace will vanish unless the total magnetic charge vanishes and magnetic charge is additive under multiplication.
If $a$ has magnetic charge $m$ and $b$ has magnetic charge $-m$, we expect the comparison between $\Tr \,a b$ and $\Tr\, \rho^2 (b)a$ to require a contour integral shift of $\zeta \to \zeta \fq^{2m}$. When checking positivity for $a$ of magnetic charge $m$, we expect that contour integral for the inner product can be shifted to an intermediate contour $\zeta \to \zeta \fq^{m}$ so that 
\begin{equation}
	\Tr_{\cT/G} \rho(a) a  = \oint \prod_i \frac{(\fq^2;\fq^2)_\infty^2 d\zeta_i}{2 \pi i \zeta_i}  \Tr^{\fq^{-m}\zeta}_{\cT,G} \rho(a) a = \oint \frac{(\fq^2;\fq^2)_\infty^2 d\zeta_i}{2 \pi i \zeta_i} \Tr^{\fq^{m}\zeta}_{\cT,G}  a \rho^{-1}(a)
\end{equation}
has a positive integrand. Accordingly, we expect a positive-definite inner product
\begin{equation}
\Tr^{\fq^{-B}\mu}_{\cT,G} \rho(a) a = \langle a|b\rangle_{\mu,B}
\end{equation}
on $M^{(m)}_\fq[\cT,G]$, leading to the definition of Hilbert spaces $\cH_\fq[\cT;G]_{\mu,B}$ via $L^2$ completion.\footnote{As in the case of sphere quantization, a less hand-waving demonstration of positivity can likely be given by identifying elements of $M^{(m)}_\fq[\cT,G]$
as K-theory classes of line defects which end a ``vortex'' surface defect and the inner product as a Schur correlation function decorated by the vortex defect.}

In practice, we have re-written the inner product in $\cH_\fq[\cT/G]$ as a direct sum/integral 
\begin{equation}
	\langle a|b\rangle = \sum_B  \oint \prod_i \frac{(\fq^2;\fq^2)_\infty^2 d\zeta_i}{2 \pi i \zeta_i} \langle a^{(B)}|b^{(B)}\rangle_{\zeta,B}
\end{equation}
where the superscript denotes the part of magnetic charge $B$. This gives an explicit spectral decomposition of $\cH_\fq[\cT/G]$ in eigenspaces of $G$ Wilson lines: 
\begin{equation}
	\cH_\fq[\cT/G] = \oint^{\oplus}_{(S^1 \times \bZ)^{\mathrm{rk}_G} } \prod_i \frac{(\fq^2;\fq^2)_\infty^2 d\zeta_i}{2 \pi i \zeta_i}   \cH_\fq[\cT;G]_{\zeta,B}
\end{equation}
and predicts again that the Wilson line spectrum should be supported on the sequence of circles $(S^1 \times \bZ)^{\mathrm{rk} G}$, with $w_R = \chi_R(\fq^{B}\mu)$ and $\wt w_R = \chi_R(\fq^{-B}\mu)$.

If $G$ is not Abelian, we still expect an Abelianized presentation of $A_\fq[\cT/G]$ to be available, where operators are written as difference operators in $v$ whose coefficients are some sort of meromorphic elements in $M^{(m)}_\fq[\cT,H]$, with $H$ being the Cartan subgroup of $G$. We also expect a spectral decomposition of $\cH_\fq(\cT/G)$ under the action of $G$ Wilson lines, with a spectrum supported on $\frac{T\times \Lambda}{W_G}$  and eigenspaces built from states formally associated to Weyl-invariant combinations of $u^m v^e$ Abelian operators. The contour integral computing $\Tr_{\cT/G}$ from $\Tr^{\mu}_{\cT,G}$ should be identified with the spectral decomposition of the inner product as a direct sum/integral of inner products in individual eigenspaces.  

\subsection{Dualities and spectral problems}
Supersymmetric gauge theories often enjoy dualities, relating the same or different theories at different values of the couplings. 
Dualities typically reorganize the line defects of the theory, resulting in non-trivial algebra morphisms between the associated $A_\fq$
algebras.

The Schur index is independent of couplings and thus the identifications extend to identifications between traces and associated Hilbert spaces $\cH_\fq$. In particular, the Wilson lines of one theory will map to some collection of non-trivial commuting difference operators 
in the dual theory. Integrable systems which arise in such manner include the relativistic open Toda chain and the trigonometric quantum Ruijsenaars-Schneider model.

As we know the spectrum of Wilson lines in one description, we immediately gain a prediction for the joint spectrum of the dual collection of commuting difference operators, thus completely solving the spectral problem for these complex quantum integrable systems.

\subsection{Boundary conditions and states}
The definition of Schur index and Schur correlators can be extended to a situation where an half-BPS boundary is present. The Schur ``half-index'' counts BPS boundary local operators and is associated to an $S^1_{\fq^2} \times HS^3$ partition function, where $HS^3$ is an hemisphere. Half-BPS line defects can be added at points on a half-great circle in $HS^3$ intersecting the boundary $S^2$
at the poles. Quarter-BPS boundary line defects can also be added at the poles of the boundary $S^2$. 

The boundary line defects for a given choice of boundary give a left module $M_\fq$ and a right module $\wt M_\fq$ for $A_\fq$. 
A Schur correlation function $I\!\!I_{\wt m_0 a_1 \cdots a_n m_{n+1}}$ will depend on a sequence of wrapped lines of the form $\wt m_0 a_1 \cdots a_n m_{n+1}$ 
which is consistent with the algebra and module operations. In other words, it gives some linear map $\wt M_\fq \otimes_{A_\fq} M_\fq \to \bC[[\fq]]$. 

By definition, the $I\!\!I_{\wt m a m}$ correlation function gives a collection of distributional states $\langle m;\wt m|$ in $A^\vee_\fq$ such that the correlation function equals $\langle m;\wt m|a \rangle$. Also by definition, 
\begin{equation}
\langle b m;\wt m c|a \rangle = \langle m;\wt m|cab \rangle = \langle m;\wt m|c \wt b|a \rangle 
\end{equation}
so this definition a collection of distributional ``boundary states'', a map $M_\fq \otimes \wt M_\fq \to A^\vee_\fq$ which commutes appropriately with the $A_\fq \times A^\op_\fq$ action.

In a Lagrangian gauge theory with a Lagrangian boundary condition, these Schur half-indices can be readily computed. For example, for Neumann boundary conditions half of the integrand of the usual Schur index is replaced by the 3d superconformal index of the boundary degrees of freedom. 

For theories of class $S$, interesting boundaries and interfaces can be associated to certain three-manifolds $M_3$ with boundary $C$, possible decorated by skeins reproducing $M_\fq$ as Skein modules $\mathrm{Sk}_\fq(M_3)$ \cite{Dimofte:2013lba}. The above collection of states has the properties expected from the path integral of complex Chern-Simons theory on $M_3$, decorated with appropriate skeins $m$ and $\wt m$ of holomorphic and anti-holomorphic Wilson lines. 

\subsection{3d limits}
The 3d Coulomb branch algebra $A_\fq$ for 4d Lagrangian gauge theories is a ``trigonometric'' version of the Coulomb branch for 3d ${\cal N}=4$ gauge theories with the same 
gauge group and matter content. In practice, the difference operators which represent the Coulomb branch of the 3d theory can be obtained by a specific $\fq \to -1$ limit from these for the 4d theory. If we write 
\begin{equation}
	\fq = - e^{- \pi R} \qquad \qquad v_a= e^{-2 \pi R V_a}
\end{equation}
and take an $R \to 0$ limit at constant $V_a$, factors such as $(1- (-\fq)^n v_a)$ become 
\begin{equation}
	2 \pi R \left(V_a + \frac{n}{2} \right)  
\end{equation}
and the difference operators which would multiply $v_a$ by $\fq^n$ effectively shift $V_a$ by $\frac{n}{2}$. The ``trigonometric'' $D_a$ difference operators are thus mapped to 
``rational'' versions $D^{\mathrm{3d}}_a$. In the BFN language, this is the limit taking equivariant K-theory classes to equivariant cohomology classes. These define the quantized Coulomb branch algebra $A^{\mathrm{3d}}_{\hbar = 2 \pi R}$ for the 3d theory. 

In computing the $R \to 0$ limit of the Schur index, it is important to observe that the integrand can be expressed as a ratio of products of $\theta$ functions $\theta_4(iR z,\tau = i R)$ or 
$\theta_1(i R z,\tau = i R)$ and $\eta(\tau = i R)$, where $\zeta = e^{-2 \pi R z}$ is a product of $\zeta_a$ and $\mu$'s. These functions behave well under modular transformations $\tau \to - \tau^{-1}$, $z \to z \tau^{-1}$, so that the integrand can be re-written in terms of $\theta_2(z,i R^{-1})$ or $\theta_1(z,i R^{-1})$ and $\eta(i R^{-1})$.
These functions, in terms, have a simple $R \to 0$ behaviour at finite $z$: up to an overall $2^m \exp (2 \pi n/R)$ prefactor which we can drop, they go to 
$\cos \pi z$ or $\sin \pi z$ and $1$. These are the building blocks for a ``Coulomb branch'' protected sphere correlation function of the 3d theory.

As long as the Lagrangian gauge theory has ``enough matter'', so that the 3d limit is not ``bad'' in the sense of \cite{Gaiotto:2008ak}, the integrand is exponentially small along the $|\zeta|=1$ integration contour outside the range of finite $z$, so that the Schur correlation functions limit to the protected sphere correlation function of the 3d theory, which provide a positive twisted trace on $A^{\mathrm{3d}}_{2 \pi R}$. 

The positive trace on $A^{\mathrm{3d}}_{2 \pi R}$ can be used to define a ``Coulomb branch'' sphere quantization
associated to the 3d theory, with an Hilbert space $\cH^{\mathrm{3d}}_{2 \pi R}$ defined as the closure of $A^{\mathrm{3d}}_{2 \pi R}$ under the inner product given by the trace \cite{Gaiotto:2023hda}. We conclude that the $\fq \to 0$ limit in this situation maps the 
Schur quantization to Coulomb branch sphere quantization, in such a way that the spherical vector and the $|a\rangle$ dense basis go to the corresponding dense basis of $\cH^{\mathrm{3d}}_{2 \pi R}$.

It is often the case that a 3d ${\cal N}=4$ gauge theory admits a ``mirror'' description, with the Coulomb branch mapped to the ``Higgs branch'' of the mirror theory. Correspondingly, the Coulomb branch sphere quantization associated to the original theory maps to an ``Higgs branch'' sphere quantization in the mirror theory, which can be described geometrically. The Coulomb and Higgs presentations of the algebra $A^{\mathrm{3d}}_{2 \pi R}$ and the Hilbert space $\cH^{\mathrm{3d}}_{2 \pi R}$ are
typically very different. In particular, the Higgs branch presentation can take a geometric form, with an algebra of holomorphic differential operators acting on $L^2$-normalizable half-densities on some auxiliary space. 

Among the examples discussed in the next section, the cases of SQED$_1$, SQED$_2$ and $SU(2)$ with $N_f=4$ are particularly instructive in a 3d limit:
\begin{itemize}
	\item The SQED$_1$ sphere quantization leads to a Weyl algebra $A^{\mathrm{3d}}_{2 \pi R}$ acting as holomorphic polynomial differential operators on $L^2(\bC)$. The spherical vector becomes a Gaussian wavefunction $e^{-|x|^2}$. 
	\item The SQED$_2$ sphere quantization leads to an algebra $A^{\mathrm{3d}}_{2 \pi R}$ which is the central quotient 
	$B_m$ of $U(\mathfrak{sl}_2)$, with quadratic Casimir $-\frac14(1+m^2)$. The Hilbert space gives the corresponding irreducible spherical principal series representation of $SL(2, \bC)$, possibly realized as $L^2(\bC P^1,|K|^{1+ i m})$. The ``spherical vector'' is the unique $SU(2)$-invariant wavefunction on $\bC P^1$. 
	\begin{equation}
		(1+|x|^2)^{-2-2 i m}
	\end{equation}
	We will momentarily employ the $SL(2, \bC)$-twisted spherical vector 
	\begin{equation}
		\psi_m\left(x;{a \, b \choose c \, d}\right) \equiv (d + c x + b \bar x + a |x|^2)^{-2-2 i m}
	\end{equation}
	depending on a point ${a \, b \choose c \, d} \in \frac{SL(2, \bC)}{SU(2)}\equiv H_3^+$.
	\item The case of $SU(2)$ with $N_f=4$ is particularly rich. The algebra is the $SL(2)$ quantum Hamiltonian reduction of the product 
	\begin{equation}
	B_{m_1+ m_2} \times B_{m_1- m_2} \times B_{m_3+ m_4} \times B_{m_3- m_4}
	\end{equation}
	with elementary generators identified with products $J_i \cdot J_j$ of $\mathfrak{sl}_2$ generators from different factors. 
	Correspondingly, Hilbert space consists of twisted half-densities on the moduli space of four points on $\bC P^1$
	modulo $SL(2, \bC)$. The spherical vector is given as an average over $H^+_3$:
	\begin{equation}
		|1\rangle_{3d} = \int_{H_3^+} d\mathrm{Vol}_h \psi_{m_1+m_2}(x_1;h) \psi_{m_1-m_2}(x_2;h) \psi_{m_3+m_4}(x_3;h) \psi_{m_3-m_4}(x_4;h) 
	\end{equation}
\end{itemize}
An analogous formula holds for all theories of class $A_1$ associated to a sphere with regular punctures. Crucially, this coincides with the large $s$ ``minisuperspace approximation'' of the WZW partition function \cite{Teschner:1997fv}, which is the candidate spherical vector $|1\rangle_{\mathrm{Hol}}$. This
verifies our conjectural identification of Schur and Holomorphic quantizations in the $s \to \infty$ limit. 

\subsection{Surface defects and alternative twists.}

The Schur correlation functions could be further modified by the insertion of surface defects along a circle which links the 
great circle where the line defects are supported. In the HT twist picture, these would wrap the holomorphic plane at the origin 
of the topological plane. Localization formulae in gauge theories are modified in a minimal way by the insertion of the 
elliptic genus $\Theta(\zeta;\fq^2)$ of the extra 2d dof. 

If the 2d dof are compact, such as a collection of charged fermions, $\Theta(\zeta;\fq^2)$ will not have poles as a function of $\zeta$. 
It will also be quasi-periodic under shifts $\zeta \to \fq^2 \zeta$. Such a surface defect insertion will thus almost preserve the trace condition but 
 modify $\rho^2$ to some other $(\rho')^2$ in a manner similar to what extra 4d matter fields would accomplish.
 
There is no obvious reason for a surface defect insertion to preserve positivity. A necessary condition is likely that $\Theta(\zeta;\fq^2)$ is positive on the unit circle. We do not know a 
sufficient condition, even in physical terms. In concrete Abelian examples, positivity can be proven rigorously for certain choices of $\rho'$ and $\Theta(\zeta;\fq^2)$ \cite{Klyuev_2022}.
We will not explore the matter in depth here, but it will appear in some examples and in a comparison to the literature on complex quantum groups. 
 
\section{Examples of Schur Quantization}\label{sec:examples}
This section contains several examples of K-theoretic Coulomb branch algebras and Schur quantization. 
Further examples can be found in the Appendices. 

The first sequence of examples have $U(1)$ gauge group and a variable number of charged hypermultiplets. They illustrate the role of the quantum torus algebra and the 
effect of matter on $\rho$.  The last example, SQED$_2$, has the remarkable property that $A_\fq = U_{\fq}(\mathfrak{sl}_2)$
and thus will provide us with an interesting family of spherical unitary representations of a real form of the $*$-algebra double
\begin{equation}
	\mathfrak{D}_{\rm S} \equiv U_{\fq}(\mathfrak{sl}_2) \times U_{\fq}(\mathfrak{sl}_2)^\op \, .
\end{equation} 
defined via a specific choice of $\rho$. This theory (and analogues for other Lie algebras) helps explain the well-known appearance of quantum groups in the quantization of character varieties and Chern-Simons theories. We will discuss the quantum groups relevant for complex quantization here and in Section \ref{sec:qg}.

The second sequence of examples have $SU(2)$ gauge group. It includes class $\cS$ theories associated to the four-punctured sphere and one-punctured torus, which are the crucial examples in the quantization of character varieties. See also Section \ref{sec:charq}. We also discuss gauging some extra $U(1)$ symmetries to give
interesting quantum group representations. 
 
\subsection{Example: Pure $U(1)$ Gauge Theory}\label{sec:sqedzero}
This is a somewhat trivial example, but it introduces the quantum torus algebra $Q_\fq$, which is a building block for all UV and IR constructions. 
All fields are gauge-neutral, so the Schur index is just $I(\fq) = (\fq^2)^2_\infty$. 

The K-theoretic Coulomb branch $\cM$ of the theory is $\bC^* \times \bC^*$, parameterized by the classical vevs $u$ and $v$ of BPS 't Hooft and Wilson line defects. 
The complex symplectic form is $d \log u \wedge d \log v$. The imaginary part of the complex symplectic form 
\begin{equation}
	d \log u \wedge d \log v- d \log \bar u \wedge d \log \bar v 
	= d \left( \log |u|^2 d \log \frac{v}{|v|} - \log |v|^2  d \log \frac{u}{|u|}\right)\,.
\end{equation}
presents $\bC^* \times \bC^*$ as the cotangent bundle $T^*T^2$. 

Our circle of ideas is completed by identifying $\cM = \cM(GL(1),T^2)$ as the space of $\bC^*$ flat connections on a two-torus $C=T^2$, aka the phase space of complex Chern-Simons theory with gauge group $\bC^*$ compactified on $C=T^2$. This identification also matches the Lagrangian submanifold $\cM_c(GL(1),T^2)$ of flat $U(1)$ connections with the base $|u|=|v|=1$ of $T^*T^2$.

The natural quantization of $\cM$ is the space $\cH_\fq = L^2(T^2)$ of $L^2$-normalizable wavefunctions on $T^2$, with $\log |u|^2$ and $\log |v|^2$ acting as 
derivatives and $\cM_c(U(1),T^2)$ quantized as the constant wavefunction on $T^2$. Schur quantization will give an equivalent answer in a Fourier-transformed 
presentation $\cH_\fq = L^2(\bZ^2)$. 

Indeed, K-theory classes $x_{m,e} \equiv [L_{m,e}]$ of BPS 't Hooft-Wilson line defects in the theory are labelled by an electric charge $e$ and a magnetic charge $m$, both integral. 
The resulting algebra $A_\fq = Q_\fq[\bZ^2]$ is the quantum torus algebra 
\begin{equation}
	x_{a,b} x_{c,d}= \fq^{a d - b c} x_{a+c,b+d}
\end{equation}
We can also introduce generators 
\begin{align}
	u &= x_{1,0} \cr
	v &= x_{0,1}
\end{align}
which satisfy 
\begin{equation}
	u v = \fq^2 v u \, ,
\end{equation}
and 
\begin{equation}
	x_{a,b}= \fq^{- a b} u^a v^b = \fq^{a b} v^b u^a
\end{equation}

Following our prescription, we get 
 \begin{equation}
	\Tr \, x_{a,b}=\delta_{a,0} (\fq^2)^2_\infty \oint \frac{d \zeta}{2 \pi i \zeta} \zeta^b = (\fq^2)^2_\infty \delta_{a,0} \delta_{b,0} 
\end{equation}
with $\rho(x_{a,b}) = x_{-a,-b}$ and thus $\rho^2=1$. 

The corresponding inner product becomes 
\begin{equation}
	\langle a,b|c,d\rangle =(\fq^2)^2_\infty  \delta_{a,c} \delta_{b,d}
\end{equation}
We thus recognize $\cH_\fq = L^2(\bZ^2)$ with a constant measure $(\fq^2)^2_\infty$. The dense image of $Q_\fq$ in $\cH_\fq$ consists of compactly-supported wavefunctions in $L^2(\bZ^2)$. In particular, the spherical vector $|1\rangle = |0,0\rangle$ is supported at the origin. 

The unitary action of the $*$-algebra double $\fD[Q_\fq,\rho]$ is written explicitly as 
\begin{align}
	x_{a,b} |m,e\rangle &= \fq^{a e- b m} |m+a,e+b\rangle \cr 
	\wt x_{a,b} |m,e\rangle  &= \fq^{-a e+ b m} |m+a,e+b\rangle
\end{align}
via normal operators which satisfy $\wt x_{a,b}  = x_{-a,-b}^\dagger$. In particular,
\begin{align}
	u |m,e\rangle &= x_{1,0}  |m,e\rangle = \fq^e |m+1,e\rangle \cr 
	v |m,e\rangle &= x_{0,1}  |m,e\rangle = \fq^{-m} |m,e+1\rangle \cr 
	\wt u |m,e\rangle  &=\wt x_{1,0} |m,e\rangle  = \fq^{-e} |m+1,e\rangle \cr
	\wt v |m,e\rangle  &=\wt x_{0,1} |m,e\rangle  = \fq^{m} |m,e+1\rangle
\end{align}
In this basis, the spherical condition is clearly solved by $|1\rangle$ only.

A full Fourier transform $L^2(\bZ^2) \to L^2(T^2)$ reproduces the natural quantization of $T^* T^2$ and maps the spherical vector to the constant wave-function on $T^2$. 
It is perhaps useful to point out that the natural domain of definition of $u$ and $v$ becomes more subtle in that description and involves functions on $T^2$ which can be analytically continued to a certain domain in $\bC^* \times \bC^*$.  

Electric-magnetic duality is an important symmetry of Abelian gauge theories. Here it acts as an $SL(2,\bZ)$ transformation on the $(m,e)$ charge vector of the 
BPS line defects. It is a manifest symmetry of the quantum torus algebra and acts on $\cH_\fq = L^2(\bZ^2)$ unitarily. Via Fourier transform, it is mapped to a unitary
 mapping-class group action on $T^2$. It preserves the spherical vector. Indeed, the spherical vector is the only $SL(2,\bZ)$-invariant normalizable state. Other basis vectors belong to orbits labelled by the mcd of $(m,e)$. 

It can also be useful to do a partial Fourier-transform $L^2(\bZ^2) \to L^2(S^1 \times \bZ)$, mapping states to wavefunctions $\psi_B(\zeta)$: 
\begin{equation}
	|m,e\rangle \to \zeta^e \delta_{B,m}
\end{equation}
The spherical vector now maps to a wave-function $\delta_{B,0}$ and $A_\fq$ to wave-functions which are compactly supported on $\bZ$ and Laurent polynomials on $S^1$.
The elementary operators act as 
\begin{align}
	u \psi_B(\zeta) &=\psi_{B-1}(\fq \zeta) \cr 
	v \psi_B(\zeta)&=\fq^{-B} \zeta \psi_B(\zeta)\cr 
	\wt u \psi_B(\zeta)&= \psi_{B-1}(\fq^{-1}\zeta)\cr
	\wt v \psi_B(\zeta) &= \fq^{B} \zeta \psi_B(\zeta)
\end{align}
This representation of $Q_\fq$ via difference operators and multi-variable generalizations thereof are the basic building blocks of many constructions below.

\subsubsection{Spaces of positive traces}
It is also instructive to characterize the trace algebraically. In the absence of twist, i.e. $\rho^2=1$, the trace condition
\begin{equation}
	\Tr \, x_{a,b} x_{c,d}= \fq^{2a d - 2b c} \Tr \,x_{c,d} x_{a,b} =\fq^{2a d - 2b c} \Tr \,x_{a,b} x_{c,d}
\end{equation}
immediately implies that $\Tr \,x_{a,b} \simeq \delta_{a,0} \delta_{b,0}$, so the trace is essentially unique and it happens to be positive
if the overall coefficient is positive.

It is instructive to see what happens if we modify the choice of $\rho$. For example, consider $\rho(x_{a,b}) = x_{-a, -n a -b}$ for some integer $n$, so that
$\rho^2(x_{a,b}) = x_{a, b+ 2 n a}$.
Then it is easy to see that $\Tr \,x_{a,b} \simeq \delta_{a,0} t_b$ for some $t_b$. Furthermore, 
\begin{equation}
	t_b = \fq^{-b} \Tr' \, x_{1,b} x_{-1,0} = \fq^{-b} \Tr' \, x_{-1,- 2 n} x_{1,b} = \fq^{2 n-2b} t_{b-2 n} \, .
\end{equation}
This is solved by $t_b = t'_b \fq^{-\frac{b^2}{2n}}$ where $t'_b = t'_{b-2n}$. Notice that the behaviour of the coefficients for large $b$ is sharply different in the $n>0$ and $n<0$ cases.
The corresponding inner product is
\begin{equation}
	\Tr' \rho(x_{a,b}) x_{c,d} =\delta_{a,c} \Tr' x_{-a, -n a -b} x_{c,d} = \delta_{a,c} \fq^{\frac{n a^2}{2}-\frac{(d-b)^2}{2n}}t'_{d-b-n a}
\end{equation}
We can restrict our attention to 
\begin{equation}
	\Tr' \rho(x_{0,2 n r}) x_{0,2 n s} =\fq^{-2n (s-r)^2}t'_{0}
\end{equation}
Computing some determinants of sub-matrices easily show that this inner product fails to be positive definite if $\fq^{-2n}\geq 1$, i.e. $n>0$. For $n<0$, 
we can Fourier-transform the answer to write the inner product as an integral involving a theta function: 
 \begin{equation}
	\Tr \, x_{a,b}=\delta_{a,0} (\fq^2)^2_\infty \oint \frac{d \zeta}{2 \pi i \zeta} \zeta^b \Theta(\zeta;\fq)
\end{equation}
and express the positive-definiteness condition in terms of the location of the zeroes of $\Theta(\zeta;\fq)$ \cite{Klyuev_2022}, with families of solutions. 
This integral expression can be given an interpretation in terms of a Schur index decorated by a surface defect. We will not pursue this point further in this example, 
but it illustrates how the standard Schur trace is a unique edge case in the space of positive twisted traces.

\subsection{Example: SQED$_1$.}
This example illustrates how matter fields modify the properties of wrapped 't Hooft lines. The contribution to the Schur index of a single hypermultiplet is 
\begin{equation}
	I_{\mathrm{hyper}}(\zeta;\fq) = \frac{1}{(-\fq \zeta;\fq^2)_\infty (-\fq \zeta^{-1};\fq^2)_\infty} = \frac{1}{\prod_{n=0}^\infty (1+ \fq^{2n+1} \zeta)(1+ \fq^{2n+1} \zeta^{-1})}
\end{equation}
The Schur index itself evaluates to
\begin{equation}
	I_\fq = \oint_{|\zeta|=1} \frac{d\zeta}{2 \pi i \zeta} (\fq^2)^2_\infty I_{\mathrm{hyper}}(\zeta;\fq) = 1 - \fq^2 + \fq^6 - \fq^{12} + \fq^{20}-\fq^{30} + \cdots
\end{equation}
i.e.  \footnote{This formula and the one below is related to bosonization of a $\beta\gamma$ system.}
\begin{equation}
	I_\fq = \sum_{n=0}^\infty (-1)^n \fq^{n(n+1)}
\end{equation}

We find the expectation value of a single Wilson line $w_k$ of charge $k$ by inserting $\zeta^k$ in the integral:
\begin{equation}
	\Tr \, w_k = \sum_{n=|k|}^\infty (-1)^n \fq^{n(n+1)-k^2} = (-\fq)^{|k|} \sum_{n=0}^\infty (-1)^n \fq^{n(n+2|k| +1)} 
\end{equation}
We anticipate that $\rho$ maps Wilson lines to Wilson lines of the opposite charge. The matrix $\Tr \rho(w_i) w_j = \Tr w_{j-i}$ is positive definite by construction, as it controls integrals of the form 
\begin{equation}
	I_\fq = \oint_{|\zeta|=1} \frac{d\zeta}{2 \pi i \zeta} |f(\zeta)|^2 (\fq^2)^2_\infty I_{\mathrm{hyper}}(\zeta;\fq)\end{equation}
where $f(\zeta)$ is a Laurent polynomial in $\zeta$ and the integration measure is manifestly positive. 

\subsubsection{The algebra $A_\fq[\mathrm{SQED}_1]$.}
In order to describe the insertion of 't Hooft defects, we need an explicit description of the K-theoretic Coulomb branch algebra $A_\fq$.
We denote as $u_\pm = [L_{\pm 1,0}]$ the K-theory classes of elementary 't Hooft operators of magnetic charge $\pm 1$ and as $v=[L_{0,1}]$ the K-theory class of 
an elementary Wilson line with electric charge $1$. Then $w_n = [L_{0,n}] = v^n$ and we have relations:
\begin{align}
	u_\pm v &= \fq^{\pm 2} v u_\pm \cr
	u_+ u_- &= 1 + \fq v \cr
	u_- u_+ &= 1 + \fq^{-1} v
\end{align}
We will also use the following relations, which follow from a repeated application of the basic ones:
\begin{equation}
	u^k_+ u^k_- = (1 + \fq^{2k-1} v) \cdots (1 + \fq v) \qquad \qquad u^k_- u^k_+ = (1 + \fq^{-2k+1} v)\cdots 1 + \fq^{-1} v
\end{equation}
These relations are enough to reduce any polynomial in $u_\pm$ and $v^{\pm 1}$ to a $\fq$-dependent linear combination of 
\begin{equation}
	D_{a,b} \equiv \fq^{- a b} u_+ ^a v^b \qquad \qquad D_{-a,b} \equiv \fq^{a b} u_-^a v^b \qquad \qquad a \geq 0 \, .
\end{equation}
We identify these with K-theory classes of generic 't Hooft-Wilson lines $L_{a,b}$, giving a linear basis for $A_\fq$. 
We will describe $\rho$ momentarily.

\subsubsection{The norm of 't Hooft operators}
The trace defined by the Schur index is only non-vanishing if the total magnetic charge vanishes. We can compute 
\begin{align}
	D_{a,b} D_{-a,c} &=  \fq^{a c+ a b} (1 + \fq^{2a-1} v) \cdots (1 + \fq v) v^{b +c} \cr
	D_{-a,c} D_{a,b} &= \fq^{-a c- a b} (1 + \fq^{-2a+1} v) \cdots (1 + \fq^{-1} v) v^{b +c} \cr
	&=\fq^{-a c- a b- a^2} (1 + \fq^{2a-1} v^{-1}) \cdots (1 + \fq v^{-1}) v^{b +c+a} \, ,
\end{align}
for $a\geq 0$.

When we insert these expressions in the trace, these factors cancel factors in the denominator,
%\begin{align}
%	\Tr D_{a,b} D_{-a,c} &=  \oint_{|\zeta|=1} \frac{d\zeta}{2 \pi i \zeta} \frac{(\fq^2)^2_\infty}{\prod_{n=0}^\infty (1+ \fq^{2n+2a+1} \zeta)(1+ \fq^{2n+1} \zeta^{-1})} \fq^{a c+ a b}\zeta^{b +c} \cr
%	\Tr D_{-a,c} D_{a,b} &=  \oint_{|\zeta|=1} \frac{d\zeta}{2 \pi i \zeta} \frac{(\fq^2)^2_\infty}{\prod_{n=0}^\infty (1+ \fq^{2n+1} \zeta)(1+ \fq^{2n+2a+1} \zeta^{-1})} \fq^{-a c- a b- a^2} \zeta^{b +c+a} 
%\end{align}
and allow a shift the integration contours by a factor of $\fq^{\pm a}$ \cite{Gaiotto:2023kuc} to
\begin{align}
	\Tr D_{a,b} D_{-a,c} &=  \oint_{|\zeta|=1} \frac{d\zeta}{2 \pi i \zeta} \frac{(\fq^2)^2_\infty}{\prod_{n=0}^\infty (1+ \fq^{2n+a+1} \zeta)(1+ \fq^{2n+a+1} \zeta^{-1})} \zeta^{b +c} \cr
	\Tr D_{-a,c} D_{a,b} &=  \oint_{|\zeta|=1} \frac{d\zeta}{2 \pi i \zeta} \frac{(\fq^2)^2_\infty}{\prod_{n=0}^\infty (1+ \fq^{2n+a+1} \zeta)(1+ \fq^{2n+a+1} \zeta^{-1})} \zeta^{b +c+a} 
\end{align}
These formulae are fully compatible with positivity if we take 
\begin{equation}
	\rho(D_{-a,-b}) = D_{a,b} \qquad \qquad \rho(D_{a,b}) = D_{-a,-a-b} \qquad \qquad a \geq 0 \, .
\end{equation}
Then 
\begin{equation}
	\rho^2(D_{a,b}) = D_{a,a+b} \, ,
\end{equation}
as expected from the $U(1)_r$ anomaly and Witten effect. 

The choice of $\rho^2$ is also compatible with cyclicity:
\begin{equation}
\Tr D_{a,b} D_{-a,c} = \Tr \rho^2(D_{-a,c}) D_{a,b} = \Tr D_{-a,-a+c} D_{a,b}
\end{equation}

In a situation like this, where $\rho^2$ is not the identity, we cannot interprete the spherical vector as the quantization of an actual Lagrangian submanifold of phase space: the classical 
constraints $u_\pm = \wt u_\pm$, $v = \wt v$ and the reality conditions 
\begin{equation}
	\wt v = \bar v^{-1} \qquad \qquad \wt u_+ = \bar v^{-1} \bar u_-  \qquad \qquad \wt u_- = \bar u_+
\end{equation}
do not define a Lagrangian sub-manifold of $\cM$.\footnote{This is not uncommon: for example, the state $e^{- \frac{x^2}{\hbar}}$ in quantum mechanics on the real line 
satisfies complexified equations $p = i x$ which do not define an actual Lagrangian submanifold of phase space.}

This theory has a (somewhat subtle) class $\cS$ description where $C$ is a plane with an irregular singularity of rank $2$ at infinity. The non-trivial action of $\rho$ has a specific geometric meaning in that context, rotating the Stokes sectors at the irregular singularity by one step. 

\subsubsection{Two useful isometries}
Presenting $\cH_\fq$ as the closure of $A_\fq$ is a bit cumbersome, as the natural linear basis in $A_\fq$ is not orthogonal under the inner product. The integral expressions above and our general discussion suggest defining first an isometry $A_\fq \to L^2(\bZ \times S^1)$ by 
\begin{equation}
	|D_{a,b}\rangle = \delta_{B,a} \zeta^b \frac{(\fq^2)_\infty}{\prod_{n=0}^\infty (1+ \fq^{2n+|a|+1} \zeta)} 
\end{equation} 
These vectors are related by an invertible triangular change of basis to the orthogonal basis $\delta_{B,a} \zeta^b$ in $L^2(\bZ \times S^1)$ and thus should give an identification of $\cH_\fq$ with $L^2(\bZ \times S^1)$. 

This isometry maps the spherical vector to 
\begin{equation}
	|1\rangle = \delta_{B,0} \frac{(\fq^2)_\infty}{\prod_{n=0}^\infty (1+ \fq^{2n+1} \zeta)} 
\end{equation} 
We can now introduce the same operators $u,v$ and $\wt u, \wt v$ acting on (a dense domain in) $L^2(\bZ \times S^1)$ which we introduced in pure $U(1)$ gauge theory. 
It is easy to see that the isometry intertwines the action of ``$v$'' in $\cH_\fq$ and $L^2(\bZ \times S^1)$. We would like to relate the actions of $u_\pm$ in  $\cH_\fq$ and $u^{\pm 1}$ in $L^2(\bZ \times S^1)$. This is straightforward. If $a >0$, we have
\begin{align}
	u |D_{a,b}\rangle &= \delta_{B,a+1} \fq^b \zeta^b \frac{(\fq^2)_\infty}{\prod_{n=0}^\infty (1+ \fq^{2n+a+2} \zeta)} = \fq^b |D_{a+1,b}\rangle \cr
	u |D_{-a,b}\rangle &= \delta_{B,-a+1} \fq^b \zeta^b \frac{(\fq^2)_\infty}{\prod_{n=0}^\infty (1+ \fq^{2n+a+2} \zeta)} = \fq^b (1+ \fq v) |D_{-a+1,b}\rangle
\end{align}
i.e. 
\begin{equation}
	u |D_{a,b}\rangle = |u_+ D_{a,b} \rangle \, ,
\end{equation}
for all $a$ and $b$. On the other hand,
\begin{equation}
	 |u_- D_{a,b} \rangle = (1+ \fq^{-1} v) u^{-1} |D_{a,b}\rangle \, ,
\end{equation}
Similarly, 
\begin{align}
	 |D_{a,b}u_- \rangle  &=\wt u^{-1} |D_{a,b}\rangle \cr
	  |D_{a,b}u_+ \rangle  &=(1+ \fq^{-1} \wt v)\wt u |D_{a,b}\rangle
\end{align}
We have thus mapped the action of $A_\fq \otimes A_{\fq}^\op$ on $\cH$ to an action via difference operators on $L^2(\bZ \times S^1)$:
\begin{align}
	u_+ &= u \cr
	u_- &= (1+ \fq^{-1} v) u^{-1} \cr
	\wt u_+ &= (1+ \fq^{-1} \wt v)\wt u \cr
	\wt u_- &= \wt u^{-1}
\end{align}
These are two natural Abelianized BFN presentations of the K-theoretic Coulomb branch algebra. 
The natural domain of definition of these operators is the space of finite linear combinations of the $|D_{a,b}\rangle$. It would be interesting to compare this with natural choices of domain which could arise in a direct attempt at quantizing $\cM$.  

Observe that these expressions can be interpreted as a morphism of $*$-algebras $\fD_\fq \to \mathfrak{Q}_\fq \equiv \fD[Q_\fq,\rho]$ composed with the unitary action of 
$\mathfrak{Q}_\fq$ on $L^2(\bZ \times S^1)$. Perhaps confusingly, this is expressed as two {\it distinct} algebra morphisms $A_\fq \to Q_\fq$ and 
$A^\op_\fq \to Q^\op_\fq$. This is essentially unavoidable. This construction is a simple example of the IR formalism discussed in our companion paper \cite{next}.

Here we discussed one of two natural isometries $\cH \to L^2(\bZ \times S^1)$. There is a second isometry given by 
\begin{equation}
	|D_{a,b};-\rangle= \delta_{B,a} \zeta^{b+\mathrm{max}(a,0)} \frac{(\fq^2)_\infty}{\prod_{n=0}^\infty (1+ \fq^{2n+|a|+1} \zeta^{-1})} 
\end{equation} 
which instead satisfies 
%\begin{align}
%	u^{-1} |D_{a,b};-\rangle &= \delta_{B,a-1} \fq^{-b-a} \zeta^{b+a} \frac{(\fq^2)_\infty}{\prod_{n=0}^\infty (1+ \fq^{2n+a+2} \zeta^{-1})} = \fq^{-b}(1+ \fq^{-1} v) |D_{a-1,b};-\rangle \cr
%	u^{-1} |D_{-a,b};-\rangle &= \delta_{B,-a-1} \fq^{-b} \zeta^{b} \frac{(\fq^2)_\infty}{\prod_{n=0}^\infty (1+ \fq^{2n-a+2} \zeta^{-1})} = \fq^{-b} |D_{-a-1,b};-\rangle
%\end{align}
%i.e. 
\begin{align}
	u^{-1} |D_{a,b};-\rangle &= |u_- D_{a,b};- \rangle \cr
	(1+ \fq v) u |D_{a,b};-\rangle &= |u_+ D_{a,b};- \rangle\cr
	(1+\fq^{-1} \wt v^{-1}) \wt u^{-1}|D_{a,b};-\rangle &= |D_{a,b}u_- ;- \rangle \cr
	\fq^{-1} \wt v \wt u |D_{a,b};-\rangle &= |D_{a,b}u_+ ;- \rangle \, .
\end{align}

The manifestly unitary transformation on $L^2(\bZ \times S^1)$ defined by the  {\it complex quantum dilogarithm} multiplication kernel 
\begin{equation}
	\Phi_B(\zeta) = \zeta^{\mathrm{max}(B,0)} \prod_{n=0}^\infty \frac{1+ \fq^{2n+|B|+1} \zeta}{1+ \fq^{2n+|B|+1} \zeta^{-1}} = \prod_{n=0}^\infty \frac{1+ \fq^{2n-B+1} \zeta}{1+ \fq^{2n-B+1} \zeta^{-1}}
\end{equation}
intertwines the two isometries. 

\subsubsection{Other positive traces}
There is a general theory of positive traces for Abelian K-theoretic Coulomb branch algebras. Consider a modification of the integral formula for the 
Schur correlation function where we insert a theta function $\Theta[\zeta;\fq]$ in the measure. This modification changes sightly the behaviour of Hermitean conjugation 
on shift operators and thus gives rise to a new automorphism $\rho'$ and $(\rho')^2$. In particular, this gives 
\begin{equation}
	(\rho')^2(D_{a,b}) = \lambda^a D_{a,b- n a} \, ,
\end{equation}
for non-negative integer $n$ and appropriate constant $\lambda$. For example, the insertion of 
\begin{equation}
	\theta(\mu \zeta;\fq^2) = \prod_{n=0}^\infty (1+ \fq^{2n+1} \mu \zeta)(1+ \fq^{2n+1} \mu^{-1}\zeta^{-1})
\end{equation}
gives a trace with $n=0$ and non-trivial $\lambda$. This extra measure factor is positive either for $|\mu|=1$ or for real $\mu$. Identifying 
a range of values which gives a positive trace requires more work.  

\subsubsection{The q-deformed Weyl algebra and q-deformed metaplectic representation.}
The quantized Coulomb branch algebra for the 3d version of SQED$_1$ is the Weyl algebra. This is a key example of 3d mirror symmetry. Sphere quantization presents $L^2(\bC)$ as a spherical representation for a $*$-algebra double of the Weyl algebra \cite{Gaiotto:2023hda}. The Weyl algebra contains a specific central quotient of $U(\mathfrak{sl}_2)$ as the sub-algebra fixed by a reflection of the generators. Sphere quantization thus also provides a spherical unitary representation of a $*$-algebra double $\fD[U(\mathfrak{sl}_2),\rho]\equiv U(\mathfrak{sl}(2,\bC)_\bR)$, where $\rho$ reflects the generators. This coincides with the representation-theoretic notion of a spherical unitary representation of $\mathfrak{sl}(2,\bC)$, which contains a cyclic vector 
which is invariant under the compact $SU(2)$ subgroup of $SL(2,\bC)$. 

All of these properties persist in a q-deformed manner in the Schur quantization of SQED$_1$, with $q = \fq^2$. The algebra $A_\fq$ can be interpreted as a q-deformed version $W_\fq$ of the Weyl algebra, albeit with an extra property usually not included in the definition. Indeed, $u_\pm$ satisfy a q-deformed commutation relation:
\begin{equation}
	\fq^{-1} u_+ u_- - \fq u_- u_+ = \fq^{-1} - \fq
\end{equation}
and $v$ can be reconstructed from the combination $u_+ u_- -1$: 
\begin{equation}
	\fq^{-1}  u_+(u_+ u_- - 1) = u_+(\fq u_- u_+ - \fq)= \fq (u_+ u_- - 1) u_+
\end{equation}
The existence of an inverse $v^{-1}$ appears to extend the naive definition of $W_\fq$ in a natural manner. For example,
a typical representation of the q-deformed commutation relations involves the (Jackson) q-derivative:
\begin{equation}
	\partial_{\fq^2} f(x)\equiv  \frac{f(\fq^2 x) - f(x)}{\fq^2 x - x}
\end{equation}
i.e. $u_+ = (1-\fq^2)\partial_{\fq^2}$, $u_-=x$, and gives $v f(x) = \fq f(\fq^2 x)$ which is invertible. 

The automorphism
\begin{equation}
	\rho(v) = v^{-1} \qquad \qquad \rho(u_-) = u_+ \qquad \qquad \rho(u_+) = \fq v^{-1} u_-
\end{equation} 
defining the $*$-algebra double $\fD[W_\fq,\rho]$ explicitly uses $v^{-1}$. The action on 
$\cH_\fq$ is a q-deformation of the representation on $L^2(\bC)$. 

We can even find a q-deformation of the metaplectic representation: $W_\fq$ includes $U_{q^2}(\mathrm{sl}_2)$ generators:
\begin{align}
	E&=\frac{u_- v^{-1} u_-}{\fq^{-2} - \fq^2} \cr
	K &= v \cr
	F&= \frac{u_+^2}{\fq^2 - \fq^{-2}}
\end{align}
with fixed Casimir element
\begin{equation}
	E F + \frac{\fq^{-2}K + \fq^2 K^{-1}}{(\fq^{-2} - \fq^2)^2} = -\frac{\fq + \fq^{-1}}{(\fq^{-2} - \fq^2)^2}
\end{equation}
and 
\begin{align}
	\rho(E) &= - \fq^2 K F \cr
	\rho(K) &= K^{-1} \cr
	\rho(F) &= - \fq^{2} K^{-1} E \, ,
\end{align}
which defines a quantum group $*$-algebra double $\fD[U_{q^2}(\mathrm{sl}_2),\rho]$. Here we encounter for the first time the ``Schur'' version of a quantum group $U_{q^2}(\mathfrak{sl}(2,\bC)_\bR)$ to be associated to 
$SL(2,\bC)$ Chern-Simons theory.  

The conditions satisfied by the spherical vector $|1\rangle$ can be re-written as 
\begin{align}
	\left( \fq^2 E (K^\dagger)^{-1} + F^\dagger \right) |1\rangle &=0 \cr
	K K^\dagger |1\rangle &=|1\rangle \cr
	\left( \fq^2 F (K^\dagger)+ E^\dagger \right) |1\rangle &=0\, .
\label{qgrp-sph}\end{align}
%Essentially, the spherical vector is left invariant by the co-product of the quantum group generators and their Hermitean conjugates. 
We are going to show in Section \ref{sec:qg} that the combinations of generators appearing in \rf{qgrp-sph}
can be interpreted as quantum deformations of the  generators of the compact subgroup in a $SL(2,\bC)$ representation. 

The appearance of the quantum group in this example is somewhat exceptional. Next, we consider an example which is instead instrumental to understand the relation between $\fD[U_{q}(\mathrm{sl}_2),\rho]$ and $SL(2, \bC)$ Chern-Simons theory. 
Notice the different power of $q$ in the deformation parameter!

\subsection{Example: SQED$_2$.}
General Abelian gauge theories work in a very similar way as  SQED$_1$. The next simplest example,
$U(1)$ gauge theory with two flavours, will allow us to discuss an example with flavour. It also has a neat relation to the theory of representations of 
quantum groups. 

The Schur index is 
\begin{equation}
	I_\fq(\mu) = \oint_{|\zeta|=1} \frac{d\zeta}{2 \pi i \zeta} (\fq^2)^2_\infty I_{\mathrm{hyper}}(\mu \zeta;\fq)I_{\mathrm{hyper}}(\mu^{-1} \zeta;\fq) 
\end{equation}

\subsubsection{Algebraic structure}
The algebra $A_\fq$ is now expressed in terms of $w_n = v^n$ and two difference operators $u_+$ and $u_-$, acting as 
\begin{equation}
	u_\pm v = \fq^{\pm 2} v u_\pm \, ,
\end{equation}
which also satisfy
\begin{equation}
	u_+ u_- = (1 + \fq \mu v)(1 + \fq \mu^{-1} v) \qquad \qquad u_- u_+ = (1 + \fq^{-1} \mu v)(1 + \fq^{-1} \mu^{-1} v)
\end{equation}
We see here a factor for each hypermultiplet. This is an example of a general formula valid for all Abelian gauge theories. 

The $u_\pm$ generators represent elementary 't Hooft lines of charge $\pm 1$. The full set of 't Hooft-Wilson lines can be written as 
\begin{equation}
	D_{a,b} \equiv \fq^{- a b} u_+ ^a v^b \qquad \qquad D_{-a,b} \equiv \fq^{a b} u_-^a v^b \qquad \qquad a \geq 0 \, .
\end{equation}
This gives a linear basis for $A_\fq$. We also have $\rho(D_{-a,-b}) = D_{a,b}$ and $\rho(D_{a,b}) = D_{-a,-2 a-b}$.

\subsubsection{Schur correlators and $\cH_\fq$.}
All formulae for the Schur correlation functions are obvious variations of these for SQED$_1$. E.g. 
\begin{align}
	\Tr D_{a,b} D_{-a,c} &=  \oint_{|\zeta|=1} \frac{d\zeta}{2 \pi i \zeta} \frac{(\fq^2)^2_\infty}{\prod_{n=0}^\infty (1+ \fq^{2n+a+1} \mu^\pm \zeta^\pm)} \zeta^{b +c} \cr
	\Tr D_{-a,c- 2 a} D_{a,b} &=  \oint_{|\zeta|=1} \frac{d\zeta}{2 \pi i \zeta} \frac{(\fq^2)^2_\infty}{\prod_{n=0}^\infty (1+ \fq^{2n+a+1} \mu^\pm \zeta^\pm)} \zeta^{b +c} 
\end{align}
where the $\pm$ notation in the denominators indicates a product over four factors with all possible signs. 

We can also define an isometry $A_\fq \to L^2(\bZ \times S^1)$ by 
\begin{equation}
	|D_{a,b}\rangle = \delta_{a,B} \zeta^b \frac{(\fq^2)_\infty}{\prod_{n=0}^\infty (1+ \fq^{2n+|a|+1} \mu \zeta) (1+ \fq^{2n+|a|+1} \mu^{-1} \zeta)} 
\end{equation} 
such that $u_+$ maps to $u$ and $\wt u_-$ to $\wt u^{-1}$. More explicitly,
\begin{align}
	u_+ &= u \cr
	u_- &= (1+ \fq^{-1} \mu v)(1+ \fq^{-1} \mu^{-1} v) u^{-1} \cr
	\wt u_+ &= (1+ \fq^{-1} \mu \wt v) (1+ \fq^{-1} \mu^{-1} \wt v)\wt u \cr
	\wt u_- &= \wt u^{-1}
\end{align}
Again, the triangular form of $|D_{a,b}\rangle$ indicates that they will be dense in $L^2(\bZ \times S^1)$, identifying this auxiliary Hilbert space with $\cH_\fq$.

There are actually four natural isometries to $L^2(\bZ \times S^1)$, intertwined by $\Phi_B(\mu^{\pm 1}\zeta)$. In each isometry, denominator factors capture half of the contribution of one hypermultiplet to the full integrand in the Schur index.

\subsection{Relation to quantum groups.}
The SQED$_2$ theory has an exceptional feature: $A_\fq$ coincides with the central quotient of $U_q(\mathfrak{sl}_2)$, with quadratic Casimir controlled by $\mu$. In order to make this explicit, observe 
%\begin{equation}
%	u_+ (\fq v^{-1} u_-) = \fq^{-1} v^{-1} + \mu + \mu^{-1} + \fq v \qquad \qquad (\fq v^{-1} u_-) u_+ = \fq v^{-1} + \mu + \mu^{-1} + \fq^{-1} v
%\end{equation}
%and thus 
\begin{equation}
	[\fq v^{-1} u_-,u_+] =(\fq^{-1}- \fq)(v- v^{-1}) \, , 
\end{equation}
so that we could define, say, 
\begin{align}
	E &= \frac{\fq v^{-1} u_-}{\fq^{-1}- \fq} \cr
	K &= v \cr
	F &=\frac{u_+}{\fq-\fq^{-1}}
\end{align}
to get the standard quantum group generators. 
The remaining relation sets the Casimir to be proportional to $\mu + \mu^{-1}$. We have again
\begin{align}
	\rho(E) &=- \fq K F \cr
	\rho(K) &= K^{-1} \cr
	\rho(F) &=- \fq K^{-1} E
\end{align}
so the $*$-algebra double is the central quotient of $\fD[U_{q}(\mathrm{sl}_2),\rho]$.

We have obtained a spherical unitary representation of $\fD[U_{q}(\mathrm{sl}_2),\rho]$ on a $\cH_\fq$  
which will  be identified in Section \ref{sec:qg} as a quantum deformation of the 
spherical principal series representation of $SL(2,\bC)$. 
This is to be expected, as the latter arises from sphere correlation functions of the 3d version of SQED$_2$ \cite{Gaiotto:2023hda}. 
The spherical vector $|1\rangle$ in $\cH_\fq$  is annihilated by certain combinations of the $U_{q}(\mathfrak{sl}_2)$ and $U^\op_{q}(\mathfrak{sl_2})$ generators, cf. \rf{qgrp-sph}. It will furthermore be shown in Section \ref{sec:qg} that the combinations of generators 
annihilating the spherical vector in   \rf{qgrp-sph} generate a quantum deformation of the Lie-algebra of the
compact sub-group of $SL(2,\bC)$. 
This will be shown to imply an algebraic structure 
of the representation on $\cH_\fq$ akin to the structure of principal series representation of $SL(2,\bC)$
as direct sum of finite-dimensional representations of a compact $SU(2)$ subgroup.

We can do more. We can gauge a $U(1)$ global symmetry acting on one of the two hypermultiplets, mapping the system to two copies of SQED$_1$. 
Accordingly, we map $A_\fq[\mathrm{SQED}_2]\to A_\fq[\mathrm{SQED}_1]\times A_\fq[\mathrm{SQED}_1]$, with $\mu$ mapping to a Wilson line:
\begin{align}
	\mu v &= v_1 \cr
	\mu^{-1} v&= v_2\cr
	u_+ &= u_{+,1} u_{+,2} \cr
	u_- &= u_{-,1} u_{-,2} 
\end{align}
We can diagonalize the Wilson line $v_1 v_2$ acting $\cH_\fq[\mathrm{SQED}_1]\times \cH_\fq[\mathrm{SQED}_1]$, say in an auxiliary description as $L^2(\bZ \times S^1) \times L^2(\bZ \times S^1)$. We have eigenvalues $\mu^2 \fq^{-M}$:
\begin{equation}\label{specdec1}
	\cH_\fq[\mathrm{SQED}_1]\times \cH_\fq[\mathrm{SQED}_1] = \sum_{M \in \bZ} \int_{\mu \in S^1} \cH_{M,\mu}
\end{equation} 
Each summand $\cH_{M,\mu}$ can be identified with a copy of $L^2(\bZ \times S^1)$  equipped with the action
\begin{align}
	u_+ &= u \cr
	u_- &= (1+ \fq^{-1} \fq^{- M/2} \mu v)(1+ \fq^{-1} \fq^{M/2} \mu^{-1} v) u^{-1} \cr
	\wt u_+ &= (1+ \fq^{-1} \fq^{M/2}\mu \wt v) (1+ \fq^{-1} \fq^{- M/2}\mu^{-1} \wt v)\wt u \cr
	\wt u_- &= \wt u^{-1}\, .
\end{align}
The decomposition \rf{specdec1} is expected to be a $q$-analogue of the decomposition of $L^2(\bC^2)$ into principal series representations of $SL(2,\bC)$, each appearing twice.  
Each $\cH_{M,\mu}$ gives an unitary representation of $\fD[U_{q}(\mathrm{sl}_2),\rho]$, with with Casimirs built from $\fq^{\mp M/2}\mu$. %These representations will be further studied in Section \ref{sec:qg}.
In our companion paper \cite{next} we will describe their braided monoidal structure in analogy to \cite{kashaev2001spectrum,Nidaiev:2013bda,schrader2019cluster,schrader2017continuous}. As for the case of quantum Teichm\"uller theory, this will allow us to use quantum groups to describe the braided monoidal category of line defects in complex Chern-Simons theory. 

%We expect the coproduct of the $U_{\fq}(\mathfrak{sl}_2)$ and $U_{\fq^{-1}}(\mathfrak{sl_2})$ generators to still act as a  
%compact sub-algebra, with finite-dimensional representations each occurring once. 

This theory is an elementary building block in an important construction. Consider any theory $\hat \cT$ which contains an $SU(2)$ gauge group coupled to both SQED$_2$ and to some other degrees of freedom, described by a theory $\cT$ with $SU(2)$ global symmetry. The K-theoretic Coulomb branch algebra $\hat A_\fq$ for $\hat \cT$ will then contain both $A_\fq$ and  $U_q(\mathfrak{sl}_2)$, with the mass parameters in $A_\fq$ promoted to $SU(2)$ Wilson lines and identified with the center of $U_q(\mathfrak{sl}_2)$. The automorphism $\rho$ 
for $\hat \cT$ will act as the standard $\rho$ on both sub-algebras. Schur quantization will thus provide a simultaneous unitary representation of both $\fD[U_{q}(\mathrm{sl}_2),\rho]$
and $\fD[A_\fq,\rho]$.

In a class $S$ context, $\hat \cT$ will typically be associated to a Riemann surface with an irregular singularity of rank $1$ and $\cT$ to the same Riemann surface with the 
irregular singularity replaced by a regular singularity. The $U_\fq(\mathfrak{sl}_2)$ generators quantize the Stokes data at the puncture and the Casimir generator quantizes the holonomy around the puncture \cite{schrader2019cluster}. The statement generalizes to other ADE groups, leading to analogous consequences for the representation theory of complex quantum groups.\footnote{More precisely, one expects the existence of a family of theories $T_{4d}[\fg]$ which can play the same role for $U_q(\fg)$. They are only known for $\mathfrak{sl}_n$ as 
4d lifts of $T[SU(n)]$. See \cite{Gaiotto:2023ezy} for some details and more citations.}  This leads to a variety of constructions which give a physical interpretation to the relation between quantum groups and 
the quantization of character varieties and Chern-Simons theory, see e.g. \cite{Witten:1989rw,Gaiotto:2014lma,schrader2019cluster}. Schur quantization leads to analogous statements about 
$\fD[U_{q}(\mathrm{sl}_2),\rho]$ and complex Chern-Simons theory.

The Hilbert space $\cH_\fq[\hat T]$ will have a spectral decomposition into 
eigenspaces of Wilson lines for the new $SU(2)$ gauge group. We expect the spectral decomposition to take the form 
\begin{equation}
	\cH_\fq[\hat T] = \int_{\frac{S^1 \times \bZ}{\bZ_2}} \cH_{M,\mu} \otimes \cH^{M,\mu}_\fq[T]
\end{equation}
with $\cH_{M,\mu}$ being the above principal series representations of $\fD[U_{q}(\mathrm{sl}_2),\rho]$ and $\cH^{M,\mu}_\fq[T]$ defining a larger class of representations for 
$\fD[\fA,\rho]$. 

\subsection{Pure $SU(2)$ gauge theory}
In the Appendices we discuss the example of a pure $U(N)$ gauge theory. For $SU(2)$ or $PSU(2)$ gauge group, one encounter subtleties related to the choice of global form of the gauge group and of a collection of mutually local line defects. We will ignore these subtleties, at the price of square roots of $\fq$ entering formulae and occasional negative signs appearing is unexpected places (but not spoiling positivity). 
We will assume $\fq>0$ for simplicity, so that $\fq^{\frac12}$ is real. Essentially, we consider an algebra $A_\fq$ which has sub-algebras which correspond to the K-theoretic Coulomb branch algebras for 
either $SU(2)$ or $PSU(2)$ gauge theories. The algebra is very well-understood, allowing us to present an explicit full linear basis. 

Recall that this is a class $\cS$ example with Lie algebra $\mathfrak{sl_2}$ and $C$ being a cylinder with irregular singularities of ``rank $\frac12$'' at both ends. 
Wilson lines in the $SU(2)$ gauge theory map to traces of holonomies around the cylinder, while 't Hooft lines map to regularized holonomies from one end to the other 
of the cylinder. See \cite{Gaiotto:2010be} for details. 

The Schur index is 
\begin{align}
	I_\fq &= \frac12 \oint_{|\zeta|=1} \frac{d\zeta}{2 \pi i \zeta} (1-\zeta^2)(1-\zeta^{-2})(\fq^2)^2_\infty (\fq^2 \zeta^2;\fq^2)^2_\infty (\fq^2 \zeta^{-2};\fq^2)^2_\infty = \cr &=1 + \fq^4 + \fq^{12} + \fq^{24} + \cdots = \sum_{n=0}^\infty \fq^{2 n(n+1)}
\end{align}
The insertion of Wilson lines $w_n$ of spin $n/2$ adds a character $\zeta^n + \zeta^{n-2} +\cdots + \zeta^{-n}$ to the integrand. E.g. $\Tr \, w_1=0$ and 
\begin{align}
	\Tr \, w_1^2 &= \frac12 \oint_{|\zeta|=1} \frac{d\zeta}{2 \pi i \zeta} (\zeta + \zeta^{-1})^2 (1-\zeta^2)(1-\zeta^{-2})(\fq^2)^2_\infty (\fq^2 \zeta^2;\fq^2)^2_\infty (\fq^2 \zeta^{-2};\fq^2)^2_\infty = \cr &=2 I_\fq - \sum_{n=-\infty}^\infty \fq^{2 n^2} = \sum_{n=1}^\infty \fq^{2 n(n-1)}(1-\fq^{2n})^2 
\end{align}

\subsubsection{The algebra}
The ``Abelianized'' description of $A_\fq$ involves auxiliary generators $v^{\pm 1}$, $u_\pm$ such that Wilson lines map to characters $w_n = v^{n} + v^{n-2}+\cdots + v^{-n}$ and the following relations hold: 
\begin{align}
	u_\pm v &= \fq^{\pm 1} v u_\pm \cr
	u_+ u_- &= \frac{1}{(v - v^{-1})(\fq v - \fq^{-1} v^{-1})} \cr
	u_- u_+ &= \frac{1}{(v - v^{-1})(\fq^{-1} v - \fq v^{-1})} 
\end{align}
Notice the single factor of $\fq$ in the first relation. This is precisely due to the choice to include both ``minimal'' electric and magnetic charges. 
The algebras for $SU(2)$ or $SO(3)$ gauge theories will be obtained by dropping either 't Hooft operators of Wilson lines of odd charge. 

The 't Hooft-Wilson operators of minimal magnetic charge do not suffer of monopole bubbling effect and are simply written as 
\begin{equation}
	H_a = \fq^{\frac{a}{2}} v^a u_+ +  \fq^{\frac{a}{2}} v^{-a} u_-
\end{equation}
in terms of the auxiliary variables. The 't Hooft-Wilson line defects of higher magnetic charge have more complicated rational expressions, which can be recovered by from products of $H_a$'s. We will come back to these momentarily. For now, we compute 
\begin{equation}
	H_a H_b= \fq^{\frac{a}{2}+\frac{3b}{2}} v^{a+b} u_+^2  +  \fq^{\frac{a}{2}-\frac{b}{2}}  \frac{v^{b-a}}{(v - v^{-1})(\fq^{-1} v - \fq v^{-1})}   +\fq^{\frac{a}{2}-\frac{b}{2}}  \frac{v^{a-b} }{(v - v^{-1})(\fq v - \fq^{-1} v^{-1})} +  \fq^{\frac{a}{2}+\frac{3 b}{2}} v^{-a-b} u_-^2
\end{equation}

The two middle terms are inserted in the integral expression for $\Tr H_a H_b$, leading to two contributions:
\begin{align}
	&\frac12 \oint_{|\zeta|=1} \frac{d\zeta}{2 \pi i \zeta}(\fq^2)^2_\infty  \fq^{\frac{a}{2}-\frac{b}{2}+1} \zeta^{b-a-2} \prod_{n=0}^\infty (1-\fq^{2n} \zeta^2) (1-\fq^{2n+2} \zeta^{-2}) (1-\fq^{2n+2} \zeta^2) (1-\fq^{2n+4} \zeta^{-2}) \cr
	&\frac12 \oint_{|\zeta|=1} \frac{d\zeta}{2 \pi i \zeta}(\fq^2)^2_\infty \fq^{\frac{a}{2}-\frac{b}{2}+1} \zeta^{a-b+2}  \prod_{n=0}^\infty (1-\fq^{2n+2} \zeta^2) (1-\fq^{2n} \zeta^{-2}) (1-\fq^{2n+4} \zeta^2) (1-\fq^{2n+2} \zeta^{-2})
\end{align}
The integration contours can be shifted to 
\begin{align}
	&\frac12 \oint_{|\zeta|=1} \frac{d\zeta}{2 \pi i \zeta}(\fq^2)^2_\infty \zeta^{b-a-2} \prod_{n=0}^\infty (1-\fq^{2n+1} \zeta^2) (1-\fq^{2n+1} \zeta^{-2}) (1-\fq^{2n+3} \zeta^2) (1-\fq^{2n+3} \zeta^{-2}) \cr
	&\frac12 \oint_{|\zeta|=1} \frac{d\zeta}{2 \pi i \zeta}(\fq^2)^2_\infty \zeta^{a-b+2}  \prod_{n=0}^\infty (1-\fq^{2n+1} \zeta^2) (1-\fq^{2n+1} \zeta^{-2}) (1-\fq^{2n+3} \zeta^2) (1-\fq^{2n+3} \zeta^{-2})
\end{align}
and combined to find
\begin{equation}
	\Tr H_{a-2} H_b = \oint_{|\zeta|=1} \frac{d\zeta}{2 \pi i \zeta}(\fq^2)^2_\infty \frac12 (\zeta^{b-a}+ \zeta^{a-b} ) (\fq \zeta^2;\fq^2)(\fq \zeta^{-2};\fq^2)(\fq^3 \zeta^2;\fq^2)(\fq^3 \zeta^{-2};\fq^2)
\end{equation}
This is compatible with the expected 
\begin{equation}
\rho(H_a) = H_{a-2} \, ,
\end{equation}
which implies norms $\Tr H_{a-2} H_a$ will have a positive integrand. 

We can look more carefully at products of two $H$'s to understand 't Hooft operators of non-minimal charge. We repeat here the crucial formula:
\begin{equation}
	 \fq^{\frac{a}{2}-\frac{b}{2}}   H_a H_b= \fq^{a+b} v^{a+b} u_+^2  +  \fq^{a-b}  \frac{\fq \frac{v^{b-a+1}- v^{a-b-1}}{v - v^{-1}} +\fq^{-1}\frac{v^{a-b+1}- v^{b-a-1}}{v - v^{-1}} }{(\fq^{-1} v - \fq v^{-1})(\fq v - \fq^{-1} v^{-1})}   +  \fq^{a+b} v^{-a-b} u_-^2
\end{equation}
If we specialize to $a=b$, we get an elementary 't Hooft operator $H^{(2)}_{2a}$ of magnetic charge $2$ and even electric charge: 
\begin{equation}
	 H^2_a= \fq^{2a} v^{2a} u_+^2  +  \frac{\fq +\fq^{-1}}{(\fq^{-1} v - \fq v^{-1})(\fq v - \fq^{-1} v^{-1})}   +  \fq^{2a} v^{-2a} u_-^2 \, .
\end{equation}
If we specialize to $b=a + 1$ we get an elementary 't Hooft operator $H^{(2)}_{2a+1}$ of magnetic charge $2$ and odd electric charge: 
\begin{equation}
	 \fq^{-\frac{1}{2}}   H_a H_{a+1}= \fq^{2a+1} v^{2a+1} u_+^2  + \frac{(v+v^{-1}) }{(\fq^{-1} v - \fq v^{-1})(\fq v - \fq^{-1} v^{-1})}   +  \fq^{2a+1} v^{-2a-1} u_-^2 =  \fq^{\frac{1}{2}}   H_{a+1} H_{a} \, .
\end{equation}
In both cases, if we were to directly compute these expressions we we would easily predict the first and last term while the middle term 
would require a careful analysis of bubbling contributions as a smooth monopole configuration screens the bare magnetic charge. 

Other $H_a H_b$ products do not give anything new. For example,
\begin{equation}
	H_a H_{a-2} = 1+ \fq^{-1} H_{a-1}^2 
\end{equation} 
More generally, if $b \geq a+2$ we have
\begin{equation}
	 \fq^{\frac{a}{2}-\frac{b}{2}}   H_a H_b- \fq^{1+\frac{a}{2}-\frac{b}{2}}   H_{a+1} H_{b-1}= \fq^{a-b+1} w_{b-a-2}
\end{equation}
Conversely, if $b\leq a-2$, 
\begin{equation}
	 \fq^{\frac{a}{2}-\frac{b}{2}}   H_a H_b- \fq^{\frac{a}{2}-\frac{b}{2}-1}   H_{a-1} H_{b+1}= \fq^{a-b-1} w_{a-b-2}
\end{equation}
The simple commutation relations between $H_a$ and $H_{a+1}$ suggest considering combinations
\begin{equation}
	D_{b+c;a(b+c)+c} \equiv \fq^{- \frac12 bc} H_a^b H_{a+1}^c \sim \fq^{a (b+c)^2 + c(b+c)} v^{a (b+c) + c} u_+^{b+c} + \cdots
\end{equation}
Although we employed three integers $a,b,c$ in the definition, $D_{m,e}$ has an unique realization for 
any $m>0$: we define $c$ as $e$ modulo $m$ in the range $0,m$, $b=m-c$ and then $a=(e-c)/m$. The leading 
term in the expression identifies this with a (K-theory class of) a 't Hooft-Wilson loop of charge $(m,e)$. 

Recall that UV line defects are labelled by a pair of a magnetic weight and a weight for the gauge group modulo 
the action of the Weyl group. Here we fixed the Weyl symmetry by setting $m\geq 0$. If $m=0$, we set $D_{0,e} \equiv w_e$. 
This exhausts the space of expected charges.  Accordingly, we expect $D_{m,e}$ to be a linear basis for $A_\fq$. 

In particular, it is easy to verify that the product of any number of $H_a$'s and $w_n$'s can be recursively reduced to a finite linear combination of $D_{m,e}$'s:
$w_n H_a$ can be expanded in a linear combination of $H_{a+k}$ with $|k| \leq n$ and any $H_a H_b$ combination can be replaced by $H_{\frac{a+b}{2}}^2$ or 
$H_{\frac{a+b-1}{2}}H_{\frac{a+b+1}{2}}$ up to terms with lower magnetic charge. 

We have
\begin{equation}
	\rho(D_{m,e}) = D_{m,e-2m} 
\end{equation}
In conclusion, the algebra $A_\fq$ and the double $\fD_\fq$ are defined by the relations 
\begin{align}
   & w_1 H_a = \fq^{-\frac12} H_{a+1} + \fq^{\frac12} H_{a-1} \cr
   & H_a w_1 = \fq^{\frac12} H_{a+1} + \fq^{-\frac12} H_{a-1} \cr
   & H_a H_{a+1} = \fq H_{a+1} H_a \cr
    & H_{a+1} H_{a-1} = 1 + \fq^{-1} H_a^2 \cr
   & H_{a-1} H_{a+1} = 1 + \fq H_{a}^2 \cr
   & D_{n+m,an+am+m} \equiv \fq^{-\frac{n m}{2}}H_a^n H_{a+1}^m \cr
   & \rho(D_{m,e}) = D_{m,e-2m} \, .
\end{align}

\subsubsection{Norms and auxiliary Hilbert space}
As we compute the norm of $|D_{m,e}\rangle$, we can attempt to systematically shift the integration contours as we did above
to reach a manifestly positive expression. This is not difficult. For brevity, we integrate the analysis into the presentation of the 
isometry from $A_\fq$ to an auxiliary Hilbert space $L^2\left(\bZ \times S^1\right)^{\bZ_2}$. 

We use a magnetic Vandermonde measure
\begin{equation}
	(v^{-1}-v)(\wt v^{-1} - \wt v) \, ,
\end{equation}	
in the definition of the auxiliary space and maps
\begin{align}
	u_+&= \frac{\fq^{\frac12} v^2}{v^2-1} u \cr
	u_- &= \frac{\fq^{\frac12}}{v^2-1} u^{- 1} \cr
	\wt u_+ &= \frac{\fq^{\frac12}}{1-\wt v^2} \wt u \cr
	\wt u_-&= \frac{\fq^{\frac12} \wt v^2}{1-\wt v^2} \wt u^{- 1}
\end{align}
with half the usual normalization: $u v = \fq v u$ and expected 
\begin{align}
	\rho(u_+) &= \wt u_+^\dagger = \fq^{-1} v^2 u_- \cr
	\rho(u_-) &= \wt u_-^\dagger  = \fq^{-1} v^{-2} u_+
\end{align}

The half-index/image of the spherical vector becomes 
\begin{equation}
	I\!\!I_B(\zeta) =  \delta_{B,0} (\fq^2;\fq^2)_\infty (\fq^2 \zeta^2;\fq^2)_\infty (\fq^2 \zeta^{-2};\fq^2)_\infty
\end{equation}
which is Weyl symmetric. 

Here we encounter another manifestation of the $SU(2)/SO(3)$ subtleties. If we define the $\bZ_2$ Weyl symmetry as $B \to -B$, $\zeta \to \zeta^{-1}$, 
the minimal 't Hooft operators are odd under the Weyl symmetry. A simple way around this obstruction is to include an extra multiplicative factor 
of $(-1)^B$ in the definition of the $\bZ_2$ action, so that states of odd $B$ are odd under $B \to -B$, $\zeta \to \zeta^{-1}$. Then 
the 't Hooft operators act within $L^2\left(\bZ \times S^1\right)^{\bZ_2}$ and we obtain the desired isometry. \footnote{Notice that we cannot just change the relative sign in the definition of $u_\pm$: that would make $\Tr H_a H_{a-2}$ negative.}

We compute 
\begin{align}
	u_+ I\!\!I_B(\zeta) &= \wt u_+ I\!\!I_B(\zeta) =  \fq^{\frac12}\delta_{B,1} (\fq^2;\fq^2)_\infty (\fq^3 \zeta^2;\fq^2)_\infty (\fq^3 \zeta^{-2};\fq^2)_\infty \cr
	u_- I\!\!I_B(\zeta) &= \wt u_- I\!\!I_B(\zeta)=- \fq^{\frac12}\delta_{B,-1} (\fq^2;\fq^2)_\infty (\fq^3 \zeta^2;\fq^2)_\infty (\fq^3 \zeta^{-2};\fq^2)_\infty 
\end{align}
which verifies the spherical condition: $I\!\!I_B(\zeta)$ is the image of $|1\rangle$ in the auxiliary Hilbert space.

As discussed in the general case, the isometry diagonalizes the Wilson lines $w_n$ and $\wt w_n$, with eigenvalues $\chi_n(\fq^{-\frac{m}{2}} \zeta)$. Obviously, $L^2\left(\bZ \times S^1\right)^{\bZ_2}$ includes a single eigenstate in each eigenspace, labelled by $(m,\zeta)$ modulo $\bZ_2$.  

\subsubsection{Inverting the isometry}
This example is sufficiently simple that we can invert the isometry, by diagonalizing the action of Wilson lines directly in $\cH_\fq$.

Diagonalizing the action of Wilson lines on Wilson lines is straightforward:
\begin{equation}
	|0;\mu \rangle = \sum_n \chi_n(\mu) |w_n \rangle
\end{equation}
have the same eigenvalue $\chi_1(\mu)$ for $w_1$ and $\wt w_1$. 

The charge $1$ 't Hooft operators can be reorganized as 
\begin{equation}
	|1;\mu \rangle = \sum_a \mu^a |H_a \rangle
\end{equation}
As 
\begin{equation}
w_1 H_a = \fq^{-\frac12} H_{a+1} + \fq^{\frac12} H_{a-1}  \qquad \qquad  H_a w_1= \fq^{\frac12} H_{a+1} + \fq^{-\frac12} H_{a-1} \, ,
\end{equation}
this is a simultaneous $w_1$ eigenvector with eigenvalue $\mu \fq^{\frac12} + \mu^{-1} \fq^{-\frac12}$ and $\wt w_1$ eigenvector with eigenvalue $\mu \fq^{-\frac12} + \mu^{-1} \fq^{\frac12}$.
It is delta-function normalizable on the unit circle: 
\begin{equation}
	\langle 1;\mu|1;\nu \rangle = \sum_{a,b} \mu^{-a} \nu^b \Tr H_{a-2} H_b =  (\fq \mu^2;\fq^2)(\fq \mu^{-2};\fq^2)(\fq^3 \mu^2;\fq^2)(\fq^3 \mu^{-2};\fq^2) \sum_b \left(\mu^{-b} \nu^b \right)  \, .
\end{equation}

At magnetic charge $2$ we have a mixing with charge $0$
\begin{align}
w_1 H^{(2)}_{2a} &= \fq^{-1} H^{(2)}_{2a+1} + \fq H^{(2)}_{2a-1}  \qquad \qquad  H^{(2)}_{2a} w_1=\fq H^{(2)}_{2a+1} + \fq^{-1}  H^{(2)}_{2a-1}  \cr
w_1 H^{(2)}_{2a+1} &= \fq^{-1} H^{(2)}_{2a+2} + \fq H^{(2)}_{2a} +1 \qquad \qquad  H^{(2)}_{2a+1} w_1= \fq H^{(2)}_{2a+2} + \fq^{-1} H^{(2)}_{2a} +1 \cr
w_1 w_n &= w_{n+1} + w_{n-1} \qquad n>0
\end{align}

As for the charge $2$ sector, we can effectively strip off the bubbling contributions by defining auxiliary states 
\begin{align}
	|2;2a\rangle &\equiv | H^{(2)}_{2a}\rangle + \left|\frac{\fq + \fq^{-1}}{(\fq + \fq^{-1})^2 - w_1^2}\right\rangle  \cr
	|2;2a+1\rangle &\equiv | H^{(2)}_{2a+1}\rangle + \left|\frac{w_1}{(\fq + \fq^{-1})^2 - w_1^2}\right\rangle  
\end{align}
where the second terms are defined as sums over $(\fq + \fq^{-1})^{-b-1} |w_1^b\rangle$. Then 
$w_1 |2;a\rangle = \fq^{-1}|2;a+1\rangle+\fq |2;a-1\rangle$ and then 
\begin{equation}
	|2;\mu \rangle = \sum_a \mu^a |2;a \rangle \, ,
\end{equation}
which is an $w_1$ eigenvector with eigenvalue $\mu \fq+ \mu^{-1} \fq$ and $\wt w_1$ eigenvector with eigenvalue $\mu \fq + \mu^{-1} \fq$.

Following this route, we can build abstractly a spectral decomposition of $\cH$ over the expected $(S^1 \times \bZ)/\bZ_2$ spectrum, with one-dimensional distributional eigenspaces. 

\subsubsection{Diagonalizing 't Hooft operators}
We can give another interesting alternative description of $\cH_\fq$ by simultaneously diagonalizing $H_{0}$ and $H_{1}$.

Recall the definition of the complex quantum dilogarithm, aka tetrahedron index: 
 \begin{equation}
	\Phi_B(\zeta) = \prod_{n=0}^\infty \frac{1+ \fq^{2n+1} v}{1+ \fq^{2n+1} \wt v^{-1}}=\prod_{n=0}^\infty \frac{1+ \fq^{2n-B/2+1} \zeta}{1+ \fq^{2n-B/2+1} \zeta^{-1}}
\end{equation}

We now define a second set of variables $\sigma$, $S$, $s$, $t$, etc. analogous to $\zeta$, $B$, $v$, $u$, etc. and consider the kernel
\begin{equation}
	U_{B,S}(\zeta,\sigma) = e^{\frac{i \pi}{2}B} \Phi_{B + S}(\sigma \zeta) \Phi_{-B + S}(\sigma \zeta^{-1}) 
\end{equation}
We have 
\begin{align}
	(u_+ U)_{B,S}(\zeta,\sigma) &=i \frac{\fq^{\frac12} v(v+s)}{v^2-1} (t U)_{B,S}(\zeta,\sigma) \cr
	(u_- U)_{B,S}(\zeta,\sigma) &= -i \frac{\fq^{\frac12}(1+s v) }{v^2-1} (t U)_{B,S}(\zeta,\sigma) \cr
\end{align}
so that 
\begin{equation}
	(H_0 \,U)_{B,S}(\zeta,\sigma) = i \fq^{\frac12} (t U)_{B,S}(\zeta,\sigma)
\end{equation}
and 
\begin{equation}
	(H_{-1} \, U)_{B,S}(\zeta,\sigma) = - i (s t U)_{B,S}(\zeta,\sigma)
\end{equation}
as well as 
\begin{equation}
	(\wt H_1 \,U)_{B,S}(\zeta,\sigma) = i \fq (\wt s^{-1} \wt t^{-1} U)_{B,S}(\zeta,\sigma)
\end{equation}
and
\begin{equation}
	(\wt H_2 \, U)_{B,S}(\zeta,\sigma) = - i \fq^{\frac32} (\wt t^{-1} U)_{B,S}(\zeta,\sigma)
\end{equation}

Clearly, if $U$ is the kernel of an unitary transformation between $L^2\left(\bZ \times S^1\right)^{\bZ_2}$
and the $L^2\left(\bZ \times S^1\right)$ space of wavefunctions in $\sigma$ and $S$, these relations
will give us the spectrum of 't Hooft operators. 

In order to prove such a statement, it is useful to avoid dealing with delta-function normalizability 
by diagonalizing operators with a discrete spectrum: 
\begin{align}
	|\wt H_1|^2 = \wt H_1 \rho(H_1) &= H_{-1} \wt H_1 \cr
	|\wt H_2|^2 = \wt H_2 \rho(H_2) &= H_{0} \wt H_2
\end{align}
We have
\begin{align}
	(|\wt H_1|^2 U)_{B,S}(\zeta,\sigma)&= \fq (|st|^2 U)_{B,S}(\zeta,\sigma) \cr
	(|\wt H_2|^2 U)_{B,S}(\zeta,\sigma)&= \fq^2 (|t|^2 U)_{B,S}(\zeta,\sigma)
\end{align} 
and thus a Fourier transform in $\sigma$ gives tentative wave-functions with fixed eigenvalues for $|\wt H_1|^2$ and  $|\wt H_2|^2$:
\begin{equation}
	U_{B;S,T}(\zeta) \equiv e^{\frac{i \pi}{2}B}\oint \frac{d\sigma}{2\pi i \sigma^{T+1}} \Phi_{B + S}(\sigma \zeta) \Phi_{-B + S}(\sigma \zeta^{-1}) 
\end{equation}
The available range for the parameters $S$ and $T$ is constrained by the requirement that the integration contour can be deformed as needed to simplify the action of the 't Hooft operators. It would be nice to verify that both parameters are constrained to be integers and that this set of eigenfunctions is complete. We will continue the discussion in our companion paper \cite{next}, as this is closely related to IR formulae for the Schur index. 

The distributional kernel employed above can be identified with the contribution to Schur correlation functions of an RG interface \cite{Dimofte:2013lba}.

\subsection{$N=2^*$ $SU(2)$ gauge theory.}
The next simplest example is the case of $N=2^*$ $SU(2)$ gauge theory. This is a theory of class $\cS$ with algebra $\mathfrak{sl}_2$ for a one-punctured torus. 

The Schur index is 
\begin{align}
	I_\fq(\mu) &= \frac12 \oint_{|\zeta|=1} \frac{d\zeta}{2 \pi i \zeta} (1-\zeta^2)(1-\zeta^{-2})\frac{(\fq^2)^2_\infty (\fq^2 \zeta^2;\fq^2)^2_\infty (\fq^2 \zeta^{-2};\fq^2)^2_\infty}{(-\fq \mu^\pm;\fq^2)_\infty (-\fq \mu^\pm \zeta^2;\fq^2)_\infty (-\fq \mu^\pm \zeta^{-2};\fq^2)_\infty} 
\end{align}
where for reason of space we condensed the denominator products as $(x \mu^\pm;\fq^2)_\infty = (x \mu;\fq^2)_\infty(x \mu^{-1};\fq^2)_\infty$. 

\subsubsection{The algebra}
The insertion of Wilson lines $w_n = v^{n} + v^{n-2}+\cdots + v^{-n}$ is straightforward. In order to describe $A_\fq$, we can introduce 
\begin{equation}
	u_\pm v = \fq^{\pm 1} v u_\pm \, ,
\end{equation}
which also satisfy
\begin{equation}
	u_+ u_- = \frac{(1+\mu \fq v^2)(1+\mu^{-1} \fq v^2)}{(1-v^2)(1-\fq^2 v^2)} \qquad \qquad u_- u_+ = \frac{(1+\mu \fq^{-1} v^2)(1+\mu^{-1} \fq^{-1} v^2)}{(1-v^2)(1-\fq^{-2} v^2)} 
\end{equation}

Again, we will enlarge the $A_\fq$ algebra by including also 't Hooft operators of minimal charge, which would strictly speaking make sense only for an $SO(3)$ gauge theory. The algebras for $SU(2)$ or $SO(3)$ gauge theories will be 
obtained by dropping either 't Hooft operators of Wilson lines of odd charge. 

The 't Hooft operators of minimal charge do not suffer of monopole bubbling effect and are simply written as 
\begin{equation}
	H_a = \fq^{\frac{a}{2}} v^a u_+ +  \fq^{\frac{a}{2}} v^{-a} u_-
\end{equation}
We can compute 
\begin{align}
	H_a H_b &= \fq^{\frac{a}{2}+\frac{3b}{2}} v^{a+b} u_+^2  +  \fq^{\frac{a}{2}-\frac{b}{2}}  v^{b-a}\frac{(1+\mu \fq^{-1} v^2)(1+\mu^{-1} \fq^{-1} v^2)}{(1-v^2)(1-\fq^{-2} v^2)}  + \cr +&\fq^{\frac{a}{2}-\frac{b}{2}} v^{a-b} \frac{(1+\mu \fq v^2)(1+\mu^{-1} \fq v^2)}{(1-v^2)(1-\fq^2 v^2)} +  \fq^{\frac{a}{2}+\frac{3 b}{2}} v^{-a-b} u_-^2
\end{align}

The two terms appearing in $\Tr H_a H_b$ are 
\begin{align}
	&\frac12 \oint_{|\zeta|=1} \frac{d\zeta}{2 \pi i \zeta}\fq^{\frac{a}{2}-\frac{b}{2}}  \zeta^{b-a} \frac{(\fq^2)^2_\infty (\zeta^2;\fq^2)_\infty (\fq^2\zeta^{-2};\fq^2)_\infty(\fq^2 \zeta^2;\fq^2)_\infty (\fq^4 \zeta^{-2};\fq^2)_\infty}{(\fq \mu^\pm;\fq^2)_\infty (\fq \mu^\pm \zeta^2;\fq^2)_\infty (\fq^3 \mu^\pm \zeta^{-2};\fq^2)_\infty}  \cr
	&\frac12 \oint_{|\zeta|=1} \frac{d\zeta}{2 \pi i \zeta}\fq^{\frac{a}{2}-\frac{b}{2}} \zeta^{a-b}  \frac{(\fq^2)^2_\infty (\fq^2\zeta^2;\fq^2)_\infty (\zeta^{-2};\fq^2)_\infty(\fq^4 \zeta^2;\fq^2)_\infty (\fq^2 \zeta^{-2};\fq^2)_\infty}{(\fq \mu^\pm;\fq^2)_\infty (\fq^3 \mu^\pm \zeta^2;\fq^2)_\infty (\fq \mu^\pm \zeta^{-2};\fq^2)_\infty}
\end{align}
The integration contours can be shifted and the integrals combined 
\begin{align}
	\Tr H_a H_b &=\frac12 \oint_{|\zeta|=1} \frac{d\zeta}{2 \pi i \zeta}(\zeta^{b-a}+ \zeta^{a-b} ) \frac{(\fq^2)^2_\infty (\fq \zeta^2;\fq^2)_\infty (\fq\zeta^{-2};\fq^2)_\infty(\fq^3 \zeta^2;\fq^2)_\infty (\fq^3 \zeta^{-2};\fq^2)_\infty}{(\fq \mu^\pm;\fq^2)_\infty (\fq^2 \mu^\pm \zeta^2;\fq^2)_\infty (\fq^2 \mu^\pm \zeta^{-2};\fq^2)_\infty} 
\end{align}
The automorphism $\rho$ acts trivially here and this expression is fully compatible with positivity.

We can write some relations:
\begin{align}
	& w_1 H_a = \fq^{-\frac12} H_{a+1} + \fq^{\frac12} H_{a-1} \cr
	& H_a w_1 = \fq^{\frac12} H_{a+1} + \fq^{-\frac12} H_{a-1} \cr
	& \fq^{-\frac12} H_a H_{a+1} - \fq^{\frac12} H_{a+1}H_a = (\fq^{-1} - \fq) w_1\cr
	& H_{a-1} H_{a+1} = \fq H_a^2 + \mu + \mu^{-1} + \fq^{-1} w_1^2 - \fq^{-1}- \fq \cr
	& H_{a+1} H_{a-1} = \fq^{-1}  H_a^2 + \mu + \mu^{-1} + \fq w_1^2 - \fq^{-1}- \fq 
\end{align}
The algebra is expected to enjoy an $SL(2,\bZ)$ S-duality symmetry generated by $T: H_a \to H_{a+1}$ and $S: H_0 \leftrightarrow w_1$. 
We expect generators $D_{m,e}=D_{-m,-e}$ with an obvious  $SL(2,\bZ)$ action, organized in orbits generated from $w_n$ with $n$ being the common divisor of $m$ and $e$. 

We set $D_{0,1} = w_1$ and $D_{1,0} = H_0$. Then $D_{1,a} = H_a$. The relation 
\begin{equation}
	D_{1,0} D_{0,1} = \fq^{\frac12} D_{1,1} + \fq^{-\frac12} D_{1,-1}  
\end{equation} 
 predicts 
\begin{equation}
	D_{a,b} D_{c,d} = \fq^{\frac12} D_{a+c,b+d} + \fq^{-\frac12} D_{a-c,b-d}   \qquad \qquad a d - b c = 1
\end{equation}  
Analogously, 
\begin{equation}
	D_{1,1} D_{1,-1} = \fq^{-1}  D_{2,0} + \fq D_{0,2}  + \mu + \mu^{-1} 
\end{equation} 
predicts
\begin{equation}
	D_{a+c,b+d} D_{a-c,b-d} = \fq^{-1}  D_{2a,2b} + \fq D_{2c,2d}  + \mu + \mu^{-1}  \qquad \qquad a d - b c = 1
\end{equation}  
We can use these relations to both define $D_{m,e}$ and test the $SL(2,\bZ)$ symmetry expectations. 

For example, we can define $D_{2,2a} = H_a^2 - 1$ and 
\begin{equation}
	D_{2,2a+1} = \fq^{-\frac12} H_a H_{a+1}- \fq^{-1} w_1 \, .
\end{equation}
Analogously, we can define $D_{3,3a} = H_a^3 - 2 H_a$ and
\begin{align}
	D_{3,3a+1} &= \fq^{-\frac12} H_a D_{2,2a+1}- \fq^{-1} H_{a+1} \cr
	D_{3,3a+2} &= \fq^{-\frac12} D_{2,2a+1}H_{a+1} - \fq^{-1} H_{a} \, .
\end{align}
Etcetera. This is a well-known quantization of the $SL(2)$ character variety for a 1-punctured torus. 

\subsubsection{The auxiliary Hilbert space}
In order to give an isometry to $\cH_\fq^{\mathrm{aux}}$, we use a magnetic Vandermonde measure
\begin{equation}
	(v^{-1}-v)(\wt v^{-1} - \wt v) \, ,
\end{equation}	
and identify with some work the expressions for the generators
\begin{align}
	u_+ &= \frac{v^2+\fq^{-1}\mu}{v^2-1} u \cr
	u_- &= \frac{\mu^{-1}+\fq^{-1}v^2}{v^2-1} u^{-1}\cr
	\wt u_+ &= \frac{1+\fq^{-1} \mu \wt v^2}{1-\wt v^2} \wt u \cr
	\wt u_- &= \frac{\fq^{-1} + \mu^{-1} \wt v^2}{1-\wt v^2} \wt u^{-1}
\end{align}
compatible with $\rho$ and a candidate spherical vector: 
\begin{equation}
	I\!\!I_B(\zeta) =  \delta_{B,0} \frac{(\fq^2;\fq^2)_\infty (\fq^2 \zeta^2;\fq^2)_\infty (\fq^2 \zeta^{-2};\fq^2)_\infty}{(-\fq \mu;\fq^2)_\infty (-\fq \mu \zeta^2;\fq^2)_\infty (-\fq \mu \zeta^{-2};\fq^2)_\infty}\, .
\end{equation}

In order to have a naive action of Weyl symmetry, we would need to correct these expressions by powers of $\mu^\frac12$. Instead, we can include a factor of $(-\mu)^B$ in the definition of the $\bZ_2$ Weyl symmetry, in the same spirit (and including) the sign fix we used for pure $SU(2)$. 

\subsubsection{More on S-duality}
S-duality is a very non-trivial symmetry of Schur correlation functions. E.g. we can verify experimentally that $\Tr \, H_a^2 = \Tr \, w_1^2$. A full proof can be given with the help of S-duality interfaces \cite{Dimofte:2013lba}. 

It is worth discussing this explicitly. The S-duality kernel is a small variation of the one employed to diagonalize 't Hooft operators in pure $SU(2)$ \cite{Dimofte:2013lba}. 
We define a second set of variables $\sigma$, $S$, $s$, $t$, etc. analogous to $\zeta$, $B$, $v$, $u$, etc. and 
consider the kernel
\begin{equation}
	U_{B,S}(\zeta,\sigma) = \sigma^{-2S} \mu^{-S+B} \zeta^{-2B} \Phi_{B + S}(\sigma \zeta) \Phi_{-B + S}(\sigma \zeta^{-1}) \Phi_{-B - S}(-\mu \sigma^{-1} \zeta^{-1}) \Phi_{B - S}(-\mu \sigma^{-1} \zeta)
\end{equation}
Then 
\begin{equation}
	\mu^{-1} s (1- \mu s^{-1} v)(1- \mu s^{-1} v^{-1}) t^{-1} U_{B,S}(\zeta,\sigma) = s^{-1}(1+ s v)(1+s v^{-1}) t U_{B,S}(\zeta,\sigma)
\end{equation}
i.e. 
\begin{equation}
	(v+v^{-1}) U_{B,S}(\zeta,\sigma) = \left( \frac{1}{t^{-1} -t}s (t + \mu^{-1} t^{-1}) +  \frac{1}{t^{-1} -t} s^{-1}(t+\mu t^{-1}) \right)U_{B,S}(\zeta,\sigma)
\end{equation}
which essentially maps the Wilson line to a 't Hooft operator built from $t$ and $s$. We also have 
\begin{align}
	(1+s v)t  U_{B,S}(\zeta,\sigma)  &= ( s v-\mu) u^{-1} U_{B,S}(\zeta,\sigma) \cr
	(1+s v^{-1})t U_{B,S}(\zeta,\sigma)  &= (-s v^{-1}\mu^{-1}+1) u U_{B,S}(\zeta,\sigma) 
\end{align}
e.g. 
\begin{align}
	s v (t-u^{-1})  U_{B,S}(\zeta,\sigma)  &=(-\mu u^{-1}-t) U_{B,S}(\zeta,\sigma) \cr
	s v^{-1}(t+\mu^{-1} u) U_{B,S}(\zeta,\sigma)  &= (u-t) U_{B,S}(\zeta,\sigma)
\end{align}
which implies 
\begin{align}
	v (\fq^{-1} t-u^{-1})(u-t)  U_{B,S}(\zeta,\sigma)  &=v^{-1}(\fq^{-1} t+\mu^{-1}u) (-\mu u^{-1}-t) U_{B,S}(\zeta,\sigma)
\end{align}
i.e.
\begin{align}
	\frac{1}{v^2-1} \left[(v^2 + \fq \mu^{-1})u + (\fq v^2 + \mu)u^{-1} \right]U_{B,S}(\zeta,\sigma)&= (\fq t^{-1}+ t) U_{B,S}(\zeta,\sigma) 
\end{align}
which, up to a $\mu \to \mu^{-1}$ convention change, maps the 't Hooft loop to a simple difference operator which is diagonalized by Fourier transform. A similar formula holds for the tilde variables. 

The distributional kernel employed above can be identified with the contribution to Schur correlation functions of a duality interface 
defined via $T[SU(2)]$ \cite{Gaiotto:2008ak}.

%A fun feature of the Schur trace of a Wilson line is that it is an average on the unit circle and thus $\Tr w_R < (\mathrm{dim} R) \Tr 1$. 
%This inequality also holds on $SL(2,\bZ)$ images of Wilson loops, which correspond to $W_{\ell,R}$ for various minimal curves on 
%the punctured torus $C$. These inequalities are compatible with the classical traces of holonomies for unitary flat connections. 
%This observation extends to any theory of class $\cS$ with regular punctures.  

\subsection{Intermission: $SU(2)$ vs $U(2)$ SQCD}
The next natural set of examples would be $SU(2)$ gauge theories with $N_f = 1,2,3,4$ flavours. These have a nice class $S$ interpretation. An unpleasant challenge is that the minimal allowed charge for monopole operators is $2$, requiring one to address directly bubbling. There is a trick to sidestep this computation: consider instead $U(2)$ gauge theories, which admit 't Hooft operators of minimal charge. An important feature of gauge theories is that 't Hooft operators which are not charged under some factor of the gauge group have the same expression as difference operators as if the factor was not gauged. We can thus write down $U(2)$ 't Hooft operators of minimal charges, combine them into 't Hooft operators with $SU(2)$ charge only, and carry them over to $SU(2)$ gauge theory. We refer to the Appendices for details.

\subsection{Abelianized algebras}
For $N_f=1$ we get. 
\begin{align}
	H_{2a} &= \fq^{2a} v^{2a} u_+ + \frac{\fq+ \fq^{-1} + \mu v + \mu v^{-1} }{(\fq v - \fq^{-1}v^{-1})(\fq^{-1} v - \fq v^{-1})}  +  \fq^{2a} v^{-2a}u_- \cr
	H_{2a+1} &= \fq^{2a+1} v^{2a+1}  u_+ + \frac{(\fq + \fq^{-1})\mu + v + v^{-1} }{(\fq v - \fq^{-1}v^{-1})(\fq^{-1} v - \fq v^{-1})}  +  \fq^{2a+1} v^{-2a-1} u_-
\end{align}
Here 
\begin{equation}
	u_\pm v = \fq^{\pm 2} v u_\pm \, ,
\end{equation}
and 
\begin{align}
	u_+ u_- &= \frac{(1+\mu \fq v)(1+\mu \fq^{-1} v^{-1})}{(v-v^{-1})(\fq v - \fq^{-1}v^{-1})^2(\fq^2 v - \fq^{-2}v^{-1})} \cr u_- u_+ &=  \frac{(1+\mu \fq^{-1} v)(1+\mu \fq v^{-1})}{(v-v^{-1})(\fq^{-1} v - \fq v^{-1})^2(\fq^{-2} v - \fq^{2}v^{-1})} 
\end{align}

More generally, for $N_f$ flavours we need
\begin{align}
	u_+ u_- &= \frac{\prod_i (1+\mu_i \fq v)(1+\mu_i \fq^{-1} v^{-1})}{(v-v^{-1})(\fq v - \fq^{-1}v^{-1})^2(\fq^2 v - \fq^{-2}v^{-1})} \cr u_- u_+ &=  \frac{\prod_i (1+\mu_i \fq^{-1} v)(1+\mu_i \fq v^{-1})}{(v-v^{-1})(\fq^{-1} v - \fq v^{-1})^2(\fq^{-2} v - \fq^{2}v^{-1})} 
\end{align}
and the tentative numerator in $H_{2a}$ becomes 
\begin{equation}
	\frac{(\fq^{-1} v - \fq v^{-1}) \prod_i (1+ \fq \mu_i v) + (\fq v - \fq^{-1}v^{-1})\prod_i (1+ \fq \mu_i v^{-1})}{v-v^{-1}}
\end{equation}
and in $H_{2a+1}$ becomes 
\begin{equation}
	\fq^{-1} \frac{v^{-1} (\fq^{-1} v - \fq v^{-1}) \prod_i (1+ \fq \mu_i v) + v (\fq v - \fq^{-1}v^{-1})\prod_i (1+ \fq \mu_i v^{-1})}{v-v^{-1}}
\end{equation}
For specific $N_f$, we can simplify the expressions by subtracting some Wilson lines. 

For $N_f=2$ we get 
\begin{align}
	&H_{2a} = \fq^{2a} v^{2a} u_+ + \frac{(\fq+ \fq^{-1})(1+\mu_1 \mu_2) + (\mu_1+\mu_2) (v + v^{-1}) }{(\fq v - \fq^{-1}v^{-1})(\fq^{-1} v - \fq v^{-1})}  +  \fq^{2a} v^{-2a}u_- \cr
	&H_{2a+1} = \cr &\fq^{2a+1} v^{2a+1}  u_+ + \frac{(\fq+ \fq^{-1})  (\mu_1+\mu_2)+ (1+\mu_1 \mu_2)(v + v^{-1})}{(\fq v - \fq^{-1}v^{-1})(\fq^{-1} v - \fq v^{-1})}  +  \fq^{2a+1} v^{-2a-1} u_-
\end{align}

For $N_f=3$ we get 
\begin{align}
	H_{2a} &= \fq^{2a} v^{2a} u_+ +  \frac{(\fq+ \fq^{-1})(1+\mu_1 \mu_2+\mu_2 \mu_3+\mu_1 \mu_3) + (\mu_1+\mu_2+ \mu_3+ \mu_1 \mu_2 \mu_3) (v + v^{-1}) }{(\fq v - \fq^{-1}v^{-1})(\fq^{-1} v - \fq v^{-1})}  +\cr &+  \fq^{2a} v^{-2a}u_- \cr
	H_{2a+1} &= \fq^{2a+1} v^{2a+1}  u_+ +\frac{(\fq+ \fq^{-1})  (\mu_1+\mu_2+ \mu_3+ \mu_1 \mu_2 \mu_3)+ (1+\mu_1 \mu_2+\mu_2 \mu_3+\mu_1 \mu_3)(v + v^{-1})}{(\fq v - \fq^{-1}v^{-1})(\fq^{-1} v - \fq v^{-1})} + \cr & +  \fq^{2a+1} v^{-2a-1} u_-
\end{align}

Finally, for $N_f=4$ we get
\begin{align}
	H_{2a} &= \fq^{2a} v^{2a} u_+ + \frac{(\fq+ \fq^{-1})(1+\sum_{i<j} \mu_i \mu_j+\prod_i \mu_i) + (\sum_i \mu_i+ \sum_{i<j<k} \mu_i \mu_j \mu_k) (v + v^{-1}) }{(\fq v - \fq^{-1}v^{-1})(\fq^{-1} v - \fq v^{-1})}  + \cr &+ \fq^{2a} v^{-2a}u_- \cr
	H_{2a+1} &= \fq^{2a+1} v^{2a+1}  u_+ + \frac{(\fq+ \fq^{-1}) (\sum_i \mu_i+ \sum_{i<j<k} \mu_i \mu_j \mu_k)+ (1+\sum_{i<j} \mu_i \mu_j+\prod_i \mu_i)(v + v^{-1})}{(\fq v - \fq^{-1}v^{-1})(\fq^{-1} v - \fq v^{-1})}  + \cr &+ \fq^{2a+1} v^{-2a-1} u_-
\end{align}

These theories actually have an $SO(2 N_f)$ global symmetry. This is not completely manifest from the above expressions, but can be restored by rescaling 
$u_\pm$ and the $H_a$ operators by $\prod_i \mu_i^{\frac12}$. E.g. the numerator factors are characters for the spinor representations of $SO(2 N_f)$. 

The $N_f=4$ theory has a class $\cS$ interpretation with Lie algebra $\mathfrak{sl}_2$ and $C$ being the four-punctured sphere, with regular singularities
of monodromy parameters $\mu_1 \mu_2^\pm$ and $\mu_3 \mu_4^\pm$. We will discuss the auxiliary Hilbert space description at length in the next section, as the main example of quantization of a complex character variety. Here we can sketch the main formulae.

The Schur index is
\begin{align}
	I_\fq(\mu) &= \frac12 \oint_{|\zeta|=1} \frac{d\zeta}{2 \pi i \zeta} (1-\zeta^2)(1-\zeta^{-2})\frac{(\fq^2)^2_\infty (\fq^2 \zeta^2;\fq^2)^2_\infty (\fq^2 \zeta^{-2};\fq^2)^2_\infty}{\prod_i (-\fq \mu_i^\pm \zeta;\fq^2)_\infty (-\fq \mu_i^\pm \zeta^{-1};\fq^2)_\infty} = \cr
	&=1+ \chi_{\mathrm{Adj}}(\mu)\fq^2 + \cdots
\end{align}
Rather non-trivially, only characters of triality-invariant representations of the $SO(8)$ flavour group appear in the index. This is due to the fact that S-dualities for this SCFT act as triality 
on the flavour group. \footnote{For example, the Schur trace of a fundamental Wilson line starts with $\fq \chi_8(\mu)$, a character of the vector representation of $SO(8)$. The spinor characters in $H_a$ guarantee that the
corresponding traces start with $\fq \chi_{8_s}(\mu)$ or  $\fq \chi_{8_c}(\mu)$ for the spinor representations, compatibly with the fact that S-duality exchanges Wilson lines and 't Hooft lines 
while acting as a triality on $SO(8)$. } 

The candidate spherical vector is 
\begin{equation}
	I\!\!I_B(\zeta) =  \delta_{B,0} \frac{(\fq^2;\fq^2)_\infty (\fq^2 \zeta^2;\fq^2)_\infty (\fq^2 \zeta^{-2};\fq^2)_\infty}{\prod_i (-\fq \mu_i \zeta;\fq^2)_\infty (-\fq \mu_i \zeta^{-1};\fq^2)_\infty}
\end{equation}
in an auxiliary Hilbert space defined with the usual magnetic Vandermonde measure
\begin{equation}
	(v^{-1}-v)(\wt v^{-1} - \wt v) \, ,
\end{equation}	
as well as $u v = \fq^2 v u$ and
\begin{align}
	u_+&= \frac{\prod_i (1+\fq^{-1} \mu_i v^{-1})}{(1-\fq^{-2} v^{-2})(1-v^{-2})} u \cr
	u_- &= \frac{\prod_i (\mu_i^{-1} +\fq^{-1} v)}{(1-\fq^{-2} v^2)(1-v^{2})} u^{- 1} \cr
	\wt u_+ &= \frac{\prod_i (1+\fq^{-1} \mu_i \wt v)}{(1-\fq^{-2} \wt v^{2})(1-\wt v^{2})} \wt u \cr
	\wt u_-&=\frac{\prod_i (\mu_i^{-1} +\fq^{-1}  \wt v^{-1})}{(1-\fq^{-2} \wt v^{-2})(1-\wt v^{-2})} \wt u^{- 1}
\end{align}
The Weyl symmetry has to be adjusted by a factor of $(-\prod_i \mu_i)^B$. 

The remaining theories also have a similar class $\cS$ interpretation: $N_f=3$ has two regular punctures and one irregular of rank $1$, 
$N_f=2$ has a realization with two regular punctures and one irregular of rank $1/2$ and a realization with two irregular of rank $1$, $N_f=1$ has a realization with 
an irregular of rank $1$ and one of rank $1/2$. 

In the remainder of this section, we will discuss some interesting applications of Schur quantization of these theories to the theory of quantum groups. 

\subsection{Back to $U(2)$ with $N_f=1$.}
In order to make contact with quantum groups, we can gauge the $U(1)$ flavour symmetry of the $SU(2)$ with two flavour theory. Then the two hypermultiplets together with the $U(1)$ gauge fields give a copy of SQED$_2$, with the $SU(2)$ flavour symmetry being gauged. This gives back $U(2)$ with $N_f=1$.

The operators inherited from SQED$_2$ include the 't Hooft operators with $U(1)$ magnetic charge only and the $U(1)$ Wilson line. They give a copy of the quantum group $U_q(\mathfrak{sl}_2)$, with the Casimir coinciding with the fundamental Wilson line for $SU(2)$. According to our general discussion, the Lagrangian formulation of Schur quantization for this theory thus presents the Hilbert space $\cH_\fq$ as a spectral decomposition into principal series representations $\cH_{M;\mu}$ of $\fD[U_{q}(\mathrm{sl}_2),\rho]$. 

We expect this representation to be a fundamental ingredient of a quantum group description of an irregular singularity of rank $\frac12$ in complex Chern-Simons theory, akin to the Teichm\"uller construction of irregular conformal blocks \cite{Gaiotto:2012sf}.

\subsection{A $q$-deformation of $T^*SL(2,\bC)$.}
Recall that one of the class $\cS$ descriptions of $SU(2)$ $N_f=2$ involves a $\bC P^1$ geometry with two irregular singularities of rank $1$.
If we gauge both $U(1)$ sugbroups of the flavour symmetry, this gives a theory such that $A_\fq$ contains two commuting copies of $U_q(\mathfrak{sl}_2)$. 
It is a $q$-deformation of the left- and right- actions of two copies of $\mathfrak{sl}_2$ on $T^*SL(2,\bC)$. 

The Casimirs of the two $U_q(\mathfrak{sl}_2)$ coincide with the  fundamental Wilson line for $SU(2)$. 
The Lagrangian formulation of Schur quantization thus presents the Hilbert space $\cH_\fq$ as a direct sum/integral of products 
$\cH_{M;\mu}\times \cH_{M;\mu}$ of two principal series representations of $\fD[U_{q}(\mathrm{sl}_2),\rho]$.
This is a $q$-deformation of the Plancherel decomposition of $L^2(SL(2,\bC))$.

\subsection{The coproduct for $\mathfrak{U}_\fq(\mathfrak{sl}_2)$.}
The class $\cS$ description of $SU(2)$ $N_f=3$  only makes manifest an $SO(2) \times SO(4)$ subgroup of the flavour symmetry of the theory. 
The $SO(2)$ factor is associated to an irregular singularity of rank $1$, the $SO(4)$ to two regular singularities. 

Gauging $SO(2)$ gives a theory with a particularly important connection to the representation theory of quantum groups. 
A full discussion requires some cluster technology \cite{schrader2019cluster} and will better fit in our companion paper \cite{next}. 
Essentially, there is a copy of $U_q(\mathfrak{sl}_2)$ in $A_\fq$ but also a map $A_\fq \to U_q(\mathfrak{sl}_2) \times U_q(\mathfrak{sl}_2)$ which is essentially an isomorphism, realizing the coproduct of $U_\fq(\mathfrak{sl}_2)$. 

As in the Teichm\"uller case \cite{Nidaiev:2013bda,schrader2017continuous}, we expect this setup to give a spectral decomposition of the tensor product
$\cH_{M_1;\mu_1}\times \cH_{M_2;\mu_2}$ of two principal series representations of $\fD[U_{q}(\mathrm{sl}_2),\rho]$ into a direct sum/integral of  
$\cH_{M;\mu}$. An important difference is that here we can also ask (and answer using explicit Schur quantization formulae) how the 
spherical vector in $\cH_{M_1;\mu_1}\times \cH_{M_2;\mu_2}$ decomposes into a direct integral of spherical vectors in $\cH_{0;\mu}$.
See \cite{Gaiotto:2023hda} for an analogous statement  in sphere quantization.

\section{Quantum groups and Schur quantisation}\label{sec:qg}
The SQED$_2$ example, both in isolation and as an building block for bigger theories, has provided us with a $*$-algebra $\fD[U_{q}(\mathrm{sl}_2),\rho]\equiv U_q(\mathfrak{sl}(2,\bC)_\bR)_{\rm S}$
which we expect to play an important role in $SL(2,\bC)$ Chern-Simons theory. In this section we will compare this proposal with  previous definitions of quantum deformations of $SL(2,\bC)$, some of which have been used
to define a quantization of $SL(2,\bC)$ CS theory.  

There are two important subtleties here, as the q-deformation of a $*$-algebra may involve both a deformation of the 
underlying algebra and a deformation of the $*$-structure. For example, the standard $U_q(\mathfrak{sl}_2)$ deformation of 
$U(\mathfrak{sl}_2)$ admits distinct $*$-structures corresponding to real forms such as $SU(1,1)$ and $SL(2,\bR)$ which are classically equivalent \cite{masuda1990unitary}.\footnote{It would be interesting to explore this statement in the context of real Schur quantization. We leave that to future work.} Furthermore, the 
quantum deformation of groups like the Lorentz group $SL(2,\bC)$, which are not semi-simple, 
is not unique \cite{WorZakr94}. 
%The standard notion of unitary $SL(2,\bC)$ representation involves a representation of the 
%$\mathfrak{sl}_2$ Lie algebra as normal operators, such that the anti-Hermitean part of the generators exponentiates to an action of $SU(2)$.
%Averages over $SU(2)$ produce further operators associated to functions on $SU(2)$. 
The quantum deformation introduced in \cite{PodWoron} is characterised by having a quantum deformation
of the compact subgroup $SU(2)$ inside of it. Another quantum deformation exhibits a deformed version of the Gauss
decomposition \cite{WorZakr92}. It will turn out that not all such features can be made fully manifest in 
the quantum deformations at the same time, but may be realised 
in more subtle ways.  
It is therefore not a priori clear which of these quantum deformations is most relevant for the goal to 
define a quantization of the $SL(2,\bC)$ CS theory.  

It is therefore not surprising that the quantum group $U_q(\mathfrak{sl}(2,\bC)_\bR)_{\rm S}$ emerging from Schur quantisation
turns out to be different from the quantum deformation of $U(\mathfrak{sl}(2,\bC)_\bR)$ 
previously studied in \cite{PodWoron,Pusz}, \cite{Buffenoir:1997ih}, here denoted as $U_q(\mathfrak{sl}(2,\bC)_\bR)_{\rm PW}$.
The quantum group $U_q(\mathfrak{sl}(2,\bC)_\bR)_{\rm PW}$ has been used to develop a quantization of complex Chern-Simons theory in 
\cite{Buffenoir:2002tx}. A comparison between the quantum Lorentz group $U_q(\mathfrak{sl}(2,\bC)_\bR)_{\rm PW}$ used in \cite{Buffenoir:2002tx} and $U_q(\mathfrak{sl}(2,\bC)_\bR)_{\rm S}$
is a natural first step to compare the corresponding quantizations of complex Chern-Simons theory. 
In this Section we will exhibit some of the differences between the two approaches. 

%\subsection{Appendix: A more mathematical take}

%Starting point is an associative algebra $A$ defined over the real numbers. Given $A$, we can naturally 
%consider $\wt{A}=A_{\rm op}$, the algebra having order-reversed multiplication\footnote{To be fully explicit: If 
%$ab=m_A(a,b)$ is the multiplication of $A$, the multiplication in $\wt{A}$ is defined by $\wt{m}(a,b)=ba$.}.
%Out of $A$ and $\wt{A}$ we can form a double $D_A=A\times\wt{A}$. Our goal is to relate two types of 
%additional structure on $A$, and on $D_A$, respectively.
%\begin{itemize}
%\item
%\item
%\end{itemize}
\subsection{The principal series of $SL(2,\bC)$}\label{compprincser}

In order motivate some of the following discussions, and to facilitate the comparison with quantum group theory, 
we shall very briefly review a few basic facts about the principal series representations
of $SL(2,\bC)$ as it arises in the closely related context of sphere quantisation \cite{Gaiotto:2023hda}.

A traditional presentation of the spherical principal series representations of $SL(2,\bC)$ involves the Hilbert space 
\begin{equation}
	\mathcal{P}_\vartheta=L^2(\mathbb{P}^1,|K|^{1-\ii\vartheta}) \qquad \qquad \vartheta\in\bR
\end{equation}
of twisted half-densities on $\mathbb{P}^1$. The holomorphic differential operators
\begin{equation}\label{EFHholo}
\mathcal{E}=\partial_x,\qquad \mathcal{H}=-2x\partial_x+J,\qquad \mathcal{F}=-x^2\partial_x+2Jx.
\end{equation}
with $J=-\frac{1}{2}+i \vartheta$ generate a representation of the central quotient 
of $U(\mathfrak{sl}_2)$, with quadratic Casimir $J(J+1)$ as global conformal transformations of $\mathbb{P}^1$. 

The anti-holomorphic differential operators 
\begin{equation}\label{EFHa-holo}
\bar{\mathcal{E}}=\partial_{\bar{x}},\qquad \bar{\mathcal{H}}=-2\bar{x}\partial_{\bar{x}}+J,\qquad 
\bar{\mathcal{F}}=-\bar{x}^2\partial_{\bar{x}}+2J\bar{x},
\end{equation}
generate a second, commuting action. With some foresight, we can identify that as an action of $U(\mathfrak{sl}_2)^\op$ with generators 
\begin{equation}\label{tildedaggerbar}
\wt{\mathcal{F}}:=-{\mathcal{E}}^{\dagger}=\bar{\mathcal{E}},\qquad 
\wt{\mathcal{H}}:=-{\mathcal{H}}^{\dagger}=\bar{\mathcal{H}},\qquad
\wt{\mathcal{E}}:=-{\mathcal{F}}^{\dagger}=\bar{\mathcal{F}},
\end{equation}
The definition is justified by the observation that the combinations
\begin{equation}\label{compactgens}
e:=\mathcal{E}-\wt{\mathcal{E}},\qquad 
f:=\mathcal{F}-\wt{\mathcal{F}}, \qquad
h:=\mathcal{H}-\wt{\mathcal{H}}.
\end{equation}
actually define the sub-algebra of rotations of $\mathbb{P}^1$ and exponentiates to an $SU(2)$ action. There is an unique 
normalizable state 
\begin{equation}
	\Phi_0(x)=(1+|x|^2)^{\ii\vartheta-1}
\end{equation}
in $\mathcal{P}_\vartheta$ which is $SU(2)$ invariant, i.e. {\it spherical} in the sense of representation theory. 
It is also cyclic: the action of $U(\mathfrak{sl}_2)$ on $\Phi_0(x) $ generates a dense basis of $\mathcal{P}_\vartheta$ 
consisting of the direct sum of all finite-dimensional $SU(2)$ representations $R_j$ of integral spin:\footnote{See
e.g. \cite{MR0274663} for a detailed discussion.}
\begin{equation}\label{P-R-decomp}
	\mathcal{P}_{\vartheta}\simeq\bigoplus_{j\in\bZ_{\geq 0}}R_j.
\end{equation} 

At this point, a careful reader can probably guess an alternative, algebraic presentation of $\mathcal{P}_\vartheta$:
$\mathcal{P}_{\vartheta}$ is a spherical unitary representation of the $*$ algebra double of the central quotient 
of $U(\mathfrak{sl}_2)$, with 
\begin{equation}\label{eq:rho}
\rho(F):=-E \qquad \qquad \rho(H) = -H \qquad \qquad \rho(E) = -F\, .
\end{equation}
It is associated to the unique trace on the central quotient 
of $U(\mathfrak{sl}_2)$, which happens to be positive when $J=-\frac{1}{2}+i \bR$. This trace is the starting point of sphere quantization. 

In order to facilitate the comparison with the Schur quantization, it is useful to recall an alternative auxiliary presentation of $\mathcal{P}_{\vartheta}$ 
which arises from a Coulomb branch perspective. The presentation is essentially a spectral decomposition into one-dimensional distributional eigenspaces for $H$ 
and is related to $L^2(\mathbb{P}^1,|K|^{1-\ii\vartheta})$ by a Mellin transform. In the Coulomb presentation, $E$ and $F$ are implemented by difference operators 
which are a $\fq \to 1$ limit of these which appear in Schur quantization. 

Schur quantization of SQED$_1$ provides a positive (twisted) trace on the central quotient of the quantum group algebra $U_q(\mathfrak{sl}_2)$ which deforms the above structure into a spherical unitary representation of a $*$-algebra double $\fD[U_{q}(\mathfrak{sl}_2),\rho]$. 
We will now review in some detail the definition of $\fD[U_{q}(\mathfrak{sl}_2),\rho]$ and then study the analogue of the $SU(2)$ action on $\mathcal{P}_{\vartheta}$.

\subsection{Real forms of quantum groups from Schur quantisation}

Recall that the algebra $U_{q}(\mathfrak{sl}_2)$ is defined by the relations
\begin{subequations}\label{Urels}
\begin{equation}
\begin{aligned}
&KE=\fq^2EK,\\
&KF=\fq^{-2}FK,
\end{aligned}\qquad \big[\,E,F\,\big]=\frac{K-K^{-1}}{\fq-\fq^{-1}}.
\end{equation}
We introduced $\fq$ with $\fq^2=q$. There is a Casimir element 
\begin{equation}
	E F + \frac{\fq^{-1}K + \fq K^{-1}}{(\fq^{-1} - \fq)^2}, %=\mu + \mu^{-1}
\end{equation}
and we will sometimes take a central quotient of the algebra fixing the Casimir to a specific value proportional to 
$\mu+\mu^{-1}$. 
%with $\mu$ being a phase. This is a q-deformation of the condition $J = -\frac12+ i\bR$.

The algebra $U_{q}(\mathfrak{sl}_2)$ is a Hopf-algebra with co-product 
\begin{align}
&\begin{aligned}
&\Delta(E)=E\otimes 1+K^{-1}\otimes E,\\ 
&\Delta(F)=F\otimes K+1 \otimes F,
\end{aligned}
\qquad \Delta(K)=K\otimes K.
\end{align}
\end{subequations}
It will be important to note that there exist
very similar, but non-isomorphic,  quantum groups often 
denoted as $U_{q}(\mathfrak{sl}_2)$ as well. One of these 
variants, in the following denoted $\hat{U}_{q}(\mathfrak{sl}_2)$ uses generators $e$, $f$, and $k$, such that
mapping $E$ to $e$, $F$ to $f$, and $K$ to $k^2$ defines 
an embedding of $U_{q}(\mathfrak{sl}_2)$ into $\hat{U}_{q}(\mathfrak{sl}_2)$.
Other variants 
have an additional generator $H$ such that 
$K=q^{2H}$, with $E$, $F$ and $H$ satisfying the relations of $\mathfrak{sl}_2$. 

In order to define a $*$-algebra double $\fD[U_{q}(\mathfrak{sl}_2),\rho]$ which deforms the $*$-algebra controlling 
unitary $SL(2,\bC)$ representations in the sense described above, we need to choose an automorphism $\rho$. It turns out that there
are multiple possible choices with the same $q \to 1$ limit. Schur quantization gives a distinguished choice $\rho_{\rm S}$:
\begin{align}\label{rho-def}
&\rho_{\rm S}(E)=-\fq KF,\qquad \rho_{\rm S}(F)=-\fq K^{-1}E,\qquad \rho_{\rm S}(K)=K^{-1}.
\end{align}
leading to the $*$-algebra double $\fD[U_{q}(\mathfrak{sl}_2),\rho]$. 

We can contrast this with a naive q-deformation
\begin{equation}\label{rho0-def}
\rho_{0}(E)=-F,\qquad \rho_{0}(F)=-E,\qquad \rho_{0}(K)=K^{-1}.
\end{equation}
Other possibilities would include e.g.
\begin{equation}\label{rho0-def}
\rho_{n}(E)=-\fq^n K^n F,\qquad \rho_{0}(F)=-\fq^n K^{-n} E,\qquad \rho_{n}(K)=K^{-1}.
\end{equation}
We will see that $\rho_{\rm S}$ has some particularly nice features. For example, an analysis based on \cite{Klyuev_2022} indicates that positive twisted traces only exist 
for $n \leq 1$ and are not unique for $n \leq 0$. 

Another nice feature is that the conditions for a spherical vector can be written as
\begin{align}\label{sph-coprod}
&(E+\fq K^{-1} F^\dagger )|1\rangle =0,\qquad (F K^\dagger+\fq^{-1} E^\dagger )|1\rangle=0 ,\qquad K K^\dagger |1\rangle=0.
\end{align}
We will see later that 
the combinations appearing in \rf{sph-coprod} are related to the co-product of the $U_q(\mathfrak{sl}_2)$ generators,
and that they define a quantum deformed analog of the compact sub-algebra of $\mathfrak{sl}(2,\bC)$. 
One may expect that the rest of the representation will decompose into a direct sum of finite-dimensional 
representations of this algebra, with each integral spin appearing once. We will demonstrate at the end of this Section
that the actual story is slightly more complicated, but reduces to \rf{P-R-decomp} when $q\rightarrow 1$.

\subsection{Quantum group representations from Schur quantisation}\label{q-group-rep-Schur}
We will now review and extend the discussion of the Schur quantization representation of $\fD[U_{q}(\mathfrak{sl}_2),\rho]$. In Section \ref{sec:examples} we 
gave an auxiliary presentation of the representation by finite 
difference operators on the Hilbert space $L^2(\bZ\times S^1)$,\footnote{To simplify the 
notation we often  do not distinguish the operators representing $A_\fq$ from the generators 
of the abstract algebra $A_\fq$.} 
\begin{subequations}\label{Schur-rep}
\begin{equation}\label{uv-rep}
\pi(E)=\frac{\fq\,v^{-1}u_-}{\fq^{-1}-\fq},\qquad \pi(K)=v,\qquad \pi(F)=\frac{u_+}{\fq-\fq^{-1}}, 
\end{equation}
where $u_\pm$ can be represented as 
\begin{equation}
u_+=(1+\fq\mu v)u, \qquad u_-=u^{-1}(1+\fq\mu^{-1}v),
\end{equation}
in terms of operators  $u$, $v$ satisfying the Weyl-algebra
$uv=\fq^2vu$
defined as 
\begin{equation}
u g_n(\theta)=g_{n+1}(\theta-\ii\hbar), \qquad v g_{n}(\theta)=\fq^{{n}}e^{\ii\theta}g_n(\theta).
\end{equation} 
\end{subequations}
We are here representing elements of $L^2(\bZ\times S^1)$ by collections $(g_n)_{n\in \bZ}$ of functions $g_n\in L^2(S^1)$ such that
$\sum_{n\in\bZ}\lVert g_n\rVert^2_{L^2(S^1)}<\infty$. This should be compared with the Mellin transform of the $\mathcal{P}_{\vartheta}$ representation.

It is useful to parameterize $\mu=-\fq^{2\ii\vartheta}$, and let us note that the representation introduced above is  
equivalent to a representation of the following form
\begin{equation}\label{Qdiff-Uqsl2-mod}
{v}f_n(p)=\fq^{2m}f_n(p),
\qquad {u}f_n(p)=f_{n+1}(p-\ii ), \qquad m=\frac{1}{2}(n+\ii p).
\end{equation}
%using the notation $$.
One may note that 
the representation \rf{Qdiff-Uqsl2-mod} can be restricted to functions $f_n(p)$ which satisfy 
$f_n(p+\frac{1}{\log \fq})=f_n(p)$.
Introducing the notation $J=-\frac{1}{2}+\ii\vartheta $ leads to  a representation of $U_q(\mathfrak{sl}_2)$ by finite difference
operators of the form
\begin{equation}\label{Qdiff-Uqsl2}
\begin{aligned}
\mu\mathsf{E}_\fq f_n(p)&=\big[J+1-m\big]f_{n-1}(p+\ii),\\
\mathsf{F}_\fq f_n(p)&=\big[J+1+m\big]f_{n+1}(p-\ii),
\end{aligned} 
 \quad 
 \mathsf{K}_\fq f_n(p)=\fq^{2m}f_n(p),\quad
 [x]:=\frac{1-\fq^{2x}}{1-\fq^{2}}.
\end{equation}
It is easy to see that the representation \rf{Qdiff-Uqsl2} reduces to the representation 
\begin{equation}\label{Qdiff-Usl2}
\begin{aligned}
\mathsf{E}f_n(p)&=\big(J+1-m)f_{n-1}(p+\ii),\\
\mathsf{F}f_n(p)&=(J+1+m)f_{n+1}(p-\ii),
\end{aligned} 
 \qquad 
 \mathsf{H}f_n(p)=mg_n(p),
% \quad m:=\frac{1}{2}(\ii p+n),
\end{equation}
of $\mathfrak{sl}_2$
in the limit $\hbar\rightarrow 1$. 
In order to compare the  representation \rf{Qdiff-Usl2} with the principal series representations 
of $SL(2,\bC)$, 
let us note that the Mellin transformation
\begin{equation}
f_n(p):=\int_{\bC}d^2x\;e^{\ii n \arg(x)}|x|^{-2(j+1)+\ii p}f(x)=\int_{\bC}\frac{d^2x}{|x|^{2j+2}}\; x^{\frac{1}{2}(\ii p+n)}\bar{x}^{\frac{1}{2}(\ii p-n)} F(x),
\end{equation} 
maps the finite difference operators $\mathsf{E}$, $\mathsf{F}$ and $\mathsf{H}$ to the differential operators 
$\mathcal{E}$, $\mathcal{F}$ and $\mathcal{H}$
generating the principal series $\mathcal{P}_\vartheta$ of $SL(2,\bC)$, respectively. 

Conversely, we can do a Fourier transform on $L^2(\bZ\times S^1)$ and diagonalize the action of $u$, with $\mu v^{-1}$ acting by a rescaling. 
Then $\pi(E)$ is essentially a q-derivative with respect to $u$ and is a natural deformation of $\mathcal{E}$. The other generators are identified with q-differential operators which 
deform $\mathcal{H}$ and $\mathcal{F}$.

\subsubsection{Spherical vectors}

Suppose that we were simply given the representation of $E$, $F$ and $K$ on $L^2(\bZ\times S^1)$ and a  choice of 
the automorphism $\rho$. 
We could then define the action of $\fD[U_{q}(\mathfrak{sl}_2),\rho]$ by acting on $E$, $F$ and $K$ 
with $\rho$, and taking Hermitean conjugates. The corresponding 
spherical vector can be represented by  a wave-function of the form $g_n(\theta)=\delta_{n,0}\varphi_{\rm S}^{}(\theta)$,
where  the condition on $n$ follows from $K|1\rangle=\wt{K}|1\rangle$. Choosing $\rho=\rho_{\rm S}$, we will find
\begin{align}
(\fq-\fq^{-1})F g_n(\theta)& %=(1+\fq\mu\, \fq^{{n}}e^{\ii\theta})\delta_{n+1,0}\varphi_0(\theta-\ii\hbar)
=(1+\mu\, e^{\ii\theta})\delta_{n+1,0}\,\varphi_{{\rm S}}^{}(\theta+\ii\hbar),\\
(\fq-\fq^{-1})\wt{F}g_n(\theta) &
%=\fq^{-1}\mu\,\fq^{-{n}}e^{\ii\theta}(1+\fq^{-1}\mu^{-1} \fq^{-{n}}e^{\ii\theta})\delta_{n+1,0}\varphi_0(\theta+\ii\hbar)\\ &
=\mu e^{\ii\theta}(1+\mu^{-1} e^{\ii\theta})\delta_{n+1,0}\,\varphi_{{\rm S}}^{}(\theta-\ii\hbar).
%\\&=\fq(1+\mu e^{-\ii\theta})\delta_{m+2,0}\varphi_0(\theta+\ii\hbar)
\label{wtF-act}\end{align}
The condition $Fg_n(\theta)=\wt{F}g_n(\theta)$ can  be solved by choosing
\[
\varphi_{\rm S}(\theta)= \prod_{k=0}^{\infty}\frac{1}{(1+\fq^{2k+1}\mu^{-1}\, e^{-\ii\theta})(1+\fq^{2k+1}\mu^{-1} e^{\ii\theta})}.
\]
This coincides with the standard expression for the spherical vector and we recover the structure of Schur quantization. 

The same representation of $E$, $F$ and $K$ can also be promoted to a representation of other $*$-algebra doubles such as $\fD[U_q(\mathfrak{sl}_2),\rho_0]$. 
It is not difficult to find wavefunctions which satisfy modified spherical conditions. Indeed, the theta function 
\[
\vartheta_\fq(v)=\prod_{k=0}^{\infty}{(1+\fq^{2k+1} v)(1+\fq^{2k+1} v^{-1})}.
\]
commutes with the the tilde generators and satisfies 
\[
u \vartheta_\fq(v)=\prod_{k=0}^{\infty}{(1+\fq^{2k+3} v)(1+\fq^{2k-1} v^{-1})} u = \vartheta_\fq(v) \fq^{-1} v^{-1} u .
\]
Then the wave-function $\vartheta_\fq(v)g_n(\theta)$ satisfies the constraints for a spherical vector for $\fD[U_q(\mathfrak{sl}_2),\rho_0]$.
A more general product of $\theta$ functions would be appropriate for $\rho_n$ with $n <0$. Solving the spherical conditions for $n>1$, instead, seems to require negative powers of the theta function, introducing poles into the wave-function 
of the spherical vector. It therefore seems unlikely that spherical vectors satisfying all relevant conditions can exist for 
$\fD[U_q(\mathfrak{sl}_2),\rho_n]$, with $n>1$.

\subsubsection{Positive traces}

As discussed at the beginning of Section \ref{sec:schur}, there is a direct correspondence between 
spherical unitary representations of the Schur double of an algebra $A$,
and positive traces on $A$. 
In order to represent 
the corresponding positive traces explicitly, we may introduce a grading $\nu$ on $U_{q}(\mathfrak{sl}_2)$ by counting 
the powers of $E$ positively, and the powers of $F$ negatively. The traces are supported on the component with grade zero,
which can be represented as functions $a=a_0(K)$, with $a_0$ being a Laurent polynomial. 
The positive traces associated to the different choices $\rho_{\rm S}^{}$, $\rho_{0}^{}$
can now be represented as expectation values of $a=a_0(K)$ defined by the spherical vectors 
\begin{equation}
\mathrm{Tr}(a)
=\int_{0}^{2\pi}dW(\theta)\, a_0(e^{\ii \theta}),\qquad dW(\theta)=\left\{ 
\begin{aligned} & |\varphi_{\rm S}^{}(\theta)|^2d\theta \;\,\text{for twist}\;\,\rho_{\rm S}^{},\\
&|\vartheta_\fq(e^{i \theta})\varphi_{\rm S}^{}(\theta)|^2d\theta  \;\,\text{for twist}\;\,\rho_{0}.
\end{aligned}\right.
\end{equation}
In this way it is becoming fully explicit how a change of the automorphism $\rho$ is reflected by a change of the measure
in the integral representations of the positive traces.

Positive traces on the central quotient of $U_{q}(\mathfrak{sl}_2)$ have been classified in \cite{Klyuev_2022}. 
It seems likely that these results can help classifying the spherical unitary representations of different $*$-algebra doubles. 
Physically, the extra theta functions in the integral can be interpreted as the contribution of extra surface defect insertions. 

\subsection{Comparison with other definitions of the quantum Lorentz group}

A quantum group called quantum Lorentz group was first constructed in \cite{PodWoron}. 
The classification of its unitary representation has been found in \cite{Pusz}, and the harmonic analysis of this 
quantum group was developed in \cite{Buffenoir:1997ih}. It has been demonstrated in \cite{WorZakr92}
that there exist other quantum deformations of the group $SL(2,\bC)$. A classification of quantum deformations
of the Lorentz group was given in in \cite{WorZakr94}. 

We shall here compare the quantum Lorentz group from Schur quantisation to the quantum  group
defined in \cite{PodWoron}, which is the quantum deformation of the Lorentz group that has attracted most 
attention up to now, and which has been used in a previous approach to the quantisation of complex
Chern-Simons theory \cite{Buffenoir:2002tx}.

It should be noted that previous studies of quantum Lorentz groups have often focused attention on 
quantum deformations $\mathrm{Fun}( \mathrm{SL}_\fq(2,\bC)_{\bR})$ of the algebra of functions on 
$\mathrm{SL}(2,\bC)_{\bR}$. We have so far mainly discussed the 
quantum deformations $U_q(\mathfrak{sl}(2,\bC)_{\bR})$ of the universal enveloping algebra
of the Lie algebra of $\mathrm{SL}(2,\bC)_{\bR}$. While it is certainly natural to expect that the solutions to these two problems 
are pairwise related by quantum group dualities, it will require further work to establish the relations in detail.
The following discussion will therefore restrict attention to some aspects where a direct comparison is possible 
on the basis of the known results. 

\subsubsection{The  quantum Lorentz group of Podles and Woronowicz}

The quantum Lorentz group  considered in
\cite{PodWoron,Pusz,Buffenoir:1997ih} is related by quantum group duality to 
a quantum deformation of $U_q(\mathfrak{sl}(2,\bC)_{\bR})_{\rm PW}$  which is
isomorphic to
$
U_q(\mathfrak{su}(2))\otimes \mathrm{Pol}(SU_q(2))
$ 
as a vector space, with 
\begin{itemize}
\item
$U_q(\mathfrak{su}(2))$ being the real form of the 
Hopf algebra $\hat{U}_q(\mathfrak{sl}_2)$ having
generators $e$, $f$, and $k$, and relations 
\begin{equation}\label{efk-rels-def}
\begin{aligned}
&ke=\fq\,ek,\\
&kf=\fq^{-1}fk,
\end{aligned}\qquad \big[\,e,f\,\big]=\frac{k^2-k^{-2}}{\fq-\fq^{-1}}, 
\end{equation}
 star-structure
\begin{equation}\label{efk-star}
k^\ast=k, \qquad e^\ast=\fq^{-1}f,\qquad f^\ast=\fq e,
\end{equation}
and co-product
\[
\begin{aligned}
&\Delta(e)=e\otimes k+k^{-1}\otimes e,\\
&\Delta(f)=f\otimes k+k^{-1}\otimes f,
\end{aligned}
\qquad \Delta(k)=k\otimes k,
\]
\item and $\mathrm{Pol}(SU_q(2))$ is the Hopf algebra
with generators $a$, $b$, $c$ and $d$, relations
\begin{equation}\label{abcd-rel}
\begin{aligned}
\fq\,ab=ba,\qquad \fq\, ac=ca,\\
\fq\,bd=db,\qquad \fq\, cd=dc,
\end{aligned}
\qquad
bc=cb,\qquad
\begin{aligned}
&ad-da=(\fq^{-1}-\fq)bc,\\
&ad-\fq^{-1}da=1,
\end{aligned}
\end{equation}
star-structure
\begin{equation}
\label{abcd-star}
a^\ast=d, \qquad b^\ast=-\fq^{-1}c,\qquad c^\ast=-\fq b,
\end{equation}
and
co-product
\begin{equation}
\begin{aligned}
&\Delta(a)=a\otimes a+b\otimes c,\\
&\Delta(c)=c\otimes a+d\otimes c,
\end{aligned}\qquad
\begin{aligned}
&\Delta(b)=b\otimes d+a\otimes b,\\
&\Delta(c)=c\otimes b+d\otimes d,
\end{aligned}
\end{equation}
\end{itemize}
The algebra structure on $U_q(\mathfrak{su}(2))\otimes \mathrm{Pol}(SU_q(2))$ 
defined in \cite{PodWoron} also involves
 the mixed relations
%\begin{equation}
\begin{align}\label{q-Lorentz-mixed}
&kc=\fq\,ck,\qquad kb=\fq^{-1}bk,\qquad ka=ak, \qquad kd=dk,\\
&[e,c]=0,\quad [e,b]=\fq^{-1}(ka-k^{-1}d), \quad ae-\fq ea=k^{-1}c,\quad ed-\fq\,de=ck,\notag\\
&[f,b]=0,\quad [f,c]=\fq(kd-k^{-1}a), \quad fa-\fq^{-1}af=bk,\quad df-\fq^{-1}fd=k^{-1}b.
\notag\end{align}
%\end{equation}
%The definition of the algebra structure defining the quantum Lorentz group is defined by 
%the collection of mixed relations listed in \cite[Formula (131)]{Buffenoir:1997ih}.
Central elements of this algebra can be constructed as 
\begin{equation}\label{casimirs}
\begin{aligned}
&\Omega_+=+\fq(\fq-\fq^{-1})eb+\fq^{-1}ka+\fq k^{-1}d,\\
&\Omega_-=-\fq^{-1}(\fq-\fq^{-1})fc+\fq^{-1}k^{-1}a+\fq kd.
\end{aligned}
\end{equation}
Our goal is to compare this version of the quantum Lorentz group to the quantum group 
$U_q(\mathfrak{sl}(2,\bC)_{\bR})_{\rm S}$ from Schur
quantisation.
We shall use the notation  
$\mathfrak{D}_{\rm\scriptscriptstyle PW}$ for the complex algebra having generators $a,b,c,d,e,f,k$,
and relations \rf{efk-rels-def}, \rf{abcd-rel} and \rf{q-Lorentz-mixed}, 
and $U_\fq(\mathfrak{sl}(2,\bC)_{\bR})_{\rm PW}$ for the  for 
quantum deformation of $U(\mathfrak{sl}(2,\bC)_{\bR})$ defined 
as a  real form $\mathfrak{D}_{\rm\scriptscriptstyle PW}$ using the  star structures 
\rf{efk-star} and \rf{abcd-star} above.

\subsubsection{Algebraic structure of the principal series of $U_\fq(\mathfrak{sl}(2,\bC)_{\bR})_{\rm PW}$}
\label{princserPW}

The interpretation of $U_\fq(\mathfrak{sl}(2,\bC)_{\bR})_{\rm PW}$ %$\mathfrak{D}_{\rm\scriptscriptstyle PW}$
 as a quantum deformation of the Lorentz group 
can be supported in particular by comparing the structure of its unitary representations described in \cite{Pusz}
to the algebraic  structure \rf{P-R-decomp} of the principal series representations of $SL(2,\bC)$. 
As a preparation for a similar analysis in the case of $U_\fq(\mathfrak{sl}(2,\bC)_{\bR})_{\rm S}$  
we shall here outline a simple approach for the case of spherical principal series representations.

Spherical principal series representations of $U_\fq(\mathfrak{sl}(2,\bC)_{\bR})_{\rm PW}$
%$\mathfrak{D}_{\rm\scriptscriptstyle PW}$ 
can be generated by the 
action of $\mathrm{Pol}(SU_q(2))$ on a  vector $v_0$ transforming trivially under the 
sub-algebra $U_q(\mathfrak{su}(2))$ of %$\mathfrak{D}_{\rm\scriptscriptstyle PW}$. 
$U_\fq(\mathfrak{sl}(2,\bC)_{\bR})_{\rm PW}$.
We shall be interested in representations $\mathcal{P}_{\vartheta.\fq}$ having a diagonal action of the 
Casimir generators $\Omega_{\pm}$ with eigenvalue $2\cos(2\hbar\vartheta)$.
We are going to argue that this implies a structure of the representation of the following form
\begin{equation}\label{princ-decomp}
\mathcal{P}_{\vartheta,\fq}\simeq\bigoplus_{j\in\bZ_{\geq 0}}\mathcal{R}_{j,\fq}, %\qquad k\mathcal{P}_j=\fq^{p}\mathcal{P}_p.
\end{equation}
with $\mathcal{R}_{j,\fq}$ being irreducible $(2j+1)$-dimensional  representations of $U_q(\mathfrak{su}(2))$.
To see this, one may first note that the relations \rf{q-Lorentz-mixed} imply 
that $v^j_j:=c^{j} v_0$ satisfies the highest weight condition $e v^j_j:=0$.  Acting with $f^{j-m}$ on $v^j_j$ allows one to define 
vectors $v^j_m$, $m=-j,\dots,j$, generating $\mathcal{R}_{j,\fq}$. 
This will allow us to 
establish \rf{princ-decomp} inductively. In order to understand 
the recursive structure, let us consider the subspace $\mathcal{R}_{j,\fq}^+$ generated by 
linear combinations of vectors of the form $gv^j_m$, with $g\in\{a,b,c,d\}$ and $m=-j,\dots,j$.
The space $\mathcal{R}_{j,\fq}^+$ is $3(2j+1)$-dimensional since $\Omega_+v^j_m=\omega_+v^j_m$
implies a relation between $av^j_m$, $bv^j_m$ and $dv^j_m$.
It decomposes into eigenspaces of $k$ with eigenvalue $\fq^n$, $n\in\bZ$.
The eigenspace with eigenvalue $\fq^{j+1}$ is one-dimensional, 
generated by $c^{j+1}v_0$. 
The subspace with eigenvalue $q^j$ is two-dimensional, spanned by the vectors $av^j_j$ and $dv^j_j$. 
It contains the vector $\wt{v}^j_j=\fq^{j+1} av^j_j+\fq^{-j-1} dv^j_j$ satisfying the highest weight condition 
$e\wt{v}^j_j=0$. One may note, however, that $\wt{v}^j_j=\Omega_+ {v}^j_j=\omega_+{v}^j_j$, which
is proportional to ${v}^j_j$. In a similar way one may see that the 
eigenspace with eigenvalue $\fq^{j-1}$ is three-dimensional, and contains the vector ${v}^{j-1}_{j-1}$.
Using these observations one may easily see that $\mathcal{P}_{\vartheta,\fq}$
contains each $\mathcal{R}_{j,\fq}$ only once. 

\subsection{Algebraic structure of the spherical principal series of $U_q(\mathfrak{sl}(2,\bC)_\bR)_{\rm S}$}

We are next going to investigate the algebraic structure of the subspace\footnote{Which may be expected to be dense in 
$L^2(\bZ\times S^1)$.}  $U|1\rangle$ 
generated by the action of $U= U_q(\mathfrak{sl}_2)$ on the spherical vectors $|1\rangle$ in 
$L^2(\bZ\times S^1)$ defined  in Section \ref{q-group-rep-Schur}. Comparison 
with the principal series of $SL(2,\bC)$ suggests that the subspace  $U|1\rangle$ 
is dense in $L^2(\bZ\times S^1)$. We are going to find a structure which is similar to, but also 
different from the structure of the spherical principal series of  %$\mathfrak{D}_{\rm\scriptscriptstyle PW}$.
$U_\fq(\mathfrak{sl}(2,\bC)_{\bR})_{\rm PW}$.

\subsubsection{Quantum analogs of the compact sub-algebras}
For concisenes, denote $U= U_q(\mathfrak{sl}_2)$.
To begin with, let us introduce two commuting copies $U_l$, $U_r$ of $U$, and observe that 
\begin{equation}\label{Xrl-wtX}
\begin{aligned}
&K=K_l, \qquad E=E_l,\qquad F=F_l,\\ &\wt{K}^{-1}=K_r, \qquad
\wt{E}=-E_rK_r,\qquad \wt{F}=-K_r^{-1}F_r,
\end{aligned}
\end{equation}
defines a map from $U\otimes U_{\rm op}$ into ${U}_l\otimes {U}_r$.
The definining conditions of $|1\rangle$, combined with \rf{Xrl-wtX}, imply 
%\[
%E_l|1\rangle =-E_rK_r^{2}|1\rangle,\qquad F_l|1\rangle=-{K}_r^{-2}F_r|1\rangle,\qquad 
%K_l|1\rangle=K_r^{-1}|1\rangle
%\]
%and therefore
\begin{equation}\label{EFKhat}
\begin{aligned}
&\hat{E}\,|1\rangle=(E_l ^{}+{K}_l^{-1}{E}_r^{})|1\rangle=0, \\ 
&\hat{F}\,|1\rangle=(F_l ^{}{K}_r+ {F}_r^{})|1\rangle=0,\end{aligned}
\qquad 
\hat{K}\,|1\rangle=K_lK_r|1\rangle=|1\rangle.
\end{equation}
We see that the spherical vector transforms trivially under the sub-algebra ${U}_q^+(\mathfrak{sl}_2)$
generated by $\hat{E}$, $\hat{F}$, and $\hat{K}$. One may note that $\hat{E}$, $\hat{F}$, and $\hat{K}$
are defined by taking the co-products $\Delta$ of ${E}$, ${F}$, and ${K}$, respectively. 

The opposite co-product $\Delta'$ is defined by exchanging the factors in the tensor product. We may use this
observation to identify another 
sub-algebra ${U}_\fq^-(\mathfrak{sl}_2)$ of $U_l\otimes U_r$ acting trivially on $|1\rangle$, 
generated by 
$\hat{E}'$, $\hat{F}'$, and $\hat{K}'$, 
\begin{equation}\label{EFKprimehat}
\hat{K}'=K_lK_r,\qquad
 \hat{E}'=\fq^{-1}E_r+\fq\, {E}_l^{}K_r^{-1},\qquad
 \hat{F}'=\fq\, {K}_lF_r ^{}+ \fq^{-1}{F}_l^{}.
\end{equation}
The factors of $\fq^{\pm 1}$ are needed to satisfy $ \hat{E}'|1\rangle=0=\hat{F}'|1\rangle$. They can be introduced into
the definition of the opposite co-product by means of the automorphism of $U_\fq(\mathfrak{sl}_2)$ 
scaling $E$ and $F$ inversely. 

As the definition of the generators 
$\hat{E}$, $\hat{F}$ and $\hat{K}$  is related to the co-product of ${U}_\fq(\mathfrak{sl}_2)$, while 
$\hat{E}'$, $\hat{F}'$ and $\hat{K}'$ 
are similarly related to  the opposite co-product, it follows that
the isomorphism between 
${U}_\fq^+(\mathfrak{sl}_2)$ and ${U}_\fq^-(\mathfrak{sl}_2)$
is described by the universal R-matrix. We will see that these structures 
offer a replacement for the compact sub-group in the quantum deformation of the 
Lorentz group from Schur quantisation.

\subsubsection{Module structure}

%Let us  furthermore note that replacing $E_i$ by $K_iE_i$, and $F_i$ by $F_i ^{}{K}_i^{-1}$, for
%$i=l,r$ defines an homomorphism of algebras. 
The identification of  the algebraic structure of  $U\otimes U_{\rm op}|1\rangle$ 
will be facilitated by the following observations.
While square-roots of $K_r$ and $K_s$ are not well-defined in the representation on $L^2(S^1\times \bZ)$, it 
is possible to define square-roots denoted as $k_lk_r$ and $k_lk_r^{-1}$ of $K_lK_r$ and $K_l^{}K_r^{-1}$, 
respectively.
This allows us to define 
%\begin{subequations}\label{a-to-k-def-first}
\begin{align}\label{efk-def-mod-mod}
%&k=K_l{K}_r,\qquad
%\begin{aligned}
%&%&=\fq^{\frac{1}{2}}E\wt{K}+\fq^{-\frac{1}{2}}K^{-1}\wt{E},\\
 %K_lE_l^{}{K}_r^{}+K_l^{-1}K_r{E}_r^{}=k(E_l+K_l^{-2}{E}_r^{})=:ke,\\
%&%&=\fq^{-\frac{1}{2}}F\wt{K}+\fq^{+\frac{1}{2}}K^{-1}\wt{F},\\
 % F_l ^{}{K}_l^{-1}{K}_r^{}+K_l^{-1} {F}_r^{}{K}_r^{-1}=(F_l ^{}{K}_r^{2}+ {F}_r^{})k^{-1}=:fk^{-1},
%\end{aligned}\\
%&k=k_l{k}_r,\qquad\qquad
% e =e_l+k_l^{-2}{e}_r^{},\qquad
 % f=f_l ^{}{k}_r^{2}+ {f}_r^{},\\
\begin{aligned}
&a=k_l^{}k_r^{-1},\\
&c=(1-\fq^2)k_l k_rE_r,
\end{aligned}\qquad\qquad
\begin{aligned}
&b=(1-\fq^{-2})F_l^{}(k_lk_r)^{-1},\\
&d=(k_l^{}k_r^{-1})^{-1}-(\fq-\fq^{-1})^2F_l^{}(k_l^{}k_r^{-1})^{-1}E_r^{}.
\end{aligned}%\label{abcd-EFK-mod-mod}
\end{align}
%\end{subequations}
Formulae  \rf{EFKhat} and \rf{efk-def-mod-mod} define an embedding of 
$\mathfrak{D}_{\rm PW}$ into ${U}_l\otimes{U}_r$.\footnote{A closely related
observation was made in the Appendix of \cite{Buffenoir:1997ih}.}
%In order to compare the quantum group appearing in the context of Schur quantisation 
%with the quantum Lorentz group studied in \cite{Buffenoir:1997ih}, let us note that 
It is not hard to show that 
%\[
%\begin{aligned}
%d|1\rangle&=\big[K_l^{-1}K_r^{}-(\fq-\fq^{-1})^2F_l^{}K_l^{-1}K_r^{}E_r^{}\big]|1\rangle\\
%&=\big[K^{-2}_l-(\fq-\fq^{-1})^2F_l^{}K_l^{-1}(-K_l^{2}E_l)K_l^{-1}\big]|1\rangle
%=\big[K^{-1}+\fq(\fq-\fq^{-1})^2 FE\big]|1\rangle\\ &=\big[K^{-1}+\fq (\fq-\fq^{-1})^2 C-\fq(\fq K+\fq^{-1}K^{-1})\big]|1\rangle
%=\big[\fq\omega-\fq^2 K\big]|1\rangle,\\
%c|1\rangle&=(1-\fq^2)K_l K_rE_r|1\rangle=(1-\fq^2)K_l K_r(-K_l^{2}E_l)|1\rangle=\fq(\fq^{2}-1)KE|1\rangle,\\
%b|1\rangle&=(1-\fq^{-2})F_l^{}K_l^{-1}K_r^{-1}|1\rangle=(1-\fq^{-2})F|1\rangle,\qquad a|1\rangle=K|1\rangle,
%\end{aligned}
%\]
\[
\bigg(\,
\begin{matrix}
a & b \\
c & d
\end{matrix}
\,\bigg)|1\rangle=\bigg(\,
\begin{matrix}
 K & (1-\fq^{-2})F\\
 \fq(\fq^{2}-1)KE & \fq\omega-\fq^2 K
 \end{matrix}\,\bigg)|1\rangle, 
 \]
using that the Casimir $C={(\fq-\fq^{-1})^{2}}FE+{\fq K+\fq^{-1}K^{-1}}$ acts diagonally with eigenvalue $\omega%=-\mu-\mu^{-1}
=2\cos(2\hbar\vartheta)$.
We note that only positive powers of $K$ appear in these expressions, and that 
$a |1\rangle$, $b|1\rangle$, $c|1\rangle$ and $d|1\rangle$ generate a three-dimensional representation 
of the sub-algebra ${U}_\fq(\mathfrak{sl}_2)$ generated by $e$, $f$ and $k$.

The arguments used in Section
\ref{princserPW} can easily be adapted to show that the 
subspace $\mathcal{P}_{\vartheta,\fq}^-$ of $U|1\rangle$ generated by $KE$, $F$, and $K$
decomposes as module of ${U}_\fq(\mathfrak{sl}_2)$ in the same way as the right side of 
\rf{princ-decomp}.

%\subsubsection{Change left and right}

Exchanging the indices $l$ and $r$, and taking into account the scaling by factors of 
$\fq$ noted above, defines another realisation of $\mathfrak{D}_{\rm PW}$  by combining
\begin{align}%\label{efk-def-mod-mod}
%&k=K_l{K}_r,\qquad
%\begin{aligned}
%&%&=\fq^{\frac{1}{2}}E\wt{K}+\fq^{-\frac{1}{2}}K^{-1}\wt{E},\\
% K_lE_l^{}{K}_r^{}+K_l^{-1}K_r{E}_r^{}=k(E_l+K_l^{-2}{E}_r^{})=:ke,\\
%&%&=\fq^{-\frac{1}{2}}F\wt{K}+\fq^{+\frac{1}{2}}K^{-1}\wt{F},\\
%  F_l ^{}{K}_l^{-1}{K}_r^{}+K_l^{-1} {F}_r^{}{K}_r^{-1}=(F_l ^{}{K}_r^{2}+ {F}_r^{})k^{-1}=:fk^{-1},
%\end{aligned}\\
%&k'=k_l^{-2}{k}_r^{-2},\qquad
% e' =e_r+{e}_l^{}k_r^{-2},\qquad
 % f'={k}_l^{2}f_r ^{}+ {f}_l^{},\\
&\begin{aligned}
&a'=k_l^{-1}k_r^{},\\
&c'=(1-\fq^2)  k_lk_r\fq E_l,
\end{aligned}\qquad\qquad
\begin{aligned}
&b'=(1-\fq^{-2}) \fq F_r^{}(k_lk_r)^{-1},\\
&d'=k_l^{}k_r^{-1}-(\fq-\fq^{-1})^2\fq^2k_l^{}E_l^{}F_r^{}k_r^{-1},\\
\end{aligned}\label{abcd-EFK-mod-mod}
\end{align}
with  \rf{EFKprimehat}. As above we may compute
%\end{subequations}
%Let us furthermore note that ($\wt{F}=-k_r^{-2}F_r$)
%\[
%\begin{aligned}
%d'|1\rangle&=\big[k_l^{}k_r^{-1}-(\fq-\fq^{-1})^2\fq^{2}k_l^{}E_l^{}F_r{}k_r^{-1}\big]|1\rangle
%=\big[k^{2}_l-(\fq-\fq^{-1})^2\fq^{2} k_l^{}E_l^{}k_l(-k_r^{2}\wt{F})\big]|1\rangle\\
%&=\big[k^{2}_l-(\fq-\fq^{-1})^2\fq^{2} k_l^{}E_l^{}k_l(-F_lk_l^{-2})\big]|1\rangle
%=\big[K+(\fq-\fq^{-1})^2 \fq EF\big]|1\rangle\\ &=\big[K+\fq (\fq-\fq^{-1})^2 C-\fq(\fq^{-1} K+\fq K^{-1})\big]|1\rangle
%=\big[\fq\omega-\fq^2 K^{-1}\big]|1\rangle,\\
%c'|1\rangle&=(1-\fq^2)\fq \,k_lE_lk_r |1\rangle=\fq^2(1-\fq^{2})E|1\rangle,\qquad a'|1\rangle=k_l^{-1}k_r^{}|1\rangle=K^{-1}|1\rangle, \\
%b'|1\rangle&=(1-\fq^{-2})\fq k_l^{-1}F_r^{}k_r^{-1}|1\rangle=(\fq-\fq^{-1})(-k_l^{-2}F_l^{})|1\rangle=(\fq^{-1}-\fq)K^{-1}F|1\rangle.
%\end{aligned}
%\]
\[
\bigg(\,
\begin{matrix}
a' & b' \\
c' & d'
\end{matrix}\,
\bigg)|1\rangle=\bigg(\,
\begin{matrix}
 K^{-1} & (\fq^{-1}-\fq)K^{-1}F\\
 \fq^2(1-\fq^{2})E & \fq\omega-\fq^2 K^{-1}
 \end{matrix}\,\bigg)|1\rangle
 \]
Only negative powers of $K$ appear in these expressions. 
Considering the subspace $\mathcal{P}_{\vartheta,\fq}^-$ of $U|1\rangle$ generated by $E$, $FK^{-1}$, and $K^{-1}$, 
one may again use 
the arguments from Section
\ref{princserPW}  to show that the vector space $U_-|1\rangle$ also 
decomposes as module of $\hat{U}_\fq(\mathfrak{sl}_2)$ generated by $e'$, $f'$ and $k'$ as the right side of 
\rf{princ-decomp}.

Taken together we find 
\begin{equation}\label{Princ-Rpm-dec}
U|1\rangle \simeq |1\rangle\oplus \bigoplus_{j\in\bZ_{> 0}}\big(\mathcal{R}_{j,\fq}^+\oplus\mathcal{R}_{j,\fq}^-\big),
\end{equation}
where $\mathcal{R}_{j,\fq}^+$ and $\mathcal{R}_{j,\fq}^-$ are $(2j+1)$-dimensional representations of 
the two sub-algebras  ${U}_\fq^+(\mathfrak{sl}_2)$
and ${U}_\fq^-(\mathfrak{sl}_2)$, respectively. 

As the difference between $\Delta$ and 
$\Delta'$ disappears in the classical limit, we expect that the classical limits of $\mathcal{R}_{j,\fq}^+$ 
and $\mathcal{R}_{j,\fq}^-$ will coincide, reproducing the direct summands $R_j$ in the decomposition 
\rf{P-R-decomp} of the 
spherical principal series representations of $SL(2,\bC)$.

\subsection{Existence of inequivalent quantum deformations of $SL(2,\bC)$}

Our results above already 
reveal both similarities and differences between the two quantum deformations 
$U_\fq(\mathfrak{sl}(2,\bC)_{\bR})_{\rm PW}$  
and $U_q(\mathfrak{sl}(2,\bC)_\bR)_{\rm S}$ 
of $U(\mathfrak{sl}(2,\bC)_{\bR})$ discussed in this paper. 
The quantum groups $U_\fq(\mathfrak{sl}(2,\bC)_{\bR})_{\rm PW}$  and 
$U_q(\mathfrak{sl}(2,\bC)_\bR)_{\rm S}$  
preserve different features of the classical Lie-algebra $\mathfrak{sl}(2,\bC)_{\bR}$. 
While $U_\fq(\mathfrak{sl}(2,\bC)_{\bR})_{\rm PW}$ preserves many features following from the Iwasawa decomposition of $\mathrm{SL}(2,\bC)$, 
the algebra $U_q(\mathfrak{sl}(2,\bC)_\bR)_{\rm S}$  from Schur quantisation is naturally associated to 
the representation of $\mathfrak{sl}(2,\bC)_{\bR}$ as a real form 
of $\mathfrak{sl}_2\oplus \mathfrak{sl}_2$. While a quantum analog of the Lie algebra $\mathfrak{su}(2)_{\bR}$ 
of the compact subgroup of $\mathrm{SL}(2,\bC)$ is built into the definition of
$U_\fq(\mathfrak{sl}(2,\bC)_{\bR})_{\rm PW}$, 
it has a more subtle counterpart  in the case of $U_q(\mathfrak{sl}(2,\bC)_\bR)_{\rm S}$. 
One may note, on the other hand, that the star structure representing 
$\mathfrak{sl}(2,\bC)_{\bR}$ as a real form of $\mathfrak{sl}(2,\bC)_{\bC}$ has a very simple 
counterpart in the definition of  $U_q(\mathfrak{sl}(2,\bC)_\bR)_{\rm S}$, while the star structure 
defining
$U_\fq(\mathfrak{sl}(2,\bC)_{\bR})_{\rm PW}$ is quite different.\footnote{To see the difference clearly, one may note
that the star structure defining $U_\fq(\mathfrak{sl}(2,\bC)_{\bR})_{\rm S}$ maps 
the generator $a$ defined in \rf{efk-def-mod-mod} to its inverse, while 
the star structure of $U_\fq(\mathfrak{sl}(2,\bC)_{\bR})_{\rm PW}$
maps the generator $a$ to the generator $d$ which does not commute with $a$.}

Existence of inequivalent quantum deformations 
of $U(\mathfrak{sl}(2,\bC)_{\bR})$ is a phenomenon that we expect to be related by
quantum group duality to the existence of the 
inequivalent deformations of the algebra of functions on $SL(2,\bC)$
classified in \cite{WorZakr94}. The deformed algebras of functions $\mathrm{Pol}(SL_\fq(2,\bC))$
considered in \cite{WorZakr94} have generators $\alpha$, $\beta$, $\gamma$, $\delta$ associated to the matrix elements
of the two-dimensional representation of $SL(2,\bC)$. 
The deformations classified in  \cite{WorZakr94} differ only in the mixed relations
between $\alpha$, $\beta$, $\gamma$, $\delta$ and $\alpha^\ast$, $\beta^\ast$, $\gamma^\ast$, $\delta^\ast$. 
The family of star-algebras denoted $G_{q,t}$ in \cite[Section 3.1]{WorZakr94} contains
a very natural candidate for the quantum group dual 
to  $U_q(\mathfrak{sl}(2,\bC)_\bR)_{\rm S}$  %arising in the context of Schur quantisation, 
associated to the parameter value $t=1$, and characterised by mutual 
commutativity of the sub-algebras generated by $\alpha$, $\beta$, $\gamma$, $\delta$, and
 $\alpha^\ast$, $\beta^\ast$, $\gamma^\ast$, $\delta^\ast$, respectively.
This feature strongly suggests that the star-algebras $G_{\fq,1}$ are the quantum deformations 
of $\mathrm{Pol}(SL_\fq(2,\bC))$ which are 
relevant in the context of Schur quantisations. The corresponding quantum 
groups clearly deserve further study.

\section{Schur quantization as complex quantization of a character variety.} \label{sec:charq}

The relations with Kapustin-Witten theory reviewed in the Introduction suggest a dual description of the 
Schur indices of theories of class $\cS$ in terms of the quantisation of character varieties. The goal of this section is 
to present a self-contained discussion of the complex quantization of $\mathfrak{sl}_2$ character varieties $\cM(SL(2),C)$ in Fenchel-Nielsen coordinates and a comparison with Schur quantization of the corresponding class $\cS$ Lagrangian gauge theories. We will review how the complex quantisation of character varieties is related to complex Chern-Simons theory 
in Section \ref{sec:hol}.

The quantization of character varieties is well-understood at the algebraic level. Observables are built from the quantum skein algebra $\mathrm{Sk}_\fq(C,G)$. 
The theory of unitary representations of $*$-algebras which can be built from $\mathrm{Sk}_\fq(C,G)$ is much less understood, though one should recall that
the KW lift of brane quantization \cite{Witten:2010cx} provides an useful perspective on the various available options. See Section \ref{sec:hol} for a discussion.
A possibility which has been explored in depth is quantum Teichm\"uller theory, available for $|\fq|=1$, which quantizes the Teichm\"uller locus in the character variety. Here we are instead interested in the case where $\fq$ is real, which has been studied less, and the phase space is the whole complex character variety, treated as a real phase space. 

The $*$-algebra of observables is thus the $*$-algebra double
\begin{equation}
\mathfrak{D}_\fq(C)=\mathrm{Sk}_\fq(C,SL(2))\times
\mathrm{Sk}_{\fq}(C,SL(2))^\op \, ,
\end{equation}
with a $*$ structure which exchanges the two factors. In the language of the rest of the paper, we consider examples where $\rho=1$.

The main new features of the representations to be studied here  originate from the existence of a spherical vector. 
This section will offer a self-contained perspective on the construction of the spherical vector
in a  representative example.

\subsection{Complex quantisation of the character variety -- Case of $C=C_{0,4}$} 

In order to illustrate the main new features arising in the regime $-1<\fq<1$ of interest here, we
shall pick a sufficiently typical example associated to $C=C_{0,4}$, allowing us to be reasonably 
brief and explicit at the same time. 

\subsubsection{Background on the character variety}

Recall that a set of generators for the algebra of holomorphic 
functions on the character variety is provided by
the trace functions $W_{R,\ell}$. This algebra carries a canonical Poisson structure. 

In order to prepare the discussion of the quantisation for the case of 
$C=C_{0,4}=\mathbb{P}^1\setminus\{z_1,z_2,z_3,z_4\}$,
let us note that the algebra of trace functions has three generators in this case, 
denoted $W$, $H$, and $D$, and associated to simple 
closed curves encircling only $(z_1,z_2)$, $(z_1,z_3)$ and $(z_2,z_3)$, respectively, 
The trace functions $W$, $H$, and $D$ satisfy the equation of the the Klein cubic
$P_{K}(W,H,D)=0$, with $P_{K}$ being a cubic polynomial. While the precise form of
$P_K$ will not be needed explicitly, one should bear in mind that the coefficients 
of $P_K$ depend on four complex numbers $\mu_r$ parameterising the traces
of the holonomies $L_r$ around the punctures $z_r$ as $L_r=\mu_r+\mu_r^{-1}$ 
for  $r=1,2,3,4$, respectively.  

Rational parameterisations of the Klein cubic can be associated to pants decompositions
of $C_{0,4}$. Considering the pants decomposition defined by a curve separating $z_1$ and $z_2$
from $z_3$ and $z_4$, for example, one can solve the equation $P_{K}(W,H,D)=0$ in terms of
two parameters $u$ and $v$ by setting
\begin{subequations}\label{lineC04-cl}
\begin{align}
& W=v+v^{-1},\label{Wilson-cl}\\
\label{tHooft-cl}
&H=c_+(v)\,u^2+c_0(v)+c_-(v)\,u^{-2},\\
&D=c_+(v)\,v\,u^2+c_0(v)+ c_-(v)\,v^{-1}u^{-2},
\end{align}
using the functions $c_+$, $c_0$ and $c_-$ defined as
\begin{align*}
&c_+(v)=1,\qquad\quad c_-(v)=
\frac{\prod_{s,s'=\pm}(1+m_1^{s}m_2^{s'}v)(1+m_3^{s}m_4^{s'}v)}{(1-v^2)^4},\\
&c_0(v)=\frac{(v+v^{-1})(L_1L_3+L_2L_4)-2(L_2L_3+L_1L_4)}{(v-v^{-1})^{2}},
\end{align*}
\end{subequations}
It will be useful to note that replacing $v$ by $v^{-1}$ and 
$u$ by $u^{-1}(c_-(v)/c_+(v))^{1/2}$ leaves the expressions for $W$, $H$ and $D$ invariant. This means 
that an open dense set in $\cM(C_{0,4},SL(2))$ can be parameterised by a 
$\bZ_2$-quotient of the space $\bC^2$ with coordinates $u$ and $v$. 
Let us furthermore note that $u$ and $v$ can be represented as exponential functions of 
Darboux coordinates for the canonical Poisson structure of 
$\cM(C_{0,4},SL(2))$ often referred to as coordinates of Fenchel-Nielsen type.

\subsubsection{Quantisation, the algebraic level}\label{alg-quant}

The algebraic level of the 
quantisation of the character varieties has been extensively studied, prompting us to be 
brief. The algebra $\mathrm{Sk}_\fq(C_{0,4},SL(2))$ has generators denoted as $W$, $H$, $D$, 
satisfying a deformed version of the equation of the Klein cubic of the form 
$
P_{K,\fq}(W,H,D)=0
$,
with $P_{K,\fq}$ being a polynomial in non-commutative variables 
which is known explicitly. 

We may start by introducing the algebra $\cW\otimes\cW^\op$, defined by the relations
\begin{equation}\label{Weyl-Weylop}
uv=\fq\,vu,\qquad \bar{u}\bar{v}=\fq^{-1}\bar{v}\bar{u}, \qquad v\bar{u}=\bar{u}v,\qquad \bar{v}u=u\bar{v}.
\end{equation}
Out of these generators we can formally\footnote{We are postponing a discussion of the analytic aspects for a moment.} 
construct a representation of $\mathrm{Sk}_\fq(C_{0,4},SL(2))$ by 
defining
\begin{subequations}\label{lineC04}
\begin{align}
& W=v+v^{-1},\label{Wilson}\\
\label{tHooft}
&H=u \,C_+(v)\,u+C_0(v)+u^{-1} C_-(v)\,u^{-1},\\
&D=u \,v\,C_+(v)\,u+C_0(v)+u^{-1} v^{-1} C_-(v)\,u^{-1},
\end{align}
using the functions $C_+$, $C_0$ and $C_-$ defined as
\begin{align*}
&C_+(v)=1,\qquad\; C_-(v)=
\frac{\prod_{s,s'=\pm}(1+m_1^{s}m_2^{s'}v)(1+m_3^{s}m_4^{s'}v)}{(1-\fq^2v^2)(1-v^2)^2(1-\fq^{-2}v^2)},\\
&C_0(v)=\frac{(v+v^{-1})(L_1L_3+L_2L_4)-(\fq+\fq^{-1})(L_2L_3+L_1L_4)}{(\fq v-\fq^{-1}v^{-1})(\fq^{-1}v-\fq v^{-1})}.
\end{align*}
\end{subequations}
These formulae are related by a similarity transformation 
to the difference operators representing the action of 
Verlinde line operators on Virasoro conformal blocks \cite{Alday:2009fs,Drukker:2009id}. 
It can be verified directly that the
relations $
P_{K,\fq}(W,H,D)=0
$ are satisfied.  

A representation of the algebra $\mathrm{Sk}_\fq(C_{0,4},SL(2))^\op$ can furthermore be generated by the 
operators $\widetilde{W}$, $\widetilde{H}$ and $\widetilde{D}$ defined by replacing 
$u,v$ by $\bar{u},\bar{v}$ in the formulae \rf{lineC04}.  The generators 
$\widetilde{W}$, $\widetilde{H}$ and $\widetilde{D}$ clearly  commute with $W$, $H$ and $D$.

One should note that the formulae \rf{lineC04} can be used to define operators $W$, $H$ and $D$ that are 
formally normal in any unitary representation of $\cW\otimes\widetilde{\cW}$ representing
the generators $u$, $v$, $\bar{u}$, $\bar{v}$ such that
$u^\dagger=\bar{u}$, $v^{\dagger}=\bar{v}$. Combined with 
\rf{lineC04} we then find the relations $W^\dagger=\widetilde{W}$, $H^\dagger=\widetilde{H}$
and $D^\dagger=\widetilde{D}$.

\subsubsection{Definition of the Hilbert space}

%\subsubsection{Auxiliary Hilbert space representation of the Weyl algebra}

The basis of our construction will be a representation of an auxiliary algebra of Weyl-type $\cW$, introduced 
in \cite{Dimofte:2011py} in a closely related context.
We are going to  define a representation of the algebra $\cW\otimes\widetilde{\cW}$, defined by the relations
\rf{Weyl-Weylop},
represented by densely defined unbounded normal operators ${v}$, $u$, $\bar{v}$, and $\bar{u}$
on the  Hilbert space $\cH_\cW=L^2(S^1\times \bZ)$, defined as
\[
\Phi=\big\{f_m\in L^2(S^1);m\in\bZ\big\}\qquad\text{such that}\qquad 
|\!| \Phi|\!|^2_\cW:=\sum_{m\in\bZ}|\!| f_m|\!|^2_{L^2(S^1)}<\infty.
\]
Operators ${v}$, $u$, $\bar{v}$, and $\bar{u}$ representing \rf{Weyl-Weylop} can be defined as 
\begin{equation}\label{W-rep}
\begin{aligned}
&u f_m(\theta)=f_{m+1}\big(\theta-\fr{\hbar}{2\ii}\big),\\
&{v} f_m(\theta)=\fq^{\frac{m}{2}}e^{\ii\theta}f_m(\theta),
\end{aligned}\qquad
\begin{aligned}
&\bar{u}f_m(\theta)=f_{m-1}\big(\theta+\fr{\hbar}{2\ii}\big).\\
&\bar{v}f_m(\theta)=\fq^{\frac{m}{2}}e^{-\ii\theta}f_m(\theta),
\end{aligned}\qquad \fq=e^{-\hbar}.
\end{equation}
One may note that the operators defined in \rf{W-rep}
satisfy ${v}^\dagger=\bar{v}$, $u^\dagger=\bar{u}$.

The auxilliary Hilbert space $\cH_\cW$ will be used to define the Hilbert space
$\cH(C_{0,4})$ by taking a $\bZ_2$-quotient of $\cH_\cW$ representing a quantised version of 
the redundancy of the parameterisation of $\cM(C_{0,4},SL(2))$ in terms of
the Fenchel-Nielsen type coordinates $u$ and $v$. In order to find the proper quantised
analog of the symmetry $v\mapsto v^{-1}$ and 
$u^2\mapsto u^{-2}c_-(v)/c_+(v)$ reflecting this redundancy, let us note that
 the formulae \rf{lineC04} are invariant under the symmetry 
\begin{equation}\label{parity}
{v}\mapsto {v}^{-1},\qquad 
u^2\mapsto u^{-1}\cdot\frac{C_-({v})}{C_+({v})}\cdot u^{-1}.
\end{equation}
This symmetry is generated by a  unitary operator on $\cH_{\cW}$
\begin{equation}
\label{refop}
\sR:=\frac{\varphi_0({v})}{\varphi_0(\bar{v})}\cdot\varpi,
\end{equation}
where $\varpi$ is the parity operator satisfying $\varpi=\varpi^{-1}$, 
$\varpi\cdot{v}\cdot\varpi={v}^{-1}$, $\varpi\cdot u\cdot\varpi=u^{-1}$,
and $\varphi_0$ is a function satisfying the difference equation
\begin{equation}\label{diffeq-phi0}
\frac{\varphi_0(\fq \,v)}{\varphi_0(\fq^{-1}v)}=\frac{C_-(v)}{C_+(v)}=%\frac{1}{d_-(qu)}=
\frac{\prod_{s,s'=\pm}(1+vm_1^{s}m_2^{s'})(1+vm_3^{s}m_4^{s'})}{(1-\fq^2v^2)(1-v^2)^2(1-\fq^{-2}v^2)}.
\end{equation}
Indeed, using the relations \rf{Weyl-Weylop} we find 
\[
\begin{aligned}
&\sR^{-1}\cdot u^{2}\cdot\sR=
\varpi\cdot u\cdot \frac{\varphi_0(\fq{v})}{\varphi_0(\fq^{-1}\,{v})}\cdot u\cdot\varpi=u^{-1}\cdot 
\frac{C_-({v})}{C_+({v})}\cdot u^{-1},
\\&\sR^{-1}\cdot \bar{u}^{2}\cdot\sR=
\varpi\cdot\bar{u}\cdot \frac{\varphi_0(\fq\bar{v})}{\varphi_0(\fq^{-1}\,\bar{v})}\cdot\bar{u}\cdot\varpi=
\bar{u}^{-1}\cdot \frac{C_-(\bar{v})}{C_+(\bar{v})}\cdot\bar{u}^{-1},
\end{aligned}
\]
using $C_+(v)=C_+(v^{-1})$. 
Note that the solution to \rf{diffeq-phi0} is given by the function
\begin{equation}\label{psi0-expl}
\varphi_0(v)=\frac{(\fq^2v^2;\fq^2)_{\infty}(v^2;\fq^2)_{\infty}}{\prod_{s,s'=\pm}(-\fq v m_1^{s}m_2^{s'};\fq^2)_\infty(-\fq vm_3^{s}m_4^{s'};\fq^2)_{\infty}}.
\end{equation}
These preparations allow us to complete the definition of  
the representation by setting
\begin{equation}
\cH(C_{0,4}):=\big\{\,\Psi\in\cH_{\cW}\,;\,\sP\Psi=0\,\big\}, \qquad \sP=\frac{1}{\sqrt{2}}(1-\sR).
\end{equation}
%[{\it desrcribe in terms of wave-functions satisfying reflection relation?}]
%The invariance of the expressions in \rf{lineC04} under the 
%symmetry \rf{parity} ensures that the projector $\sP$ defining $\cH_{C}$ for $C=C_{0,4}$ from 
%$\cH_{\cW}$ is a map of $\cO_q(\mathfrak{X}_{C})\times\cO_q(\mathfrak{X}_{C})_{\rm op}$-modules.

It seems very likely that formulae \rf{lineC04}, \rf{W-rep} define a representation of $\mathrm{Sk}_\fq(C_{0,4},SL(2))$ 
on $\cH(C_{0,4})$. In order to establish this claim one needs to address the unboundedness of 
the operators generating $\mathrm{Sk}_\fq(C_{0,4},SL(2))$.
This unboundedness not only comes from the 
unboundedness of the operators representing
$u$ and $v$, one also needs to take into account singularities % behaviour 
from vanishing denominators in the formulae for $C_{i}(v)$, $i=-,0,+$.
We will later demonstrate that there exists  a dense domain within
$\cH(C_{0,4})$ on which the unbounded operators generating $\mathfrak{D}_\fq(C_{0,4},SL(2))$
can be defined. The operators  defined in this way 
admit a normal extension to a  dense domain within $\cH(C_{0,4})$, satisfying
\begin{equation}\label{normality}
W_a^{\dagger}=\widetilde{W}_a\,.
\end{equation}
One may, in particular, be worried that the poles of $C_{i}(v)$, $i=-,0,+$, could spoil normalisability of $H\Psi$, 
for example.  In this regard
it seems encouraging to note
that the wave-functions representing elements of $\cH(C_{0,4})$ satisfy certain 
vanishing conditions at $v=1$. 
Existence of a normal extension is easy to prove in the case of $W$, being realised as a 
pure multiplication operator in the representation \rf{lineC04}, \rf{W-rep}.

\subsubsection{Dependence on choice of pants decomposition}

The representation \rf{lineC04} clearly depends on a choice of a pants decomposition. 
There are three basic pants decompositions of $C_{0,4}$, defined by contours
separating the pairs $(z_1,z_2)$, $(z_2,z_3)$ and $(z_1,z_3)$ from the remaining 
two punctures, respectively.  For each of these pants decompositions one can define
representations of $\mathfrak{D}_\fq(C_{0,4})$
by using formulae obtained from  \rf{lineC04} by appropriate 
permutations of the indices $1,2,3,4$. We conjecture that these three representations 
are all unitarily equivalent to each other. 

The next paper in this series \cite{next} will outline the construction of unitary operators relating the three
representations obtained in this way. For now one may note that 
this amounts to the solution of the spectral problems for the operators $H$ and $D$ 
within the representation above. We may anticipate, in particular that 
the unitary operator diagonalising $H$, for example, 
can be represented in the form 
\begin{equation}\label{F-move}
\Psi_s(\theta,m)=\int_{S^1}d\theta\sum_{m'\in\bZ}F_{{\mu}}
\left(\begin{smallmatrix}
\theta & \theta'\\
m & m'
\end{smallmatrix}\right)%\!(\theta,\theta')_{mm'}^{}
\Psi_t(\theta',m'),
\end{equation}
with $\mu=(\mu_1,\dots,\mu_4)$, and
$\mathcal{F}_{\theta',m'}^\mu(\theta,m)=F_{{\mu}}
\left(\begin{smallmatrix}
\theta & \theta'\\
m & m'
\end{smallmatrix}\right)$ being an eigenfunction of $H$ with eigenvalue $h=v'+1/{v'}$, where
$v'=\fq^{\frac{m'}{2}}e^{\ii\theta'}$.
Existence of this unitary operator implies that $H$ is normal, as Conjecture 1 predicts. 

In a way that is analogous to the quantum Teichm\"uller theory, one may 
use the
unitary operators representing the changes of pants decomposition in order to
define a representation of the braid group of $C_{0,4}$, and an analog of a modular functor.

%\noindent [{\it Discuss dependence on choice of coordinates.}]

\subsubsection{Spherical vector}

A central role is played in this representation by the spherical vector $|1\rangle\in\cH_\cW$ satisfying
\begin{equation}\label{H-barH}
W_a^{}\,|1\rangle={W}_a^\dagger|1\rangle, \qquad \forall\;a\in\mathrm{Sk}_\fq(C_{0,4},SL(2)).
\end{equation}
We will represent $|1\rangle$ by wave-functions 
$f_m(\theta)=\langle \theta,m |1\rangle$.
We claim that the unique solution to these conditions is of the form 
%\begin{subequations}
\begin{align}\label{Phi0C04}%\label{Phi0-diag}
&f_m(\theta)=\delta_{m,0}\phi_0(\theta),\quad\text{where}\quad \phi_0(\theta)=\varphi_0(e^{\ii\theta}),%\\
%&\varphi_0(u)=\frac{(q^2u^2;q^2)_{\infty}(u^2;q^2)_{\infty}}{\prod_{s,s'=\pm}(-quM_1^{s}M_2^{s'};q^2)_\infty(-quM_3^{s}M_4^{s'};q^2)_{\infty}}.
%\\
%&=(1-u^4)\frac{\big((q^2u^2;q^2)_{\infty}\big)^2}{\prod_{s,s'=\pm}(-quM_1^{s}M_2^{s'};q^2)_\infty(-quM_3^{s}M_4^{s'};q^2)_{\infty}}
\end{align}
with function $\varphi_0$ defined in \rf{psi0-expl}.
%\end{subequations}

In order to verify that  \rf{Phi0C04} solves \rf{H-barH} let us note, on the one hand,
\[
\begin{aligned}
Hf_m(\theta)&=C_-(\fq^{-1}{v})\delta_{m-2,0}\phi_0(\theta-\ii\hbar)+C_0({v})\delta_{m,0}\phi_0(\theta)+
\delta_{m+2,0}C_+(\fq{v})\phi_0(\theta+\ii\hbar)\\
&=\delta_{m-2,0}C_-(e^{\ii\theta})\phi_0(\theta-\ii\hbar)+\delta_{m,0}C_0(e^{\ii\theta})+
\delta_{m+2,0}C_+(e^{\ii\theta})\phi_0(\theta+\ii\hbar),
\end{aligned}
\]
using ${v} f_m(\theta)=\fq^{\frac{m}{2}}e^{\ii\theta}f_m(\theta)$. We have, on the other hand, 
%Hermitian conjugation is most easily described in representation 2,
\[
H^{\dagger}= C_+(\fq^{-1}\bar{v})\,\bar{u}^2 +C_0(\bar{v})+ C_-(\fq\bar{v})\,\bar{u}^{-2},\\
\]
Using  $C_\pm(e^{-\ii\theta})=C_\pm(e^{\ii\theta})$ 
this implies
\[
\begin{aligned}
&H^{\dagger}f_m(\theta)=\delta_{m+2,0}C_-(e^{\ii\theta})\phi_0(\theta-\ii\hbar)+\delta_{m,0}C_0(e^{\ii\theta})
+\delta_{m-2,0}C_+(e^{\ii\theta})\Phi_0(\theta+\ii\hbar).
\end{aligned}
\]
Equation \rf{H-barH} is therefore equivalent to 
\begin{equation}\label{diffeq-c}
C_-(e^{\ii\theta})\phi_0(\theta-\ii\hbar)=C_+(e^{\ii\theta})\phi_0(\theta+\ii\hbar).
\end{equation}
Representing $\phi_0(\theta)$ as $\phi_0(\theta)=\varphi_0(e^{\ii\theta})$, and using the 
explicit expressions for $C_\pm(v)$ given above, 
we find the following 
difference equation for $\varphi_0$:
\begin{equation}\label{diffeq-c}
\frac{\varphi_0(\fq^2v)}{\varphi_0(v)}=C_-(\fq v)=%\frac{1}{d_-(qu)}=
\frac{\prod_{s,s'=\pm}(1+\fq v m_1^{s}m_2^{s'})(1+\fq vm_3^{s}m_4^{s'})}{(1-\fq^4v^2)(1-\fq^{2}v^2)^2(1-v^2)}.
\end{equation}
It remains to notice that equation \rf{diffeq-c} is solved by \rf{Phi0C04}. 

%\subsubsection{Key features of the complex representation}
The spherical vector is indeed contained in $\cH(C_{0,4})$, as
 follows from $\mathsf{R} |1\rangle=|1\rangle$, and the finiteness of the norm
\begin{align}\label{Schur}
\lVert\Phi_0\lVert^2=\ii\int_{S^1}\frac{dv}{v}(v-v^{-1})^2
&\frac{(\fq^2v^2;\fq^2)_{\infty}^2(\fq^2v^{-2};\fq^2)_{\infty}^2}
       {\prod\limits_{s,s_1,s_2=\pm}(-\fq v^{s}m_1^{s_1}m_2^{s_2};\fq^2)_\infty(-\fq v^{s}m_3^{s_1}m_4^{s_2};\fq^2)_{\infty}}.
   %    \\
 % \times     &\frac{(q^2u^{-2};q^2)_{\infty}^2}{\prod_{s,s'=\pm}(-qu^{-1}M_1^{s}M_2^{s'};q^2)_\infty(-qu^{-1}M_3^{s}M_4^{s'};q^2)_{\infty}}.
  % \notag    
   \end{align}
   
 We claim that  the vector $|1\rangle$ is in the domain of all $W_a\in \mathrm{Sk}_\fq(C_{0,4},SL(2))$.
 This can be verified for $W$, $H$, and $D$ noting that the measure defined 
 by the functions $\phi_0$ cancels potentially non-integrable factors in $C_i$, $i=-,0,+$. 
 We conjecture that
 this holds in general. A somewhat non-trivial claim is formulated as the following
   \begin{equation*}%\label{mainconj}
\text{\it Conjecture: The  vectors $W_a|1\rangle$, $a\in \mathrm{Sk}_\fq(C_{0,4},SL(2))$, 
 span a dense subspace in $\cH(C_{0,4})$. }
\end{equation*}
This conjecture is not at all obvious at this stage. 
Our next paper will introduce techniques for addressing this issue.

\subsection{Relation to the Schur quantisation}

There is a considerable freedom in the choice of representation of the skein algebra 
introduced above. Similarity transformations can be used to modify the representation
by finite difference operators. This can be useful to reveal certain properties of the 
representation. We will here consider the example which facilitates the comparison 
with the Schur quantisation.  

%\subsubsection{Algebraic level}

We shall consider the similarity transformation
$W_a'=S^{-1}\cdot W_a\cdot S$, with
\begin{equation}\label{similarity}
S={(v\bar{v})^{-\frac{1}{2\hbar}\log(m_2m_3)}}\frac{(v^2;\fq^2)_{\infty}}{(\fq^2\bar{v}^2;\fq^2)_{\infty}}
\prod_{s=\pm}
\frac{(-\fq\,m_2^{-1}m_1^s\,\bar{v};\fq^2)_{\infty} (-\fq\,m_3^{-1}m_4^s\,\bar{v};\fq^2)_{\infty}}
 {(-\fq\,m_2^{}m_1^s\,v;\fq^2)_{\infty} (-\fq\,m_3^{}m_4^s\,v;\fq^2)_{\infty}}.
\end{equation}
It is not hard to verify that the generators $W'$, $H'$, $D'$, and 
$\widetilde{W}'$, $\widetilde{H}'$, $\widetilde{D}'$ 
defined in this way are represented by finite difference operators having a similar form 
as in \rf{lineC04}, but with modified coefficient functions. Note, in particular that
\[
\begin{aligned}
&\frac{1}{(v^2;\fq^2)_{\infty}}\cdot u^2 \cdot (v^2;\fq^2)_{\infty}
=u\cdot\frac{(\fq^{2}v^2;\fq^2)_{\infty}}{(\fq^{-2}v^2;\fq^2)_{\infty}}\cdot u
=u\cdot\frac{1}{(1-\fq^{-2}v^2)(1-v^2)}\cdot u\,, \\
&(\bar{v}^2;\fq^2)_{\infty}\cdot \bar{u}^2 \cdot \frac{1}{(\fq^2\bar{v}^2;\fq^2)_{\infty}}
=\bar{u}\cdot\frac{(\fq^{4}\bar{v}^2;\fq^2)_{\infty}}{(\bar{v}^2;\fq^2)_{\infty}}\cdot \bar{u}
=\bar{u}\cdot\frac{1}{(1-\bar{v}^2)(1-\fq^{2}\bar{v}^2)}\cdot \bar{u}\,.
\end{aligned}
\]
For $H'$ and $\widetilde{H}'$ one thereby finds the expressions
\begin{subequations}\label{lineC04-AS}
\begin{align}
&H'=D_+(v)\,u^2+C_0(v)+ D_-(v)\,u^{-2},\\
&\widetilde{H}'=\bar{D}_+(\bar{v})\,\bar{u}^2+C_0(\bar{v})+\bar{D}_-(\bar{v})\,\bar{u}^{-2},
\end{align}
using the functions $D_{\pm}$ and $\bar{D}_\pm$ defined as
\begin{equation}\label{Dpm-def}
\begin{aligned}
&D_+(v)
%= \frac{\prod_{s=\pm}(1+\fq^{-1}m_1^{s}m_2^{-1}v^{-1})(1+\fq^{-1}m_3^{-1}m_4^{s}\,v^{-1})}{(1-\fq^{-2}v^{-2})(1-v^{-2})}
=\frac{\prod_{s=\pm}(1+\fq \,m_1^{s}m_2^{}\,v)(1+\fq\,m_3^{}m_4^{s}\,v)}{ m_2^{}m_3^{}(1-\fq^{2}v^2)(1-v^2)}
=D_-(v^{-1}),\\
%&\phantom{D_+(v)}=\frac{\prod_{s=\pm}(1+\fq^{-1} \,m_1^{s}m_2^{-1}\,v^{-1})(1+\fq^{-1}\,m_3^{-1}m_4^{s}\,v^{-1})}{m_2^{-1}m_3^{-1}(1-\fq^{-2}v^{-2})(1-v^{-2})},\\
%&D_-(v)=
%\frac{\prod_{s=\pm}(1+\fq^{-1}m_1^{s}m_2^{-1}\,v)(1+\fq^{-1}m_3^{-1}m_4^{s}\,v)}{m_2^{-1}m_3^{-1}(1-\fq^{-2}v^2)(1-v^2)}
%=\frac{\prod_{s=\pm}(m_1^{s}m_2^{}+\fq^{-1}\,v)(m_3^{}m_4^{s}+\fq^{-1}\,v)}{(1-\fq^{-2}v^2)(1-v^2)},
%\\
%&\phantom{D_-(v)}=
%\frac{\prod_{s=\pm}(m_1^{s}m_2^{-1}+\fq^{}\,v^{-1})(m_3^{-1}m_4^{s}+\fq^{}\,v^{-1})}{m_2^{1/2}m_3^{1/2}(1-\fq^{2}v^{-2})(1-v^{-2})}=D_+(v^{-1}),\\
&\bar{D}_+(\bar{v})= 
\frac{\prod_{s=\pm}(1+\fq^{-1}\,m_1^{s}m_2^{-1}\,\bar{v})(1+\fq^{-1}\,m_3^{-1}m_4^{s}\,\bar{v})}{m_2^{-1}m_3^{-1}(1-\fq^{-2}\bar{v}^{2})
(1-\bar{v}^{2})}
%\\
%&\bar{D}_-(\bar{v})= 
%\frac{\prod_{s=\pm}(1+\fq\,m_1^{s}m_2^{}\,\bar{v})(1+\fq\,m_3^{}m_4^{s}\,\bar{v})}{m_2^{}m_3^{}(1-\fq^{2}\bar{v}^{2})
%(1-\bar{v}^{2})}
=\bar{D}_{-}(\bar{v}^{-1}). %\\
%&\qquad\quad=\frac{\prod_{s=\pm}(1+\fq^{-1}\,m_1^{s}m_2^{-1}\,\bar{v}^{-1})(1+\fq^{-1}\,m_3^{-1}m_4^{s}\,\bar{v}^{-1})}{m_2^{-1}m_3^{-1}(1-\fq^{-2}\bar{v}^{-2})
%(1-\bar{v}^{-2})}=\bar{D}_{+}(\bar{v}^{-1}).
\end{aligned}
\end{equation}
\end{subequations}

It had been argued in  \cite{Dimofte:2011py,Gang:2012yr} that the difference operators 
appearing in the integral formulae for Schur indices of line operators should coincide with the 
operators representing the insertion of the same
line operators in four-ellipsoid partition functions \cite{Gomis:2011pf}. 
The latter are known to to be related to the difference operators
representing the action of Verlinde line operators 
on Virasoro conformal blocks \cite{Alday:2009fs,Drukker:2009id}.

In order to compare the Hilbert space realisations of $\mathfrak{D}_\fq(C)$ coming from Schur 
quantisation and complex CS-theory, one may first compare the explicit formula \rf{Schur} for
the norm of $|1\rangle$ with the Schur index. 
This coincides with the  UV formula for the Schur-index in the $N_f=4$-theory 
as given in \cite{Cordova:2016uwk}. 

It has been shown in \cite{Allegretti:2024svn} that the K-theoretic Coulomb branches of 
theories of class $\cS$ coincide with the skein algebra $\mathrm{Sk}_\fq(C_{0,4},SL(2))$. 
This has been verified in \cite{Allegretti:2024svn} by comparing the representation of
$\mathrm{Sk}_\fq(C_{0,4},SL(2))$  generated by the difference operators $W'$, $H'$, $D'$ 
introduced above with the relevant special case of the 
more general formulae for the generators 
of the K-theoretic Coulomb branches obtained in \cite{finkelberg2019multiplicative}.

\section{Schur quantization of complex Chern-Simons theory}\label{sec:hol}
The main topic of this section is Chern-Simons theory with complex (say simply-laced in this paper) 
gauge group $G_\bC$ and imaginary level $\kappa = i s$, with action \cite{Witten:1989ip}:
\begin{equation}
	\frac{i s}{2} S_{CS}(\cA) - \frac{i s}{2} S_{CS}(\bar \cA) \, .
\end{equation}
Here $\cA$ is a $G_\bC$ connection and $S_{CS}$ the standard Chern-Simons action.\footnote{One can consider more general levels $(k+i s)/2$ and $(k-i s)/2$ for integer $k$. Some of the constructions in this paper can be extended to that case. See e.g. \cite{Han:2024nkf}.} The choice of imaginary level means that we use as the symplectic form the imaginary part of the natural complex symplectic form on $\cM(G,C)$. As described in the introduction, we aim to describe this theory via Schur quantization of theories of class $\cS$ at $\fq = e^{-\pi s^{-1}}$.

The classical equations of motion of Chern-Simons theory require the complex connection $\cA$ to be flat. If space is a compact two-dimensional surface $C$, the theory has a finite-dimensional phase space:
the moduli space $\cM(C,G)$ of flat $G_\C$ connections on $C$, equipped with a symplectic form proportional to 
\begin{equation}
	i \int_C \left[\delta \cA \wedge \delta \cA - \delta \bar \cA \wedge \delta \bar \cA \right]
\end{equation}
The classical phase space carries several structures reflecting the topological nature of the theory and which we would like to persist in the quantum theory:
\begin{itemize}
	\item The solutions of the equations of motions on a three-dimensional space-time $M_3$ with boundary $C$ give a Lagrangian 
	submanifold $\cL(M_3,G)$ consisting of flat $G_\bC$ connections on $C$ which extend to $M_3$. 
	\item The special case of $M_3$ being a mapping cylinder gives a representation of the mapping class group of 
$C$ as Lagrangian correspondences.
	\item Wilson lines for $\cA$ computed at fixed time along a path $\ell$ on $C$ give classical observables 
	\begin{equation}
		W_{R,\ell} \equiv \Tr_R \,\mathrm{Pexp} \oint_\ell \cA
	\end{equation}
	labelled by $ \ell$ and a finite-dimensional representation $R$ of $G$. These are holomorphic functions on $\cM(C,G)$. Other holomorphic functions $W_a$
	can be realized as ``skeins'' $a$ of Wilson lines on $C$ joined by intertwining tensors. 
	The product and Poisson brackets on the phase space are local on $C$ and closes within this class of functions.
	If we invert the direction of the path $\ell$, 
	we dualize the representation: $W_{R,\ell} = W_{R^\vee, \ell^{-1}}$. 
	\item Wilson lines for $\bar \cA$ give a second collection of classical observables 
	\begin{equation}
		\wt W_{R,\ell} \equiv \Tr_R \,\mathrm{Pexp} \oint_\ell \bar \cA
	\end{equation}
	These are anti-holomorphic functions on $\cM(C,G)$. We use conventions where $\overline{W_{R,\ell}}= \wt W_{R,\ell^{-1}}$. Other anti-holomorphic functions $\wt W_a$ can be realized as skeins $a$. We have $\overline{W_{\rho(a)}} = \wt W_{a}$ for an appropriately defined dual skein $\rho(a)$. In these conventions, $\wt W_a = W_a$ if the connection is unitary. The Poisson bracket closes within this class of functions, which Poisson-commute with the holomorphic functions.
\end{itemize}
All of this data only depends on the topology of the (sub)manifolds involved. 

The quantum theory should associate to $C$ some Hilbert space $\cH_s(C,G)$ which quantizes $\cM(C,G)$. This Hilbert space 
should carry compatible actions of:
\begin{itemize}
\item The mapping class group of $C$.
\item The quantized algebra of holomorphic Wilson line networks in $C \times \R$, isomorphic to the Skein algebra $\mathrm{Sk}_{\fq}(C,G)$. Here we defined $\fq = e^{-\frac{\pi}{s}}$ and we will sometimes employ the notation $\cH_\fq(C,G)$ for the Hilbert space $\cH_s(C,G)$. See \cite{2023arXiv230214734J} for a modern discussion. 
\item The quantized algebra $\mathrm{Sk}_{\fq^{-1}}(C,G) = \mathrm{Sk}_{\fq}(C,G)^{\mathrm{op}}$ of anti-holomorphic Wilson lines in $C \times \R$, commuting with $\mathrm{Sk}_{\fq}(C,G)$. We should have $W_{\rho(a)}^\dagger = \wt W_{a}$, so that the algebras are realized by normal operators. 
Here $\rho$ is an automorphism % anti-holomorphic involution 
of the Skein algebra, extended anti-linearly over the complex numbers.
\end{itemize}
The quantization procedure should also produce a canonical collection of (possibly distributional) states $|M_3\rangle$ in $\cH_s(C,G)$ given by a path integral over three-manifolds with boundary $C$, compatible with the above actions. More precisely, there is a combinatorial way to build ``Skein modules'' $\mathrm{Sk}_{\fq}(M_3,G)$ for $\mathrm{Sk}_{\fq}(C,G)$ which literally encode skeins $W_m$ of Wilson lines in $M_3$ and their relations to skeins in $C$. Quantization should provide a state for every decoration of $M_3$ by a skein, i.e. a module map 
\begin{equation}
\mathrm{Sk}_{\fq}(M_3,G) \times \mathrm{Sk}_{\fq}(M_3,G)^{\mathrm{op}} \to \cH_s(C,G)
\end{equation}
though the images $|M_3;m,\wt m\rangle$ could include distributional states.\footnote{In particular, there is no guarantee that the partition function on a three-manifold will be finite: the TFT is not fully extended.}

Schur quantization in class $\cS$ precisely provide all of this data:
\begin{itemize}
	\item The algebra $A_\fq$ coincides with $\mathrm{Sk}_{\fq}(C,G)$ and thus the Hilbert space $\cH_\fq$ carries the desired actions of the Skein algebra.
	\item The space of couplings of $\cT$ coincides with the space of complex structures of $C$ and thus $\cH_\fq$ carries an unitary action of the mapping class group compatible related to the natural permutation action on $\mathrm{Sk}_{\fq}(C,G)$.
	\item Boundary conditions $B(M_3,G)$ labelled by three-manifolds \cite{Dimofte:2013lba} give the desired states $|B(M_3)\rangle$ in $\cH_\fq$. Decorations by boundary line defects provide $|M_3;m,\wt m\rangle$ with the expected properties.
\end{itemize} 

We will denote this theory as ``complex Chern-Simons theory'' or as ``$G_\bC$ Chern-Simons theory''. We should list other variants of Chern-Simons theory which may be confused with this theory:
\begin{itemize}
	\item Standard Chern-Simons theory with unitary gauge group $G_c$, the compact form of $G_\bC$. This TFT has a quantized level $k$ and a phase space $\cM(G_c,C)$ which consists of unitary flat connections. Here we only encounter $\cM(G_c,C)$ as a Lagrangian sub-manifold of $\cM(G,C)$ which we aim to quantize to a special state in the Hilbert space.
	\item Chern-Simons theory with $G_\bR$ gauge group, with $G_\bR$ being some other real form of $G_\bC$ and phase space $\cM(G_\bR,C)$. A full definition of this theory, possibly extending the % distinct from 
	quantum  Teichm\"uller theory \cite{Mikhaylov:2017ngi,Collier:2023fwi}, is not quite available at this point. It will not play a role in this paper.\footnote{The space $\cM(G_\bR,C)$ as a Lagrangian sub-manifold of $\cM(G,C)$ and its connected components have applications in (Lorentzian, positive curvature) 3d quantum gravity. The role of $\cM_c(G,C)$ is less clear.} 
	\item Analytically continued Chern-Simons theory with gauge group $G$. This is not actually a 3d theory: it describes general properties of path integrals with an $S_{CS}(\cA)$ Chern-Simons action but no specified reality condition on $\cA$. It can be formulated as a relative theory, living at the boundary of 4d Kapustin-Witten theory \cite{Witten:2010cx}. 
	Analytically continued Chern-Simons theory can be an ingredient in the analysis of all the other Chern-Simons theories mentioned above, allowing one to embed them in 4d KW theory.
\end{itemize}

The literature presents two very different quantization strategies \cite{Witten:1989ip,Dimofte:2011py} for complex Chern-Simons theory, which have important limitations and are difficult to compare with each other. See \cite{Dimofte:2016pua} for a review of the problem. These strategies are akin to the two sides of the Riemann-Hilbert correspondence: one describes the phase space in terms of bundles equipped with holomorphic connections (``de Rham'') and the other in terms of the associated representation of the fundamental group (``Betti''), in some respects following the paradigm of quantum Teichm\"uller theory. As discussed in the introduction, Schur quantization applied to theories of class $\cS$  is a variant of the second option which can bridge the gap between these two descriptions. Indeed, Schur quantization provides an Hilbert space $\cH$ equipped with the  spherical cyclic vector $|1\rangle \in \cH$, which we identify as the quantization of the Lagrangian submanifold of unitary flat connections $\cM_c(G,C) \subset \cM(G,C)$. This vector acts as a Rosetta stone: it allows us relate this quantization strategy to the previous two by identifying analogous spherical vectors in the respective Hilbert spaces. 

\subsection{A topological boundary condition from unitary flat connections}
We should briefly review the Chern-Simons interpretation of the state $|1\rangle $ as being created by a topological boundary condition $B_c$.
 We define the boundary condition by restricting both the connection $\cA$ and the gauge transformations to lie in the maximal compact 
subgroup $G_c$ at the boundary. This is possible because the potential boundary gauge anomaly restricted to the $G_c$ subgroup is the difference between the holomorphic and anti-holomorphic levels and thus vanishes. 

At the level of the phase space, the boundary condition defines the Lagrangian submanifold $\cM_c$ of unitary (i.e. $G_c$) flat connections 
inside the moduli space of complex flat connections. Semiclassically, the state associated to this submanifold can be represented by
the intersection of $\cM_c$ with the space of flat connections on a given bundle. By Narasimhan-Seshadri and generalizations, the intersection exists and is essentially unique if the bundle is stable. Locally, it %this means that ${\cal U}$ 
can be described as the graph of a generating function. 

The intersection can be described in terms of local data as follows. Pick an unitary flat connection $a$ and solve locally $a_{\bar z} = g^{-1} \bar \partial g$. The solutions associated to different local frames are related by left action of the transition functions of the $G$-bundle associated to the connection. The combination $\rho = g g^\dagger$ gives a map from the surface to the $G_\C/G_c$ homogeneous space, twisted by the transition functions on the left and their hermitian conjugate on the right. 

We find that $\rho$ satisfies the $G_\C/G_c$ WZW equations of motion 
\begin{align}
	\partial \bar \partial \rho &= \partial \rho \rho^{-1} \bar \partial \rho
\end{align}
which imply conservation of holomorphic and anti-holomorphic currents  
\begin{align}
	\partial \left(\rho^{-1} \bar \partial \rho \right)&= 0 \cr
	\bar \partial \left(\partial \rho \rho^{-1}\right)&= 0
\end{align}
Conversely, given a bundle we can solve for such a $\rho$ and then the current $\partial \rho \rho^{-1}$ gives a holomorphic connection with unitary monodromy. 

The $G_\C/G_c$ WZW action $S_{\mathrm{WZW}}$, which is just the action for a sigma model with target $G_\C/G_c$, evaluated on a solution 
of the equations of motion as a function of the choice of bundle, gives the generating function for the space of unitary flat connections. Indeed, essentially by definition,
\begin{equation}
	\frac{\delta}{\delta \cA_{\bar z}} S_{\mathrm{WZW}}= J_z[\rho] = - i \frac{s}{8 \pi} \cA_z[\rho] \qquad \qquad \frac{\delta}{\delta \bar \cA_{z}}S_{\mathrm{WZW}} = J_{\bar z}[\rho] =- i \frac{s}{8 \pi} \bar \cA_{\bar z}
\end{equation}

It is easy to argue that these statements will persist quantum-mechanically: the partition function $Z_{\mathrm{WZW}}$ for a $G_\C/G_c$ WZW sigma model at imaginary level coincides with the wave-function which quantized $\cM_c$.\footnote{For real level, this is a well-studied 2d CFT. We need to analytically continue these results.} 

The argument is simple: the wavefunction can be computed from a slab geometry, with $B_c$ at one end and Dirichlet boundary conditions at the other end. 
The computation will not depend on the thickness of the slab, as the 3d theory is topological. When the slab is very thin, the 3d Chern-Simons theory reduces precisely to the 2d WZW model with target $G/G_\C$ and level $is/2$. A similar argument can be used to study the pairing of $B_c$ to oper boundary conditions, leading to the partition function of the 2d Toda CFT.

We expect the state $|1 \rangle$ created by $B_c$ to be normalizable, as $\cM_c$ is compact. In terms of the WZW model, this means that the integral of $|Z_{\mathrm{WZW}}|^2$
over $\Bun(C,G)$ should converge. The WZW partition function is expected to be singular at the ``wobbly locus'' of $\Bun(C,G)$, so this statement is rather non-trivial. 
 
The $B_c$ boundary conditions support topological Wilson lines $W_{R,\ell}$ labelled by finite-dimensional representations $R$ of $G_c$. These coincide with the image of both holomorphic and anti-holomorphic bulk lines associated to the same data. 
 Skeins of Wilson lines added to the  boundary define a more general collection of states $|a\rangle$, coinciding with $W_a |1\rangle=\wt W_a |1\rangle$. 
 These states also do not depend on the complex structure on $C$. The mapping class group should act on them as it acts on 
 the skeins themselves. %, modulo topological manipulations. 
In the language of the WZW model, the states $|a\rangle$ should be given explicitly  by partition functions of the WZW theory decorated by skeins $W_a$ of Verlinde lines.

Irregular singularities on $C$ featured prominently in some of our examples and in quantum group applications. They complicate the semi-classical interpretation of the  
state $|1\rangle$. Indeed, $\rho^2$ acts on the Stokes data of irregular singularities by rotating the Stokes lines by one full sector. In particular, it does not square to $1$. 
The classical equation $W_{\rho(a)} = \wt W_a^\ast= W_a^\ast$  implies the rather restrictive condition $W_a = W_{\rho^2(a)}$ and thus does not appear to describe 
a real Lagrangian manifold in the real phase space. Instead, the condition $W_a = \wt W_a$  describes some complexified Lagrangian manifold. 

At first sight, there is some tension between this statement and the desired definition of $|1\rangle$ as being created by the $B_c$ boundary condition. The tension is only ''local'' in $C$, though: the condition $W_a = \wt W_a$ is compatible with the monodromy data away from irregular singularities being unitary. Only the Stokes data at the irregular singularity 
is affected by  $\rho^2$. Irregular singularities on $C$ can be thought of as some intricate disorder line defect in the 3d Chern-Simons theory. The state $|1\rangle$ prescribes some specific behaviour at the point where the disorder line meets the $B_c$ boundary. It would be interesting to understand this point better.

It is natural to wonder if alternative options could be available. In the specific case where $A_\fq$ is $U_{\fq}(\mathfrak{sl}_2)$, we saw that the Schur correlation functions define a positive-definite inner product associated to a specific $\rho$, but other options are available (and previously studied) which employ a different choice $\rho'$
and may be associated to Schur indices modified by surface defects. Such alternative options could be used 
at any rank 1 irregular singularity, just by employing the same surface defect for SQED$_2$. It seems plausible that 
a range of alternative options would be independently available at each irregular puncture. We leave this point to future work.  

\subsection{$GL(1)$ on $T^2$}
As a final toy model, consider a $GL(1)$ Chern-Simons theory compactified on an elliptic curve $E_\tau$. We can gauge-fix the flat connection to be constant,
\begin{equation}
\cA_{\bar z} = \frac{2 \pi i a}{\tau - \bar \tau}
\end{equation}
The normalization is chosen so that a gauge transformation by 
\begin{equation}
	e^{2 \pi i (\delta \bar z + \bar \delta z)} 
\end{equation}
with 
\begin{equation}
	\delta = \frac{n \tau + m}{\tau - \bar \tau} 
\end{equation}
shifts 
\begin{equation}
	a \to a + n \tau + m
\end{equation}
and identifies $\Bun$ with the elliptic curve $E_\tau$.

If we denote $\cA_z = p$, the momentum conjugate to $a$, then the gauge transformations also shift 
\begin{equation}
	p \to p - 2 \pi i \frac{n \bar \tau + m}{\tau - \bar \tau}
\end{equation}
so the phase space is a twisted cotangent bundle of $E_\tau$ and the Hilbert space will be given by $L^2$ normalizable sections of a bundle on 
$E_\tau$. When we quantize $p = - i s^{-1} \partial_a$, the gauge transformations will have to be accompanied by multiplication of the wavefunctions by 
\begin{equation}
	c_{n,m} \exp \left[2 \pi s \frac{n \bar \tau + m}{\tau - \bar \tau} a + 2 \pi s \frac{n \tau + m}{\tau - \bar \tau} \bar a \right]
\end{equation}
for some constant $c_{n,m}$. 

A prototypical wave-function is the Gaussian
\begin{equation}
	|1 \rangle = \exp 2 \pi s \frac{|a|^2}{\tau - \bar \tau}
\end{equation}
so that $c_{n,m} = \exp 2 \pi s \frac{|n \tau + m|^2}{\tau - \bar \tau}$. 

The notation anticipates that this wavefunction has a nice behaviour under the action of the quantum holonomies:
\begin{equation}
	u = \exp \left[- i s^{-1} \partial_a + \frac{2 \pi i a}{\tau - \bar \tau}\right] \qquad \qquad v = \exp  \left[- i s^{-1}\tau  \partial_a + \frac{2 \pi i a \bar \tau}{\tau - \bar \tau} \right]
\end{equation}
Indeed, the quantum holonomies act with norm $1$
\begin{align}
	u |1 \rangle &= \exp \left[\frac{2 \pi i (a - \bar a)}{\tau - \bar \tau}+ s^{-1}\frac{\pi}{\tau - \bar \tau}\right]|1 \rangle = (u^\dagger)^{-1}|1 \rangle \cr
	v |1 \rangle &= \exp  \left[\frac{2 \pi i (a \bar \tau- \bar a \tau)}{\tau - \bar \tau}+ s^{-1}\frac{ \pi  |\tau|^2}{\tau - \bar \tau} \right]|1 \rangle =  (v^\dagger)^{-1}|1 \rangle 
\end{align}
If we define more general quantum holonomies 
\begin{equation}
	x_{m,n} = \exp \left[- i s^{-1} (n \tau + m) \partial_a + \frac{2 \pi i a (n \bar \tau + m) }{\tau - \bar \tau}\right]
\end{equation}
then we can  produce a dense basis of states 
\begin{equation}
	x_{m,n}|1 \rangle = \exp \left[ 2 \pi s \frac{|a -i s^{-1} (n \tau + m))|^2}{\tau - \bar \tau} -s^{-1} \frac{\pi |n \bar \tau + m|^2 }{\tau - \bar \tau}\right]
\end{equation}
which identify the Hilbert space with $L^2(\bZ^2)$, the natural quantization of the space $\C^* \times \C^*$ of holonomies. 

Up to a $\tau$-dependent pre-factor, the wavefunction $|1 \rangle$ we proposed precisely matches the analytically continuation to imaginary $\kappa$ of the partition function of a non-compact free boson of level $\kappa$, which is a WZW model with target $GL(1,\C)/U(1) = \R$. By construction, it satisfies the KZ equations. 

The trace associates to the state $|1\rangle$ is $\Tr x_{m,n} = \delta_{n,0} \delta_{m,0}$.

\subsection{Outlook: A new quantization of complex Chern-Simons theory}

Developing the quantisation of complex Chern-Simons theory more deeply and in larger generality
will require more powerful instruments. In previously studied cases of quantum Chern-Simons
theory there were two main instruments which have turned out to be very useful,
one being quantum group theory, the other being cluster algebra technology. 
Quantum groups can in particular represent  a residual gauge symmetry
in the quantisation of Chern-Simons theory, allowing one to construct quantum Chern-Simons theory 
from quantum group representation theory.

The existence of different quantum deformations of $\mathrm{SL}(2,\bC)$ suggests that there may
exist quantisations of complex Chern-Simons theory which differ from the one previously constructed in \cite{Buffenoir:2002tx}.
To close this section we'd like to explain why we expect that the quantum Lorentz group from 
Schur quantisation is particularly well-suited for developing a new quantization of complex Chern-Simons theory.
Roughly, it seems particularly well-suited for the use of a powerful blend of 
cluster algebra and quantum group theory, generalising the paradigm 
provided by quantum Teichm\"uller theory. 

It has been observed in \cite{kashaev2001spectrum}
that the co-product of $U_q(\mathfrak{sl}_2)$ is related to the quantum cluster algebra associated 
to the marked twice punctured disk in the context of quantum Teichm\"uller theory. It follows that the 
braiding in quantum Teichm\"uller theory is naturally related to the R-matrix of the modular
double of $U_q(\mathfrak{sl}_2)$ constructed in \cite{FaddeevMD}. These 
observations have been generalised to higher Teichm\"uller 
theory in \cite{schraderCluster,IpCluster}.  The relation 
to quantum Teichm\"uller theory helps to compute the Clebsch-Gordan decomposition of 
tensor products of modular double representations \cite{Nidaiev:2013bda}.
These connections represent key ingredients in the passage from the cluster algebra
structures originally defining quantum Teichm\"uller theory to the modular functor structure
associated to pants decompositions  \cite{Teschner:2013tqy}.

In the forthcoming companion paper we will study relations between cluster algebras and Schur quantisation. 
It  will  turn out that the quantum group from Schur quantisation defined here 
has a natural relation to cluster algebras which 
generalises  the relations known from quantum Teichm\"uller theory. This should 
be a  key ingredient for a new quantisation of complex Chern-Simons theory which has a natural relation 
to Schur quantisation, as predicted by the dualities discussed in the introduction. % of this paper. 

\section{Real quantization} \label{sec:real}
In this Section we discuss some evidence for the existence of a real version of Schur quantization: an Hilbert space $\cH_\fq^\bR$ 
equipped with an unitary action of a $*$-algebra $\fA_\fq^\bR$ obtained by equipping $A_\fq$ with a star structure 
\begin{equation}
a^* = \tau(a) \,.
\end{equation} The Hilbert space will be defined as 
the $L^2$ closure of an auxiliary $A_\fq$ module $M_\fq$ equipped with a certain inner product. We will be schematic and leave many details to future work. 

As a quick motivation, consider the standard notion of unitary representations of real forms $\fg_\bR$ of Lie algebras. Such representation can be 
thought of as unitary representations $\cH$ of a $*$-algebra $\fU(\fg_\bR)$ defined by equipping $U(\fg)$ with a $*$-structure $\tau$, an automorphism of the Lie algebra 
fixing $\fg_\bR$. The Lie algebra $\fg_\bR$ has a maximal compact sub-algebra which exponentiates to a compact Lie group $K$ acting on the representation $\cH$. 
Defining $\rho$ as before as a reflection of the generators of $\fg$, the compact sub-algebra is fixed by $\rho \circ \tau$. We can decompose both $U(\fg)$ and $\cH$ into 
finite-dimensional irreps of $K$. It is easy to see that this gives a dense basis $M$ in $\cH$ which behaves as a module for $U(\fg)$, sometimes denotes as a $(\fg,K)$-module.
Conversely, $\cH$ can be recovered as the $L^2$ closure of $M$ under an inner product. The notion of $K$-invariant, e.g. spherical vectors in $\cH$ is also important. 

These structures are expected to arise naturally in sphere quantizations, though a systematic analysis is still not available. The mathematical theory is rich. See e.g. \cite{2021arXiv210803453L}. We expect real Schur quantization to 
provide analogous structures for $*$-algebras $\fU_\fq(\fg_\bR)$ defined by a similar star structure $\tau$ on $U_q(\fg)$.

\subsection{A Chern-Simons motivation}
An analogue real version of sphere quantization is reasonably well understood in the 3d setup \cite{Gaiotto:2023hda} and employs correlation functions on hemispheres. The choice of boundary conditions for the hemisphere determines the structure of the module and of the inner product. The $*$ structure on the algebra and the positivity property of the inner product are obtained as a generalization of certain properties \cite{Ishtiaque:2017trm} of for protected sphere two-point functions for $(2,2)$ SCFTs \cite{Benini:2012ui,Doroud:2012xw}. They are typically obscure, and 
identifying a boundary condition which gives a given $\tau$ is challenging. 

We are not aware of K-theoretic generalizations of \cite{Ishtiaque:2017trm} which could be relevant in the current context. Indeed, we doubt they exits. As a result, the real Schur quantization procedure we sketch below will only work for a certain class boundary conditions, which at the moment we do not know how to characterize. Fortunately, 
some considerations about complex Chern-Simons theory, KW theory and 2d CFT lead to the definition of a bounty of boundary conditions suitable for the real 
Schur quantization procedure, at least for theories of class $\cS$. Indeed, it may well be the case that such constructions for $\fU_\fq(\fg_\bR)$ may shed light on the 
representation theory of both quantum and classical Lie algebras, after a judicious 3d limit.

In complex CS theory, we can attempt quantization in a situation where the surface $C^\bR$ has a boundary and/or is non-orientable. In order to do so (in a topological way), we should specify a topological boundary condition at each boundary components of $C^\bR$. We have already encountered a natural set of options associated to real forms $G_\bR$ of $G$, such that the complex connection restricts to a  $G_\bR$ connection at the boundary. We will employ these. 

Sometimes, a boundary condition can be defined via a reflection trick. This is the case here. We can describe $C^\bR$ as a quotient $C/\tau$, where $C$ is the orientation cover of $C^\bR$ and $\tau$ is an anti-holomorphic involution of $C$. The boundaries of $C^\bR$ lift to the locus of points in $C$ fixed by the action of $\tau$. As $\tau$ flips the orientation on $C$, we can keep the complex CS action invariant if we define an action of $\tau$ which complex-conjugates the connection. At a boundary component, the action of $\tau$ is such that the component of the connection parallel to the boundary is $\tau$-invariant if it lies in (the Lie algebra of) $G_\bR$.

Correspondingly, we have a lift of $\tau$ to an anti-holomorphic involution of the space $\cM(C)$ of complex flat connections on $C$. The $\tau$-fixed locus $\cM(C^\bR)$ in $\cM(C)$ gives a real phase space for the system, which we aim to quantize. A point in $\cM(C^\bR)$ gives, in particular, a complex flat connection on $C$. We thus have classical observables $W_a$ and $\wt W_a$ labelled by a skein $a$ in $C$. Restricted to $\cM_\bR$, these observables satisfy a reality condition we write as:
\begin{equation}
	\wt W_{a} =W^\dagger_{\rho(a)} = W_{\tau(\rho(a))} \, .
\end{equation}

This complex CS setup has a lift to KW theory which modifies slightly a construction from \cite{Gaiotto:2021tsq}. We can define a three-manifold 
\begin{equation}
	U_\tau \equiv \frac{C \times [-1,1]}{\bZ_2}
\end{equation}
where $\bZ_2$ acts as a combination of $\tau$ and a reflection of the segment. The resulting manifold has a co-dimension $1$ singular locus 
associated to boundaries of $C^\bR$. In \cite{Gaiotto:2021tsq}, a prescription was given to smoothen the singular locus in a manner depending on the choice of $G_\bR$. 
Then the space of states of KW theory on $U_\tau$ gives a tentative definition for the space of states of complex CS theory on $C^\bR$. 

Next, we can introduce the real analogue of the $B_c$ boundary condition. Away from boundaries of $C^\bR$, we can restrict the connection to be unitary, i.e. be a $G_c$ connection. At the boundary, we need to further select some junction between the $G_\bR$ boundary condition and the $G_c$ boundary condition. Both boundary conditions can be implemented in KW theory by a reflection trick \cite{Gaiotto:2024tpl}. We thus quotient $C \times [-1,1] \times \bR$ by the above $\bZ_2$ and a $\bZ_2$ reflecting both factors in $[-1,1] \times \bR$.

The resulting geometry is a bit complicated. The two reflections combine in particular to a $\bZ_2$ quotient of the boundary $C \times \bR$ factors by a simultaneous reflection. This 
can be smoothened to a manifold $V_\tau$ akin to $U_\tau$, but with a semi-infinite cylindrical region. As a consequence, we can create states $|m;\bR\rangle$ labelled by skeins $m$ in the Skein Module $M_\fq$ associated to $V_\tau$.  These are our tentative dense collection of states for 3d CS theory on $C_\bR$, to be completed to an Hilbert space $\cH_\fq^\bR$
by computing somehow the inner products
\begin{equation}
	\langle m,\bR|m',\bR \rangle \,.
\end{equation}
with an action of the Skein algebra $A_\fq$ equipped with a $*$ structure by the action of $\tau$ on skeins.

Next, we conjecture that the inner products can be computed by a careful deformation and a chain of dualities mapping them to Schur half-indices, aka a twisted partition functions on $HS^3 \times S^1$. We will not attempt to prove this fact. In the bulk of $HS^3$ we place the class $\cS$ theory associated to $C$ and at the boundary of $HS^3$ we place the boundary condition defined by $V_\tau$ according to the 3d-3d correspondence \cite{Dimofte:2011py,Dimofte:2013lba}. As a check, the boundary lines give indeed elements of $M_\fq$ with the correct action of 
the Skein algebra $A_\fq$ and we can thus write a meaningful equality
\begin{equation}
	\langle m,\bR|m',\bR \rangle = I\!\!I_{m,m'}(\fq)
\end{equation}
which can be further decorated by Skein algebra elements/bulk line defects:
\begin{equation}
	\langle m,\bR|W_a|m',\bR \rangle = I\!\!I_{m,a m'}(\fq)
\end{equation}
This identification implies positivity properties for half-indices of boundary conditions which arise from $V_\tau$.

We should elaborate on the conjectural algebraic structures which appear in this construction. Suppose that we are given a (left) module $M_\fq$ of $A_\fq$ and we are looking for a positive-definite inner product on $M_\fq$ compatible with a $*$-structure  $W^\dagger_{a} = W_{\tau(a)}$. The map $\tau$ is an algebra morphism $A_\fq \to A_\fq^\op$. It maps $M_\fq$ to 
a right module $\tau(M_\fq)$. The inner product gives, in particular, a pairing $(\bullet, \bullet)$ between $\tau(M_\fq)$ and $M_\fq$, i.e. a linear functional on the tensor product $\tau(M_\fq) \otimes_{A_\fq} M_\fq$. 

In general, given $\fA_\fq$ and $M_\fq$, one can find a finite-dimensional space of such linear functionals, which may or not include a cone of positive-definite ones. Again, it would be nice to find a way to characterize the functional provided by the Schur half-indices counting local operators between 
boundary lines.

A key property of the Schur half-indices, of course, is that they will only depend on the theory and boundary condition and not on the 
specific duality frame used to describe either of them. 

\subsection{Example: free boundary conditions in pure U(1) gauge theory}
Consider the case of the quantum torus. Up to $SL(2,\bZ)$ re-definitions, there are two natural choices for $\tau$: 
\begin{itemize}
	\item The choice $\tau: (u,v)\to (u,v^{-1})$ classically fixes the locus $u^\dagger =u$, $|v|^2=1$. This locus has two components: $u$ can be positive or negative. The locus fixed by $\rho \circ \tau$ is $u= \pm 1$. The corresponding unitary representations have unitary $v$ and self-adjoint $u$.
	\item The choice $\tau: (u,v)\to (v,u)$ classically fixes the locus $u = v^\dagger$. The locus fixed by $\rho \circ \tau$
	is $u v=1$. The corresponding unitary representations have $u$, $v$ adjoint to each other. In particular, they are not normal operators. Instead, $u v$ is self-adjoint and 
	$\fq u v^{-1}$ is unitary. 
\end{itemize}
A prototypical representation of the first type involves the Hilbert space $L^2(\bZ) \simeq L^2(S^1)$. 
In the first description, $v$ is a translation operator in $\bZ$ and $u$ is a multiplication operator $\fq^{2n}$. 
In particular, the state $|1,\bR\rangle$ supported at the origin is a reasonable quantization of the $u=1$ locus 
and generates a basis $|v^n,\bR\rangle$ of the whole Hilbert space under the action of $v$. The images define a simple module for the quantum torus algebra, consisting of powers of $v$. 

Expectation values
\begin{equation}
\langle 1;\bR| \fq^{-ab} u^a v^b |1;\bR\rangle = \delta_{b,0}
\end{equation}
can be identified, up to a $(\fq^2)_\infty$ normalization factor, with half-indices for Neumann boundary conditions in the 
4d pure $U(1)$ gauge theory. The $v^n$ module elements represent K-theory classes of boundary Wilson lines. 
Dirichlet boundary conditions would exchange the role of $u$ and $v$. 
\footnote{An alternative realization would employ the same Hilbert space, but $u$ acting as $- \fq^{2n}$. At the level of 4d gauge theory, the difference between these two choices is a bit subtle. Essentially, it has to do with a choice of fermion parity for the local operator representing the endpoint of a bulk 't Hooft line at the Neumann boundary. }

A prototypical representation of the second type also involves the Hilbert space $L^2(\bZ) \simeq L^2(S^1)$. 
We can define the action of $u$ and $v$ as a combination of the translations in either direction along $\bZ$ and multiplication by $\fq^{n}$. Again, the $|1,\bR\rangle$ supported at the origin is a reasonable quantization of the $u v=1$ locus and generates a basis of the whole Hilbert space. In the 4d gauge theory description, the relevant boundary condition 
is a Neumann boundary condition equipped with one unit of Chern-Simons coupling. 

If we identify $\bC^* \times \bC^*$ with the moduli space of $\bC^*$ flat connections on $T^2$, we could attempt to match the above involutions with geometric involutions of the $T^2$. Denote as $\sigma_1$, $\sigma_2$ the two angular coordinates on $T^2$. A reflection $\sigma_2 \to - \sigma_2$, $\sigma_1 \to \sigma_1$ has fixed loci $\sigma_2 = 0$ and $\sigma_2 = \pi$ and the quotient of $T^2$ gives an annulus. A reflection $\sigma_1 \to \sigma_2$ gives the Mo\"ebius strip and $\sigma_2 \to - \sigma_2$, $\sigma_1 \to \sigma_1 + \pi$ is a Klein bottle. 

It is pretty clear that the annulus and Klein bottle will give involutions of the first type and the Mo\"ebius strip of the second type. 
We leave a detailed identification of different quantizations and choices of real forms at boundaries to future work.

\subsection{Some comments on representations of the $U_q(\mathfrak{sl}_2)$ quantum group}
Already for the case of Abelian gauge theories, there is a large collection of boundary conditions which 
are compatible with some involution $\tau$ and may give non-trivial representations of $A_\fq$ with 
the corresponding Hermiticity properties. For SQED$_2$, we can correspondingly engineer a variety of unitary representations of $U_q(\mathfrak{sl}_2)$. 
For the case of SQED$_1$ we similarly expect $q$-deformed versions of representations of the Weyl algebra. 
In particular we expect $q$-deformations of several unitary representations of $U(\mathfrak{sl}_2)$
encountered in (hemi)sphere quantization: principal series representations, discrete series and finite-dimensional representations. 

One may note, on the other hand, that the $\ast$-algebra structures on $U_q(\mathfrak{sl}_2)$
have been classified in \cite{masuda1990unitary}. For real $\fq$ one only finds quantum 
deformations of $SU(2)$ and $SU(1,1)$. It would be interesting to clarify if these representations 
can recovered within Schur quantisation. 
 We leave details to future work. 

\subsection{Some comments about 2d CFT constructions.}
For a real analogue  of a 2d CFT analysis, we can approximate $\cM(C^\bR)$ as the twisted cotangent bundle of a space of 
``real bundles'' $\mathrm{Bun}_\bR$. A real bundle should be understood as the data necessary to define a 2d CFT with Kac-Moody symmetry on a Riemann surface $C$ with boundaries or cross-caps. The $G_\bR$ data at boundaries controls the gluing condition for chiral and anti-chiral currents. We get a candidate Hilbert space $\cH_\bR$ as a space of $L^2$-normalizable twisted half-densities on $\mathrm{Bun}_\bR$. The geometry of the problem may allow a greater choice of twists than in the complex case. We will not attempt to characterize them here.

Holomorphic quantization depends on a choice of complex structure on $C$
via the KZ equations. With some work, it should be possible to extend the definition of Verlinde lines to include the quantum analogues of the classical observables defined above, satisfying again $W^\dagger_{a} = W_{\tau(a)}$ for bulk observables. 

From the 2d perspective, it is natural to consider the definition of a $G_\bC/G_c$ WZW model on $C$. This will require a choice of boundary conditions and cross-cap states for the WZW model, which in turn may allow for a variety of extra parameters 
in the construction. We will not attempt to characterize them here. Up to this hidden data, the partition function should define a 
state $|1;\bR\rangle \in \cH_\bR$. We can produce further states by acting with quantum observables, defining the image in $\cH_\bR$ of some module $M_\fq$ for $A_\fq$ which, in a sense, quantizes $\cL$. 

We expect this to provide a dictionary between real Schur and 2d CFT constructions.
\section*{Acknowledgements}
This research was supported in part by a grant from the Krembil Foundation. DG is supported by
the NSERC Discovery Grant program and by the Perimeter Institute for Theoretical Physics.
Research at Perimeter Institute is supported in part by
the Government of Canada through the Department of Innovation, Science and Economic
Development Canada and by the Province of Ontario through the Ministry of Colleges and
Universities.

The work of J.T. was funded by the Deutsche Forschungsgemeinschaft (DFG, German Research Founda-
tion) – SFB-Gesch\"aftszeichen SFB 1624/1 2024 – Projektnummer 506632645, and furthermore
supported by the Deutsche Forschungsgemeinschaft under Germany's
Excellence Strategy EXC 2121 "Quantum Universe" 390833306.

\appendix

\section{Some $U(N)$ examples}
\subsection{Pure $U(N)$ gauge theory}
We now discuss briefly the gauge theory with $U(N)$ gauge group and no matter fields in order to illustrate the general combinatorics of the Schur correlation functions and the isometry to an auxiliary Hilbert space. Furthermore, the description of $U(2)$ gauge theory is helpful in setting up conventions for an $SU(2)$ gauge group, which is our next example.
 
The Schur index becomes 
\begin{align}
	I_\fq &= \frac{1}{N!} \oint_{|\zeta|=1} \prod_i \frac{(\fq^2)^2_\infty d\zeta_i}{2 \pi \i \zeta_i} \prod_{i\neq j}(1-\zeta_i \zeta_j^{-1}) (\fq^2 \zeta_i \zeta_j^{-1};\fq^2)^2_\infty 
\end{align}
We will not attempt to describe the full algebra $A_\fq$. Recall that its generators are expected to be classes $[L_{\lambda_m, \lambda_e}]$ of 
't Hooft-Wilson loops, with labels defined up to the action of the Weyl group. Here $\lambda_m$ and $\lambda_e$ are vectors in $\bZ^N$, with an $S_N$ Weyl group action. 

There is an useful notion of ``minuscule'' magnetic charge, $\lambda_m = (1, \cdots, 1,0\cdots, 0)$ up to an overall shift by the diagonal magnetic charge $(1,\cdots, 1)$. The 't Hooft operators of ``minuscule'' charge do not suffer from bubbling and thus the corresponding difference operators are readily written. They can be dressed by generic electric charges. The residual $S_k \times S_{N-k}$ action, where $k$ is the number of ``1'' entries, reduces the electric charge to a choice of an $U(k) \times U(N-k)$ weight. 

 We will present the difference operators in a form adapted to an isometry to $L^2((S^1 \times \bZ)^N)^{S_N}$ 
with spherical vector image
\begin{align}
	I\!\!I_B(\zeta) &= \delta_{B,0}(\fq^2)^N_\infty\prod_{i\neq j}(\fq^2 \zeta_i \zeta_j^{-1};\fq^2)_\infty 
\end{align}
We define $N$ copies $(u_i,v_i)$ and $(\wt u_i, \wt v_i)$ of the standard set of multiplication and shift operators. 

In order to avoid some square roots of phases in the formulae below, we will use a slightly modified magnetic Vandermonde measure 
\begin{equation}
 	\Delta_B(\zeta) = \prod_{j < i}(v_i-v_j)(\wt v_i^{-1} -\wt v_j^{-1})
\end{equation}
which differs from the standard one by a factor of $\fq^{(N-1) \sum_i B_i}$.\footnote{The price of this modification is some factors of $\fq^{N-1}$ which may have to be added to our expressions of 't Hooft line insertions below to match the correct answer for half-BPS line defects. }

An important ingredient in the presentation of 't Hooft operators of minimal charge are the combinations 
\begin{align}
	u_{+,i} &\equiv \frac{\fq^{N-1}}{\prod_j^{j \neq i}(1-v_j v_i^{-1})}u_i \cr
	u_{-,i} &\equiv \frac{1}{\prod_j^{j \neq i}(v_i v_j^{-1}-1)}u^{-1}_i \cr
	\wt u_{+,i} &\equiv \frac{\fq^{N-1}}{\prod_j^{j \neq i}(1-\wt v_i \wt v_j^{-1})}\wt u_i \cr
	\wt u_{-,i} &\equiv \frac{1}{ \prod_j^{j \neq i}(\wt v_j \wt v_i^{-1}-1)}\wt u^{-1}_i
\end{align}
The difference operators are $S_N$ -invariant combinations of these and $v_i$.

We choose the relative normalization so that 
\begin{equation}
	u_{\pm, i} I\!\!I_B(\zeta) = \wt u_{\pm, i}I\!\!I_B(\zeta)\, . 
\end{equation}. 
and we have formal adjoint relations (with respect to the above Vandermonde measure)
\begin{align}
	\rho(u_{-,i}) &= \prod_j^{j\neq i} (\fq^{-1} v_j v_i^{-1}) u_{+,i} \cr
	\rho(u_{+,i}) &= u_{-,i} \prod_j^{j\neq i} (\fq v_i v_j^{-1})
\end{align}
and the expected Witten effect:
\begin{align}
	\rho^2(u_{-,i}) &=\prod_j^{j\neq i} (v_i v_j^{-1}) u_{-,i} \prod_j^{j\neq i} (v_i v_j^{-1}) \cr
	\rho^2(u_{+,i}) &= \prod_j^{j\neq i} (v_j v_i^{-1}) u_{+,i} \prod_j^{j \neq i} (v_j v_i^{-1}) 
\end{align}
Wilson lines are realized by characters $\chi_R(v)$ of $U(N)$. 

The 't Hooft operators of minimal charge are simply the Weyl-invariant sums
\begin{equation}
H_1 \equiv \sum_i u_{+,i} \qquad \qquad H_{-1} \equiv \sum_i u_{-,i}
\end{equation}
They can be dressed by Wilson lines for $U(1) \times U(N-1)$ in a natural way by inserting appropriate characters in the sum. E.g. we can define
\begin{equation}
H_{1,n} \equiv \sum_i \fq^{n} v_i^n u_{+,i} \qquad \qquad H_{-1,n} \equiv \sum_i \fq^{-n} v_i^n u_{-,i}
\end{equation}
by inserting $U(1)$ characters evaluated on $v_i$. Characters for $U(N-1)$ will be evaluated on $v_j$ for $j \neq i$. 
 
A product of the form $w_R H_1$ thus gives a sum of minimal 't Hooft operators dressed by the $U(1) \times U(N-1)$ representations 
contained in $R$, with extra powers of $\fq$ controlled by the $U(1)$ charge. 

The 't Hooft operators of higher minuscule charge $k$ are sums of ${N \choose k}$ higher shift operators
\begin{equation}
	u_{+,I} \equiv \frac{\fq^{k(N-k)}}{\prod_{j \notin I}\prod_{i\in I}(1-v_j v_i^{-1})}u_i 
\end{equation}
where $I$ is a subset of size $k$ in $1, \cdots, N$. They can be dressed by Wilson lines for $U(k) \times U(N-k)$ in a natural way. 

The full algebra of observables can be recovered from Wilson lines and 't Hooft lines. E.g. 
\begin{equation}
	[H_1, w_1] = [H_1,\sum_i v_i] = (\fq - \fq^{-1}) H_{1,1}
\end{equation}
etcetera. 

As a richer example of the relations which appear in $A_\fq$, note 
\begin{equation}
	u_{+,i} u_{+,j}  = \frac{\fq^2}{(1-\fq^2 v_i v_j^{-1})(1-v_j v_i^{-1})}u_{+,i,j}
\end{equation}
and consider 
\begin{equation}
H_{1,n_1} H_{1,n_2} = \sum_{i} \fq^{n_1+3 n_2} v_i^{n_1+n_2} u_{+,i}^2 + \sum_{i \neq j} \frac{\fq^{n_1+n_2+2} v_i^{n_1+1}v_j^{n_2+1}}{(v_j-\fq^2 v_i)(v_i-v_j)} u_{+,i,j} 
\end{equation}
and then 
\begin{align}
&\fq^{n_1-n_2} H_{1,n_1} H_{1,n_2} - \fq^{-n_1-n_2}  H_{1,0} H_{1,n_1+n_2} = \sum_{i \neq j} \frac{v_j^{n_1} - \fq^{2 n_1} v_i^{n_1} }{v_j-\fq^2 v_i} \frac{ \fq^2 v_i v_j^{n_2+1}}{v_j-v_i} u_{+,i,j} = \cr
&= \sum_{i<j} \left[\frac{v_j^{n_1} - \fq^{2 n_1} v_i^{n_1} }{v_j-\fq^2 v_i}v_j^{n_2} -  \frac{v_i^{n_1} - \fq^{2 n_1} v_j^{n_1} }{v_i-\fq^2 v_j}v_i^{n_2} \right] \frac{\fq^2 v_i v_j}{v_j-v_i} u_{+,i,j} 
\end{align}
The coefficient of $u_{+,i,j}$ is a symmetric polynomial in $v_i$ and $v_j$, corresponding to some $U(2)$ Wilson line dressing for a 't Hooft operators of minuscule charge $2$.
E.g. 
\begin{align}
&\fq^{2} H_{1,1} H_{1,-1} -  H_{1,0}^2 = \sum_{i<j} \fq^2 u_{+,i,j} \equiv \fq^2 H_2
\end{align}

Another important relation is 
\begin{equation}
H_{1,0} H_{1,1} = \fq^2 H_{1,1} H_{1,0}
\end{equation}
which is a first step towards building a cluster structure on the K-theoretic Coulomb branch \cite{schrader2019ktheoretic}. 

The ``bare'' 't Hooft operators coincide with the Hamiltonians for the open relativistic quantum Toda chain and in particular commute with each other. This is not completely obvious from the explicit formulae: 
\begin{align}
	[H_1, H_2] = \sum_{i} \sum_{j<k} [u_{+,i}, u_{+,jk}] &= \sum_{j<k}\sum^{i \neq j, i\neq k}_{i}  \frac{\fq^3}{(1-\fq^2 v_i v_j^{-1})(1-v_j v_i^{-1})(1-\fq^2 v_i v_k^{-1})(1-v_k v_i^{-1})} u_{+,ijk} + \cr &-\frac{\fq^3}{(1- v_i v_j^{-1})(1-\fq^2 v_j v_i^{-1})(1-v_i v_k^{-1})(1-\fq^2 v_k v_i^{-1})} u_{+,ijk} 
\end{align}
vanishes only after symmetrization of $i,j,k$. On the other hand
\begin{align}
	\fq^2 H_{1,1} H_2 = H_2 H_{1,1}  
\end{align}
gives another piece of the cluster structure. The cluster structure is closely related to the IR perspective on Schur quantization discussed in the companion paper \cite{next}.
It will be used to predict the spectrum of the Toda Hamiltonians in this complex quantization scheme. 

\subsection{The ${\cal N}=2^*$ $U(N)$ gauge theory}
We now discuss very briefly the gauge theory with $U(N)$ gauge group and adjoint matter fields, the simplest conformal example. 
The Schur index becomes 
\begin{align}
	I_\fq &= \frac{1}{N!} \oint_{|\zeta|=1} \prod_i \frac{(\fq^2)^2_\infty d\zeta_i}{2 \pi \i (-\fq \mu;\fq^2)_\infty(-\fq \mu^{-1};\fq^2)_\infty \zeta_i} \prod_{i\neq j}\frac{(1-\zeta_i \zeta_j^{-1}) (\fq^2 \zeta_i \zeta_j^{-1};\fq^2)^2_\infty}{(-\fq \mu \zeta_i \zeta_j^{-1};\fq^2)_\infty {(-\fq \mu^{-1} \zeta_i \zeta_j^{-1};\fq^2)_\infty}}
\end{align}
\subsubsection{Abelianization ingredients}
The 't Hooft operators of ``minuscule'' charge do not suffer from bubbling. We will present them in a form adapted to an isometry to $L^2((S^1 \times \bZ)^N)^{S_N}$ 
with spherical vector image
\begin{align}
	I\!\!I_B(\zeta) &= \delta_{B,0}\frac{(\fq^2)^N_\infty}{(\fq \mu;\fq^2)^N_\infty}\prod_{i\neq j}\frac{(\fq^2 \zeta_i \zeta_j^{-1};\fq^2)_\infty}{(-\fq \mu \zeta_i \zeta_j^{-1};\fq^2)_\infty}
\end{align}
The presence of similar factors at numerator and denominator leads to neat simplifications of various formulae below.

Define $N$ copies $(u_i,v_i)$ and $(\wt u_i, \wt v_i)$ of the standard set of multiplication and shift operators. An important ingredient in the presentation of 't Hooft operators of minimal charge are the combinations\footnote{One may include some overall factors of $\mu^{\pm \frac 12}$ in order to restore a $\mu \to \mu^{-1}$ symmetry in the presentation of $A_\fq$ generators below. Again, we made a choice here which minimizes square roots of phases.} 
\begin{align}
	u_{+,i} &\equiv \prod_j^{j \neq i}\frac{\fq v_i+ \mu v_j }{v_i-v_j}u_i \cr
	u_{-,i} &\equiv \prod_j^{j \neq i} \frac{\mu^{-1} v_j+ \fq^{-1} v_i}{v_j-v_i }u^{-1}_i \cr
	\wt u_{+,i} &\equiv \prod_j^{j \neq i}\frac{\fq \wt v_j+ \mu \wt v_i}{\wt v_j-\wt v_i}\wt u_i \cr
	\wt u_{-,i} &\equiv \prod_j^{j \neq i}\frac{\mu^{-1} \wt v_i+ \fq^{-1} \wt v_j}{ \wt v_i-\wt v_j}\wt u^{-1}_i
\end{align}
Taking adjoints with the magnetic Vandermonde measure\footnote{Again we avoided some square roots at the price of a $\fq^{(N-1)\sum_i B_i}$ factor, leading to some $\fq^{N-1}$ factors below}
\begin{equation}
 	\Delta_B(\zeta) = \prod_{j < i}(v_i-v_j)(\wt v_i^{-1} -\wt v_j^{-1})\, ,
\end{equation}
we learn that 
\begin{align}
	\rho(u_{-,i}) &=  u_{+,i} \cr
	\rho(u_{+,i}) &= u_{-,i} 
\end{align}
so that $\rho^2=1$ as expected. 

The following intertwining relations hold for the images of the spherical vector $I\!\!I$
\begin{equation}
	u_{\pm, i} I\!\!I_B(\zeta) = \wt u_{\pm i}I\!\!I_B(\zeta) 
\end{equation}

\subsubsection{The algebra}
We can now review the presentation of the K-theoretic Coulomb branch algebra. First of all, Wilson lines are described by characters $\chi_R(v)$ for (finite-dimensional) $U(N)$ representations. 

The 't Hooft operators of minimal charge are simply
\begin{equation}
H_1 \equiv \sum_i u_{+,i} \qquad \qquad H_{-1} \equiv \sum_i u_{-,i}
\end{equation}
They can be dressed by Wilson lines for $U(1) \times U(N-1)$ in a natural way. E.g. 
\begin{equation}
H_{1,n} \equiv \sum_i \fq^{-n} v_i^n u_{+,i} \qquad \qquad H_{-1,n} \equiv \sum_i \fq^{n} v_i^n u_{-,i}
\end{equation}

The 't Hooft operators of higher minuscule charge $k$ are sums of ${N \choose k}$ higher shift operators
\begin{equation}
	u_{+,I} \equiv \prod_{j \notin I}\prod_{i\in I}\frac{\fq v_i - \mu v_j}{v_i-v_j}u_i 
\end{equation}
where $I$ is a subset of size $k$ in $1, \cdots, N$. They can be dressed by Wilson lines for $U(k) \times U(N-k)$ in a natural way. 

Again, ``bare'' 't Hooft lines commute, via miraculous-looking simplifications of the commutators. 
They coincide with the Hamiltonians for a trigonometric quantum Ruijsenaars-Schneider model. 

%As an example of the relations which appear in $A_\fq$, note 
%\begin{equation}
%	u_{+,i} u_{+,j}  = \fq \frac{(v_j+ \mu \fq v_i)( \fq v_i- \mu v_j)}{(v_j-\fq^2 v_i)(v_i-v_j)}u_{+,i,j}
%\end{equation}
%and consider 
%\begin{equation}
%H_{1,n_1} H_{1,n_2} = \sum_{i} \fq^{-n_1+n_2} v_i^{n_1+n_2} u_{+,i}^2 + \sum_{i \neq j} \fq^{-n_1-n_2+1} v_i^{n_1}v_j^{n_2} \frac{(v_j+ \mu \fq v_i)( \fq v_i+ \mu v_j)}{(v_j-\fq^2 v_i)(v_i-v_j)} u_{+,i,j} 
%\end{equation}
%and then 
%\begin{align}
%&\fq^{2 n_1} H_{1,n_1} H_{1,n_2} -  H_{1,0} H_{1,n_1+n_2} = \sum_{i \neq j} \frac{v_j^{n_1} - \fq^{2 n_1} v_i^{n_1} }{v_j-\fq^2 v_i} \frac{ \fq^{-n_1-n_2+1} v_j^{n_2}(v_j+ \mu \fq v_i)( \fq v_i+ \mu v_j)}{v_j-v_i} u_{+,i,j} = \cr
%&= \sum_{i<j} \left[\frac{v_j^{n_1} - \fq^{2 n_1} v_i^{n_1} }{v_j-\fq^2 v_i}v_j^{n_2}(v_j+ \mu \fq v_i)( \fq v_i+ \mu v_j) -  (i \leftrightarrow j) \right] \frac{ \fq^{-n_1-n_2+1}}{v_j-v_i} u_{+,i,j} 
%\end{align}
%The coefficient of $u_{+,i,j}$ is a symmetric polynomial in $v_i$ and $v_j$, corresponding to some $U(2)$ Wilson line dressing for a 't Hooft operators of minuscule charge $2$.
%E.g. 
%\begin{align}
%&\fq^{2} H_{1,1} H_{1,-1} -  H_{1,0}^2 = -\fq \mu \sum_{i<j} (-\fq \mu^{-1}- \fq \mu +v_i v_j^{-1}+ 2 + v_j v_i^{-1})u_{+,i,j} 
%\end{align}

\subsubsection{Comments on S-duality}
This theory is endowed with S-duality, acting as $SL(2,\bZ)$ on the magnetic and electric labels of 't Hooft-Wilson loops. E.g. the $S$ transformation permutes $H_1$ 
and the Wilson line $w_1$ in the fundamental representation. The Schur index and Schur quantization are invariant under $SL(2,\bZ)$, but this is far from obvious from the 
above presentation. An immediate consequence of S-duality is that the joint spectrum of the bare minuscule 't Hooft operators, which commute with each other, must coincide with that of the minuscule Wilson lines, i.e. Wilson lines for antisymmetric powers of the fundamental representation. This result fully characterizes the spectrum of this complex quantization of the 
trigonometric Ruijsenaars-Schneider model.

One can produce a formally unitary integral kernel on the auxiliary Hilbert space implementing $S$ as superconformal index of a certain $T[U(N)]$ 
theory. The construction is actually best understood in a recursive way, in terms of S-dual interfaces between $U(N)$ and $U(N-1)$ gauge theories.

One interface simply reduce the gauge group from $U(N)$ to $U(N-1)$ by a partial Dirichlet boundary condition. Concretely, that means a 
$U(N)$ Wilson line, say, brought to the interface is decomposed into $U(N-1) \otimes U(1)$ Wilson lines and the latter are evaluated on 
a fixed value of the $v_N$ fugacity. The Schur index in the presence of the interface takes the form of a pairing in $\cH^{\mathrm{aux}}_\fq[N-1]$ 
of a Dirichlet wavefunction for $U(N-1)$ and the restriction of a Dirichlet wavefunction for $U(N)$ in $\cH^{\mathrm{aux}}_\fq[N]$ to fixed values of $\zeta_N$ and $B_N$. 

The S-dual interface couples both $U(N)$ and $U(N-1)$ gauge fields to two sets of ``bifundamental'' 3d free chiral fields. Concretely, the 
Schur index in the presence of the interface takes the form of a pairing with a (distributional) kernel in $\cH^{\mathrm{aux}}_\fq[N-1] \times \cH^{\mathrm{aux}}_\fq[N]$
which is a product of $2N(N-1)$ complex quantum dilogarithms. Elementary 't Hooft operators for $U(N)$ acting on the kernel can be traded for the
a linear combination of 't Hooft operators for $U(N-1)$ which is analogous to the decomposition of Wilson lines. 

A convolution of $N-1$ such kernels fully diagonalized 't Hooft operators. We leave details of the construction to an enthusiastic reader, referring to \cite{Gaiotto:2013bwa} for a classical version of the construction.  

\subsection{$U(N)$ SQCD with $N_f$ flavours.}
This is our final general example. 
The Schur index becomes 
\begin{align}
	I_\fq &= \frac{1}{N!} \oint_{|\zeta|=1} \prod_i \frac{(\fq^2)^2_\infty d\zeta_i}{2 \pi \i  \zeta_i} \frac{\prod_{i\neq j}(1-\zeta_i \zeta_j^{-1})(\fq^2 \zeta_i \zeta_j^{-1};\fq^2)^2_\infty}{\prod_{i}\prod_{r=1}^{N_f} (-\fq \mu_r \zeta_i;\fq^2)_\infty {(-\fq \mu_r^{-1} \zeta_i^{-1};\fq^2)_\infty}}
\end{align}
The 't Hooft operators of ``minuscule'' charge do not suffer from bubbling. We will present them in a form adapted to an isometry to $L^2((S^1 \times \bZ)^N/S_N)$ 
with spherical vector image
\begin{align}
	I\!\!I_B(\zeta) &= \delta_{B,0}(\fq^2)^N_\infty\frac{\prod_{i\neq j}(\fq^2 \zeta_i \zeta_j^{-1};\fq^2)_\infty}{\prod_i \prod_{r=1}^{N_f} (-\fq \mu_r \zeta_i;\fq^2)_\infty}
\end{align}
Define $N$ copies $(u_i,v_i)$ and $(\wt u_i, \wt v_i)$ of the standard set of multiplication and shift operators. An important ingredient in the presentation of 't Hooft operators of minimal charge are the combinations 
\begin{align}
	u_{+,i} &\equiv \frac{\fq^{N-1}}{\prod_j^{j \neq i}(1-v_j v_i^{-1})}u_i \cr
	u_{-,i} &\equiv \frac{\prod_{r=1}^{N_f}(1+\fq^{-1} \mu_r v_i)}{\prod_j^{j \neq i}(v_i v_j^{-1}-1)}u^{-1}_i \cr
	\wt u_{+,i} &\equiv \frac{\fq^{N-1} \prod_{r=1}^{N_f}(1+\fq^{-1} \mu_r \wt v_i)}{\prod_j^{j \neq i}(1-\wt v_i \wt v_j^{-1})}\wt u_i \cr
	\wt u_{-,i} &\equiv \frac{1}{ \prod_j^{j \neq i}(\wt v_j \wt v_i^{-1}-1)}\wt u^{-1}_i
\end{align}
Taking adjoints, we get
\begin{align}
	\rho(u_{-,i}) &= \prod_j^{j\neq i} (\fq^{-1} v_j v_i^{-1}) u_{+,i} \cr
	\rho(u_{+,i}) &= u_{-,i} \prod_j^{j\neq i} (\fq v_i v_j^{-1})\prod_{r=1}^{N_f}(\fq^{-1} \mu^{-1}_r v^{-1}_i)
\end{align}
and thus the expected Witten effect:
\begin{align}
	\rho^2(u_{-,i}) &=\prod_j^{j\neq i} (v_i v_j^{-1}) u_{-,i} \prod_j^{j\neq i} (v_i v_j^{-1}) \prod_{r=1}^{N_f}(\fq^{-1} \mu^{-1}_r v^{-1}_i)\cr
	\rho^2(u_{+,i}) &= \prod_j^{j\neq i} (v_j v_i^{-1}) u_{+,i} \prod_j^{j \neq i} (v_j v_i^{-1}) \prod_{r=1}^{N_f}(\fq^{-1} \mu_r v_i)
\end{align}
controlled by the anomaly $2 N- N_f$.

Again, 't Hooft operators of minuscule charge are built from the $u_{\pm,i}$ and analogous $u_{\pm,I}$. 

For $N_f = 2 N$, the theory is conformal and is endowed with a non-trivial S-duality which re-arranges the $U(1)$ and $SU(N)$ parts of the 't Hooft charges and maps Wilson lines to dyonic lines of even magnetic charge.  

\section{From $U(2)$ to $SU(2)$ 't Hooft operators}
First, we can illustrate the construction of pure $U(2)$ 't Hooft operators from $SU(2)$ and $U(1)$ expressions. We introduce symbols $v_1$ and $v_2$, as well as $u_{1,\pm}$ and $u_{2,\pm}$ which multiplicatively shift $v_1$ and $v_2$ by $\fq^2$, as for two copies of $U(1)$ gauge theory. We should think about $u_{i,\pm}$ as combinations of $U(1)$ and $SU(2)$ generators, so that 
\begin{align}
	u_{1,+} &= u^{SU(2)}_+ u^{U(1)}_+ \cr
	u_{1,-} &= u^{SU(2)}_- u^{U(1)}_- \cr
	u_{2,+} &= u^{SU(2)}_- u^{U(1)}_+ \cr
	u_{2,-} &= u^{SU(2)}_+ u^{U(1)}_-
\end{align}
and correspondingly $v_1 = v_{SU(2)} v_{U(1)}^{\frac12}$ and $v_2 = v_{SU(2)}^{-1} v_{U(1)}^{\frac12}$, so that $v_{U(1)} = v_1 v_2$.

Correspondingly, we have product rules such as 
\begin{align}
	u_{1,+} u_{1,-} &= \frac{v_1 v_2}{(v_1 - v_2)(\fq v_1 - \fq^{-1} v_2)} \cr
	u_{1,+} u_{2,+} &= \frac{v_1 v_2}{(v_1 - v_2)(\fq v_1 - \fq^{-1} v_2)} (u^{U(1)}_+)^2 \cr
	u_{1,+} u_{2,-} &=  u_{2,-} u_{1,+} 
\end{align}
etcetera.

The elementary 't Hooft operators in the $SU(2)$ gauge theory can be promoted to 
\begin{align}
	&\fq^{\frac{a}{2}} \fq^b v_{U(1)}^b u^{U(1)}_+ v_{SU(2)}^a u^{SU(2)}_+ +  \fq^{\frac{a}{2}} \fq^b v_{U(1)}^b u^{U(1)}_+ v_{SU(2)}^{-a} u^{SU(2)}_- \cr
	& \fq^{\frac{a}{2}} \fq^{-b} v_{U(1)}^b u^{U(1)}_- v_{SU(2)}^a u^{SU(2)}_+ +  \fq^{\frac{a}{2}}\fq^{-b} v_{U(1)}^b u^{U(1)}_- v_{SU(2)}^{-a} u^{SU(2)}_- 
\end{align} 
i.e. 
\begin{align}
	H_{1,0;a,b} = H_{0,1;b,a}&= \fq^{a} v_1^a v_2^b u_{1,+}  +  \fq^{a} v_2^a v_1^b  u_{2,+}  \cr
	H_{-1,0;a,b}=  H_{0,-1;b,a}  &= \fq^{-a} v_1^a v_2^b u_{1,-} + \fq^{-a}  v_2^a v_1^b  u_{2,-}
\end{align} 

Next, we can present the difference operators for $U(2)$ with a single flavour. The fundamental flavour modifies the product rules as 
\begin{align}
	u_{1,+} u_{1,-} &= \frac{v_1 v_2}{(v_1 - v_2)(\fq v_1 - \fq^{-1} v_2)} (1 + \fq v_1) \cr
	u_{1,+} u_{2,+} &= \frac{v_1 v_2}{(v_1 - v_2)(\fq v_1 - \fq^{-1} v_2)} (u^{U(1)}_+)^2 \cr
	u_{1,+} u_{2,-} &=  u_{2,-} u_{1,+} 
\end{align}
etcetera. The expressions for the elementary 't Hooft operators is unchanged: 
\begin{align}
	H_{1,0;a,b} = H_{0,1;b,a}&= \fq^{a} v_1^a v_2^b u_{1,+}  +  \fq^{a} v_2^a v_1^b  u_{2,+}  \cr
	H_{-1,0;a,b} H_{0,-1;b,a}  &= \fq^{-a} v_1^a v_2^b u_{1,-} + \fq^{-a}  v_2^a v_1^b  u_{2,-}
\end{align}

We can now consider the product 
\begin{align}
	&H_{1,0;a,b} H_{-1,0;c,d} = (\fq^{a} v_1^a v_2^{b} u_{1,+}  +  \fq^{a} v_2^a v_1^b  u_{2,+})(\fq^{-c} v_1^c v_2^d u_{1,-} + \fq^{-c}  v_2^c v_1^d  u_{2,-}) \cr
	&= \fq^{a-c+2 d } v_2^{a+d} v_1^{b+c} u_{1,-} u_{2,+} +\fq^{a+c } v_1^{a+c} v_2^{b+d} u_{1,+} u_{1,-}  + \cr&+ \fq^{a+c} v_2^{a+c} v_1^{b+d}  u_{2,+} u_{2,-} + \fq^{a-c+2 d} v_1^{a+d} v_2^{b+c}  u_{1,+}  u_{2,-} \cr
	&= \fq^{a-c+2 d } v_2^{a+d} v_1^{b+c} (u^{SU(2)}_-)^2 +\fq^{a+c } v_1^{a+c} v_2^{b+d}  \frac{v_1 v_2}{(v_1 - v_2)(\fq v_1 - \fq^{-1} v_2)} (1 + \fq v_1) + \cr &+ \fq^{a+c} v_2^{a+c} v_1^{b+d}   \frac{v_1 v_2}{(v_1 - v_2)(\fq^{-1} v_1 - \fq v_2)} (1 + \fq v_2) + \fq^{a-c+2 d} v_1^{a+d} v_2^{b+c}  (u^{SU(2)}_+)^2
\end{align}
In particular, we get 
\begin{align}
	& H_{1,0;a,b} H_{-1,0;-a,-b} = \fq^{2a - 2b} v_{SU(2)}^{2a-2b} (u^{SU(2)}_+)^2 + \frac{v_1 v_2}{(v_1 - v_2)(\fq v_1 - \fq^{-1} v_2)} (1 + \fq v_1) + \cr &+ \frac{v_1 v_2}{(v_1 - v_2)(\fq^{-1} v_1 - \fq v_2)} (1 + \fq v_2) +  \fq^{2a-2b } v_{SU(2)}^{-2a+2b}(u^{SU(2)}_-)^2 \cr
	&= \fq^{2a - 2b} v_{SU(2)}^{2a-2b} (u^{SU(2)}_+)^2 + \frac{(\fq+ \fq^{-1} + v_1 + v_2)}{(\fq v_{SU(2)} - \fq^{-1} v_{SU(2)}^{-1})(\fq^{-1} v_{SU(2)} - \fq v_{SU(2)}^{-1})}  + \cr &+ \fq^{2a-2b } v_{SU(2)}^{-2a+2b}(u^{SU(2)}_-)^2 
\end{align}
and 
\begin{align}
	& H_{1,0;a,b} H_{-1,0;-a,1-b} =  \fq^{2a-2b +2} v_1^{a-b+1} v_2^{b-a}  (u^{SU(2)}_+)^2 +  v_2  \frac{v_1 v_2}{(v_1 - v_2)(\fq v_1 - \fq^{-1} v_2)} (1 + \fq v_1) + \cr &+ v_1   \frac{v_1 v_2}{(v_1 - v_2)(\fq^{-1} v_1 - \fq v_2)} (1 + \fq v_2) +\fq^{2a-2b+2 } v_2^{a-b+1} v_1^{b-a} (u^{SU(2)}_-)^2 \cr
	&=  \fq^{2a-2b +2} v_{SU(2)}^{2a-2b+1} v_{U(1)}^{\frac12} (u^{SU(2)}_+)^2 +\fq \frac{((\fq+ \fq^{-1} )v_1 v_2+ v_1 + v_2)}{(\fq v_{SU(2)} - \fq^{-1} v_{SU(2)}^{-1})(\fq^{-1} v_{SU(2)} - \fq v_{SU(2)}^{-1})}+\cr &+\fq^{2a-2b+2 } v_{SU(2)}^{2b-2a-1} v_{U(1)}^{\frac12} (u^{SU(2)}_-)^2 
\end{align}

\bibliographystyle{JHEP}

\bibliography{mono}

\providecommand{\href}[2]{#2}\begingroup\raggedright\begin{thebibliography}{100}

\bibitem{Kinney:2005ej}
J.~Kinney, J.~M. Maldacena, S.~Minwalla, and S.~Raju, {\it {An Index for 4
  dimensional super conformal theories}},  {\em Commun. Math. Phys.} {\bf 275}
  (2007) 209--254, [\href{http://arxiv.org/abs/hep-th/0510251}{{\tt
  hep-th/0510251}}].

\bibitem{Gadde:2011ik}
A.~Gadde, L.~Rastelli, S.~S. Razamat, and W.~Yan, {\it {The 4d Superconformal
  Index from q-deformed 2d Yang-Mills}},  {\em Phys. Rev. Lett.} {\bf 106}
  (2011) 241602, [\href{http://arxiv.org/abs/1104.3850}{{\tt
  arXiv:1104.3850}}].

\bibitem{Gadde:2011uv}
A.~Gadde, L.~Rastelli, S.~S. Razamat, and W.~Yan, {\it {Gauge Theories and
  Macdonald Polynomials}},  {\em Commun. Math. Phys.} {\bf 319} (2013)
  147--193, [\href{http://arxiv.org/abs/1110.3740}{{\tt arXiv:1110.3740}}].

\bibitem{Dimofte:2011py}
T.~Dimofte, D.~Gaiotto, and S.~Gukov, {\it {3-Manifolds and 3d Indices}},  {\em
  Adv. Theor. Math. Phys.} {\bf 17} (2013), no.~5 975--1076,
  [\href{http://arxiv.org/abs/1112.5179}{{\tt arXiv:1112.5179}}].

\bibitem{Fukuda:2012jr}
Y.~Fukuda, T.~Kawano, and N.~Matsumiya, {\it {5D SYM and 2D q-Deformed YM}},
  {\em Nucl. Phys. B} {\bf 869} (2013) 493--522,
  [\href{http://arxiv.org/abs/1210.2855}{{\tt arXiv:1210.2855}}].

\bibitem{Gang:2012yr}
D.~Gang, E.~Koh, and K.~Lee, {\it {Line Operator Index on $S^{1}\times
  S^{3}$}},  {\em JHEP} {\bf 05} (2012) 007,
  [\href{http://arxiv.org/abs/1201.5539}{{\tt arXiv:1201.5539}}].

\bibitem{Beem:2013sza}
C.~Beem, M.~Lemos, P.~Liendo, W.~Peelaers, L.~Rastelli, and B.~C. van Rees,
  {\it {Infinite Chiral Symmetry in Four Dimensions}},  {\em Commun. Math.
  Phys.} {\bf 336} (2015), no.~3 1359--1433,
  [\href{http://arxiv.org/abs/1312.5344}{{\tt arXiv:1312.5344}}].

\bibitem{Lemos:2014lua}
M.~Lemos and W.~Peelaers, {\it {Chiral Algebras for Trinion Theories}},  {\em
  JHEP} {\bf 02} (2015) 113, [\href{http://arxiv.org/abs/1411.3252}{{\tt
  arXiv:1411.3252}}].

\bibitem{Beem:2014rza}
C.~Beem, W.~Peelaers, L.~Rastelli, and B.~C. van Rees, {\it {Chiral algebras of
  class S}},  {\em JHEP} {\bf 05} (2015) 020,
  [\href{http://arxiv.org/abs/1408.6522}{{\tt arXiv:1408.6522}}].

\bibitem{Drukker:2015spa}
N.~Drukker, {\it {The $ \mathcal{N}=4 $ Schur index with Polyakov loops}},
  {\em JHEP} {\bf 12} (2015) 012, [\href{http://arxiv.org/abs/1510.02480}{{\tt
  arXiv:1510.02480}}].

\bibitem{Tachikawa:2015iba}
Y.~Tachikawa and N.~Watanabe, {\it {On skein relations in class S theories}},
  {\em JHEP} {\bf 06} (2015) 186, [\href{http://arxiv.org/abs/1504.00121}{{\tt
  arXiv:1504.00121}}].

\bibitem{Cecotti:2015lab}
S.~Cecotti, J.~Song, C.~Vafa, and W.~Yan, {\it {Superconformal Index, BPS
  Monodromy and Chiral Algebras}},  {\em JHEP} {\bf 11} (2017) 013,
  [\href{http://arxiv.org/abs/1511.01516}{{\tt arXiv:1511.01516}}].

\bibitem{Cordova:2015nma}
C.~Cordova and S.-H. Shao, {\it {Schur Indices, BPS Particles, and
  Argyres-Douglas Theories}},  {\em JHEP} {\bf 01} (2016) 040,
  [\href{http://arxiv.org/abs/1506.00265}{{\tt arXiv:1506.00265}}].

\bibitem{Cordova:2016uwk}
C.~Cordova, D.~Gaiotto, and S.-H. Shao, {\it {Infrared Computations of Defect
  Schur Indices}},  {\em JHEP} {\bf 11} (2016) 106,
  [\href{http://arxiv.org/abs/1606.08429}{{\tt arXiv:1606.08429}}].

\bibitem{Watanabe:2016bwr}
N.~Watanabe, {\it {Wilson punctured network defects in 2D q-deformed Yang-Mills
  theory}},  {\em JHEP} {\bf 12} (2016) 063,
  [\href{http://arxiv.org/abs/1603.02939}{{\tt arXiv:1603.02939}}].

\bibitem{Watanabe:2017bmi}
N.~Watanabe, {\it {Schur indices with class S line operators from networks and
  further skein relations}},  \href{http://arxiv.org/abs/1701.04090}{{\tt
  arXiv:1701.04090}}.

\bibitem{Cordova:2017ohl}
C.~Cordova, D.~Gaiotto, and S.-H. Shao, {\it {Surface Defect Indices and 2d-4d
  BPS States}},  {\em JHEP} {\bf 12} (2017) 078,
  [\href{http://arxiv.org/abs/1703.02525}{{\tt arXiv:1703.02525}}].

\bibitem{Neitzke:2017cxz}
A.~Neitzke and F.~Yan, {\it {Line defect Schur indices, Verlinde algebras and
  $U(1)_r$ fixed points}},  {\em JHEP} {\bf 11} (2017) 035,
  [\href{http://arxiv.org/abs/1708.05323}{{\tt arXiv:1708.05323}}].

\bibitem{Fluder:2019dpf}
M.~Fluder and P.~Longhi, {\it {An infrared bootstrap of the Schur index with
  surface defects}},  {\em JHEP} {\bf 09} (2019) 062,
  [\href{http://arxiv.org/abs/1905.02724}{{\tt arXiv:1905.02724}}].

\bibitem{Pan:2021mrw}
Y.~Pan and W.~Peelaers, {\it {Exact Schur index in closed form}},  {\em Phys.
  Rev. D} {\bf 106} (2022), no.~4 045017,
  [\href{http://arxiv.org/abs/2112.09705}{{\tt arXiv:2112.09705}}].

\bibitem{Hatsuda:2023iwi}
Y.~Hatsuda and T.~Okazaki, {\it {Exact $ \mathcal{N} $ = 2$^{*}$ Schur line
  defect correlators}},  {\em JHEP} {\bf 06} (2023) 169,
  [\href{http://arxiv.org/abs/2303.14887}{{\tt arXiv:2303.14887}}].

\bibitem{Gaiotto:2023hda}
D.~Gaiotto, {\it {Sphere quantization of Higgs and Coulomb branches and
  Analytic Symplectic Duality}},  \href{http://arxiv.org/abs/2307.12396}{{\tt
  arXiv:2307.12396}}.

\bibitem{Kapustin:2006hi}
A.~Kapustin, {\it {Holomorphic reduction of N=2 gauge theories, Wilson-'t Hooft
  operators, and S-duality}},  \href{http://arxiv.org/abs/hep-th/0612119}{{\tt
  hep-th/0612119}}.

\bibitem{Kapustin:2007wm}
A.~Kapustin and N.~Saulina, {\it {The Algebra of Wilson-'t Hooft operators}},
  {\em Nucl. Phys. B} {\bf 814} (2009) 327--365,
  [\href{http://arxiv.org/abs/0710.2097}{{\tt arXiv:0710.2097}}].

\bibitem{Pestun:2007rz}
V.~Pestun, {\it {Localization of gauge theory on a four-sphere and
  supersymmetric Wilson loops}},  {\em Commun. Math. Phys.} {\bf 313} (2012)
  71--129, [\href{http://arxiv.org/abs/0712.2824}{{\tt arXiv:0712.2824}}].

\bibitem{Drukker:2009tz}
N.~Drukker, D.~R. Morrison, and T.~Okuda, {\it {Loop operators and S-duality
  from curves on Riemann surfaces}},  {\em JHEP} {\bf 09} (2009) 031,
  [\href{http://arxiv.org/abs/0907.2593}{{\tt arXiv:0907.2593}}].

\bibitem{Alday:2009fs}
L.~F. Alday, D.~Gaiotto, S.~Gukov, Y.~Tachikawa, and H.~Verlinde, {\it {Loop
  and surface operators in N=2 gauge theory and Liouville modular geometry}},
  {\em JHEP} {\bf 01} (2010) 113, [\href{http://arxiv.org/abs/0909.0945}{{\tt
  arXiv:0909.0945}}].

\bibitem{Drukker:2009id}
N.~Drukker, J.~Gomis, T.~Okuda, and J.~Teschner, {\it {Gauge Theory Loop
  Operators and Liouville Theory}},  {\em JHEP} {\bf 02} (2010) 057,
  [\href{http://arxiv.org/abs/0909.1105}{{\tt arXiv:0909.1105}}].

\bibitem{Gaiotto:2010be}
D.~Gaiotto, G.~W. Moore, and A.~Neitzke, {\it {Framed BPS States}},  {\em Adv.
  Theor. Math. Phys.} {\bf 17} (2013), no.~2 241--397,
  [\href{http://arxiv.org/abs/1006.0146}{{\tt arXiv:1006.0146}}].

\bibitem{Gomis:2011pf}
J.~Gomis, T.~Okuda, and V.~Pestun, {\it {Exact Results for 't Hooft Loops in
  Gauge Theories on $S^4$}},  {\em JHEP} {\bf 05} (2012) 141,
  [\href{http://arxiv.org/abs/1105.2568}{{\tt arXiv:1105.2568}}].

\bibitem{Ito:2011ea}
Y.~Ito, T.~Okuda, and M.~Taki, {\it {Line operators on $S^1 \times \mathbb
  {R}^3$ and quantization of the Hitchin moduli space}},  {\em JHEP} {\bf 04}
  (2012) 010, [\href{http://arxiv.org/abs/1111.4221}{{\tt arXiv:1111.4221}}].
  [Erratum: JHEP 03, 085 (2016)].

\bibitem{Saulina:2011qr}
N.~Saulina, {\it {A note on Wilson-'t Hooft operators}},  {\em Nucl. Phys. B}
  {\bf 857} (2012) 153--171, [\href{http://arxiv.org/abs/1110.3354}{{\tt
  arXiv:1110.3354}}].

\bibitem{2015arXiv150303676N}
H.~{Nakajima}, {\it {Towards a mathematical definition of Coulomb branches of
  $3$-dimensional $\mathcal N=4$ gauge theories, I}},  {\em arXiv e-prints}
  (Mar., 2015) arXiv:1503.03676, [\href{http://arxiv.org/abs/1503.03676}{{\tt
  arXiv:1503.03676}}].

\bibitem{Braverman:2016wma}
A.~Braverman, M.~Finkelberg, and H.~Nakajima, {\it {Towards a mathematical
  definition of Coulomb branches of $3$-dimensional $\mathcal{N} = 4$ gauge
  theories, II}},  {\em Adv. Theor. Math. Phys.} {\bf 22} (2018) 1071--1147,
  [\href{http://arxiv.org/abs/1601.03586}{{\tt arXiv:1601.03586}}].

\bibitem{finkelberg2019multiplicative}
M.~Finkelberg and A.~Tsymbaliuk, {\it Multiplicative slices, relativistic toda
  and shifted quantum affine algebras},  {\em Representations and Nilpotent
  Orbits of Lie Algebraic Systems: In Honour of the 75th Birthday of Tony
  Joseph} (2019) 133--304.

\bibitem{schrader2019k}
G.~Schrader and A.~Shapiro, {\it $ k $-theoretic coulomb branches of quiver
  gauge theories and cluster varieties},  {\em arXiv preprint arXiv:1910.03186}
  (2019).

\bibitem{Drukker:2010jp}
N.~Drukker, D.~Gaiotto, and J.~Gomis, {\it {The Virtue of Defects in 4D Gauge
  Theories and 2D CFTs}},  {\em JHEP} {\bf 06} (2011) 025,
  [\href{http://arxiv.org/abs/1003.1112}{{\tt arXiv:1003.1112}}].

\bibitem{Dedushenko:2018tgx}
M.~Dedushenko, {\it {Gluing II: boundary localization and gluing formulas}},
  {\em Lett. Math. Phys.} {\bf 111} (2021), no.~1 18,
  [\href{http://arxiv.org/abs/1807.04278}{{\tt arXiv:1807.04278}}].

\bibitem{Witten:1997sc}
E.~Witten, {\it {Solutions of four-dimensional field theories via M theory}},
  {\em Nucl. Phys. B} {\bf 500} (1997) 3--42,
  [\href{http://arxiv.org/abs/hep-th/9703166}{{\tt hep-th/9703166}}].

\bibitem{Gaiotto:2009hg}
D.~Gaiotto, G.~W. Moore, and A.~Neitzke, {\it {Wall-crossing, Hitchin systems,
  and the WKB approximation}},  {\em Adv. Math.} {\bf 234} (2013) 239--403,
  [\href{http://arxiv.org/abs/0907.3987}{{\tt arXiv:0907.3987}}].

\bibitem{Gaiotto:2009we}
D.~Gaiotto, {\it {N=2 dualities}},  {\em JHEP} {\bf 08} (2012) 034,
  [\href{http://arxiv.org/abs/0904.2715}{{\tt arXiv:0904.2715}}].

\bibitem{Chacaltana:2012zy}
O.~Chacaltana, J.~Distler, and Y.~Tachikawa, {\it {Nilpotent orbits and
  codimension-two defects of 6d N=(2,0) theories}},  {\em Int. J. Mod. Phys. A}
  {\bf 28} (2013) 1340006, [\href{http://arxiv.org/abs/1203.2930}{{\tt
  arXiv:1203.2930}}].

\bibitem{Kapustin:2006pk}
A.~Kapustin and E.~Witten, {\it {Electric-Magnetic Duality And The Geometric
  Langlands Program}},  {\em Commun. Num. Theor. Phys.} {\bf 1} (2007) 1--236,
  [\href{http://arxiv.org/abs/hep-th/0604151}{{\tt hep-th/0604151}}].

\bibitem{Witten:2010cx}
E.~Witten, {\it {Analytic Continuation Of Chern-Simons Theory}},  {\em AMS/IP
  Stud. Adv. Math.} {\bf 50} (2011) 347--446,
  [\href{http://arxiv.org/abs/1001.2933}{{\tt arXiv:1001.2933}}].

\bibitem{Gaiotto:2021tsq}
D.~Gaiotto and E.~Witten, {\it {Gauge Theory and the Analytic Form of the
  Geometric Langlands Program}},  \href{http://arxiv.org/abs/2107.01732}{{\tt
  arXiv:2107.01732}}.

\bibitem{Gaiotto:2024tpl}
D.~Gaiotto and J.~Teschner, {\it {Quantum Analytic Langlands Correspondence}},
  \href{http://arxiv.org/abs/2402.00494}{{\tt arXiv:2402.00494}}.

\bibitem{Turaev:1992hq}
V.~G. Turaev and O.~Y. Viro, {\it {State sum invariants of 3 manifolds and
  quantum 6j symbols}},  {\em Topology} {\bf 31} (1992) 865--902.

\bibitem{Levin:2004mi}
M.~A. Levin and X.-G. Wen, {\it {String net condensation: A Physical mechanism
  for topological phases}},  {\em Phys. Rev. B} {\bf 71} (2005) 045110,
  [\href{http://arxiv.org/abs/cond-mat/0404617}{{\tt cond-mat/0404617}}].

\bibitem{Ponzano:461451}
G.~Ponzano and T.~E. Regge, {\it {Semiclassical limit of Racah coefficients}},
  .

\bibitem{Buffenoir:2002tx}
E.~Buffenoir, K.~Noui, and P.~Roche, {\it {Hamiltonian quantization of
  Chern-Simons theory with SL(2,C) group}},  {\em Class. Quant. Grav.} {\bf 19}
  (2002) 4953, [\href{http://arxiv.org/abs/hep-th/0202121}{{\tt
  hep-th/0202121}}].

\bibitem{Bonzom:2014bua}
V.~Bonzom, M.~Dupuis, and F.~Girelli, {\it {Towards the Turaev-Viro amplitudes
  from a Hamiltonian constraint}},  {\em Phys. Rev. D} {\bf 90} (2014), no.~10
  104038, [\href{http://arxiv.org/abs/1403.7121}{{\tt arXiv:1403.7121}}].

\bibitem{next}
G.~Gaiotto and J.~Teschner, {\it Schur quantization in the ir},  {\em To
  Appear}.

\bibitem{Witten:1989ip}
E.~Witten, {\it {Quantization of {Chern-Simons} Gauge Theory With Complex Gauge
  Group}},  {\em Commun. Math. Phys.} {\bf 137} (1991) 29--66.

\bibitem{Dimofte:2016pua}
T.~Dimofte, {\it {Perturbative and nonperturbative aspects of complex
  Chern\textendash{}Simons theory}},  {\em J. Phys. A} {\bf 50} (2017), no.~44
  443009, [\href{http://arxiv.org/abs/1608.02961}{{\tt arXiv:1608.02961}}].

\bibitem{Kapustin:2005py}
A.~Kapustin, {\it {Wilson-'t Hooft operators in four-dimensional gauge theories
  and S-duality}},  {\em Phys. Rev. D} {\bf 74} (2006) 025005,
  [\href{http://arxiv.org/abs/hep-th/0501015}{{\tt hep-th/0501015}}].

\bibitem{cautis2023canonical}
S.~Cautis and H.~Williams, {\it Canonical bases for coulomb branches of 4d
  $\mathcal{N}=2$ gauge theories},  2023.

\bibitem{Klyuev_2022}
D.~Klyuev, {\it Twisted traces and positive forms on generalized q-weyl
  algebras},  {\em Symmetry, Integrability and Geometry: Methods and
  Applications} (Jan., 2022).

\bibitem{Seiberg:1996nz}
N.~Seiberg and E.~Witten, {\it {Gauge dynamics and compactification to
  three-dimensions}},  in {\em {Conference on the Mathematical Beauty of
  Physics (In Memory of C. Itzykson)}}, pp.~333--366, 6, 1996.
\newblock \href{http://arxiv.org/abs/hep-th/9607163}{{\tt hep-th/9607163}}.

\bibitem{Gaiotto:2008ak}
D.~Gaiotto and E.~Witten, {\it {S-Duality of Boundary Conditions In N=4 Super
  Yang-Mills Theory}},  {\em Adv. Theor. Math. Phys.} {\bf 13} (2009), no.~3
  721--896, [\href{http://arxiv.org/abs/0807.3720}{{\tt arXiv:0807.3720}}].

\bibitem{Dimofte:2013lba}
T.~Dimofte, D.~Gaiotto, and R.~van~der Veen, {\it {RG Domain Walls and Hybrid
  Triangulations}},  {\em Adv. Theor. Math. Phys.} {\bf 19} (2015) 137--276,
  [\href{http://arxiv.org/abs/1304.6721}{{\tt arXiv:1304.6721}}].

\bibitem{Gukov:2008ve}
S.~Gukov and E.~Witten, {\it {Branes and Quantization}},  {\em Adv. Theor.
  Math. Phys.} {\bf 13} (2009), no.~5 1445--1518,
  [\href{http://arxiv.org/abs/0809.0305}{{\tt arXiv:0809.0305}}].

\bibitem{Kapustin:2001ij}
A.~Kapustin and D.~Orlov, {\it {Remarks on A branes, mirror symmetry, and the
  Fukaya category}},  {\em J. Geom. Phys.} {\bf 48} (2003) 84,
  [\href{http://arxiv.org/abs/hep-th/0109098}{{\tt hep-th/0109098}}].

\bibitem{Bressler:2002eu}
P.~Bressler and Y.~Soibelman, {\it {Mirror symmetry and deformation
  quantization}},  \href{http://arxiv.org/abs/hep-th/0202128}{{\tt
  hep-th/0202128}}.

\bibitem{Kapustin:2005vs}
A.~Kapustin, {\it {A-branes and noncommutative geometry}},
  \href{http://arxiv.org/abs/hep-th/0502212}{{\tt hep-th/0502212}}.

\bibitem{Pestun:2006rj}
V.~Pestun, {\it {Topological strings in generalized complex space}},  {\em Adv.
  Theor. Math. Phys.} {\bf 11} (2007), no.~3 399--450,
  [\href{http://arxiv.org/abs/hep-th/0603145}{{\tt hep-th/0603145}}].

\bibitem{Gualtieri:2007bq}
M.~Gualtieri, {\em {Branes on Poisson varieties}}.
\newblock 10, 2007.
\newblock \href{http://arxiv.org/abs/0710.2719}{{\tt arXiv:0710.2719}}.

\bibitem{Aldi:2005hz}
M.~Aldi and E.~Zaslow, {\it {Coisotropic branes, noncommutativity, and the
  mirror correspondence}},  {\em JHEP} {\bf 06} (2005) 019,
  [\href{http://arxiv.org/abs/hep-th/0501247}{{\tt hep-th/0501247}}].

\bibitem{Witten:2009at}
E.~Witten, {\it {Geometric Langlands From Six Dimensions}},
  \href{http://arxiv.org/abs/0905.2720}{{\tt arXiv:0905.2720}}.

\bibitem{Mikhaylov:2017ngi}
V.~Mikhaylov, {\it {Teichm\"uller TQFT vs. Chern-Simons theory}},  {\em JHEP}
  {\bf 04} (2018) 085, [\href{http://arxiv.org/abs/1710.04354}{{\tt
  arXiv:1710.04354}}].

\bibitem{Cordova:2013cea}
C.~Cordova and D.~L. Jafferis, {\it {Complex Chern-Simons from M5-branes on the
  Squashed Three-Sphere}},  {\em JHEP} {\bf 11} (2017) 119,
  [\href{http://arxiv.org/abs/1305.2891}{{\tt arXiv:1305.2891}}].

\bibitem{jordan2023quantum}
D.~Jordan, {\it Quantum character varieties},  2023.

\bibitem{Gaiotto:2012rg}
D.~Gaiotto, G.~W. Moore, and A.~Neitzke, {\it {Spectral networks}},  {\em
  Annales Henri Poincare} {\bf 14} (2013) 1643--1731,
  [\href{http://arxiv.org/abs/1204.4824}{{\tt arXiv:1204.4824}}].

\bibitem{Dimofte:2014ria}
T.~Dimofte and R.~van~der Veen, {\it {A Spectral Perspective on
  Neumann-Zagier}},  \href{http://arxiv.org/abs/1403.5215}{{\tt
  arXiv:1403.5215}}.

\bibitem{Neitzke:2020jik}
A.~Neitzke and F.~Yan, {\it {$q$-nonabelianization for line defects}},  {\em
  JHEP} {\bf 09} (2020) 153, [\href{http://arxiv.org/abs/2002.08382}{{\tt
  arXiv:2002.08382}}].

\bibitem{Freed:2022yae}
D.~S. Freed and A.~Neitzke, {\it {3d spectral networks and classical
  Chern-Simons theory}},  \href{http://arxiv.org/abs/2208.07420}{{\tt
  arXiv:2208.07420}}.

\bibitem{Reshetikhin:1990pr}
N.~Y. Reshetikhin and V.~G. Turaev, {\it {Ribbon graphs and their invariants
  derived from quantum groups}},  {\em Commun. Math. Phys.} {\bf 127} (1990)
  1--26.

\bibitem{Alekseev:1994pa}
A.~Y. Alekseev, H.~Grosse, and V.~Schomerus, {\it {Combinatorial quantization
  of the Hamiltonian Chern-Simons theory}},  {\em Commun. Math. Phys.} {\bf
  172} (1995) 317--358, [\href{http://arxiv.org/abs/hep-th/9403066}{{\tt
  hep-th/9403066}}].

\bibitem{Alekseev:1994au}
A.~Y. Alekseev, H.~Grosse, and V.~Schomerus, {\it {Combinatorial quantization
  of the Hamiltonian Chern-Simons theory. 2.}},  {\em Commun. Math. Phys.} {\bf
  174} (1995) 561--604, [\href{http://arxiv.org/abs/hep-th/9408097}{{\tt
  hep-th/9408097}}].

\bibitem{kashaev2001spectrum}
R.~Kashaev, {\it On the spectrum of dehn twists in quantum teichm{\"u}ller
  theory},  in {\em Physics and combinatorics}, pp.~63--81.
\newblock World Scientific, 2001.

\bibitem{Nidaiev:2013bda}
I.~Nidaiev and J.~Teschner, {\it {On the relation between the modular double of
  $U_q(sl(2,R))$ and the quantum Teichmueller theory}},
  \href{http://arxiv.org/abs/1302.3454}{{\tt arXiv:1302.3454}}.

\bibitem{Teschner:2013tqy}
J.~Teschner and G.~S. Vartanov, {\it {Supersymmetric gauge theories,
  quantization of $\mathcal{M}_{\mathrm{flat}}$, and conformal field theory}},
  {\em Adv. Theor. Math. Phys.} {\bf 19} (2015) 1--135,
  [\href{http://arxiv.org/abs/1302.3778}{{\tt arXiv:1302.3778}}].

\bibitem{schrader2019cluster}
G.~Schrader and A.~Shapiro, {\it A cluster realization of $\mathcal{U}_q
  (\mathfrak{sl}_ n)$ from quantum character varieties},  {\em Inventiones
  mathematicae} {\bf 216} (2019) 799--846.

\bibitem{schrader2017continuous}
G.~Schrader and A.~Shapiro, {\it Continuous tensor categories from quantum
  groups i: algebraic aspects},  {\em arXiv preprint arXiv:1708.08107} (2017).

\bibitem{kinnear2024nonsemisimple}
P.~Kinnear, {\it Non-semisimple crane-yetter theory varying over the character
  stack},  2024.

\bibitem{PodWoron}
P.~Podles and S.~L. Woronowicz, {\it Quantum deformation of {L}orentz group},
  {\em Comm. Math. Phys.} {\bf 130} (1990), no.~2 381--431.

\bibitem{Buffenoir:1997ih}
E.~Buffenoir and P.~Roche, {\it {Harmonic analysis on the quantum Lorentz
  group}},  {\em Commun. Math. Phys.} {\bf 207} (1999) 499--555,
  [\href{http://arxiv.org/abs/q-alg/9710022}{{\tt q-alg/9710022}}].

\bibitem{BraKazh}
A.~Braverman and D.~Kazhdan, {\it To appear}, .

\bibitem{Gawedzki:1988hq}
K.~Gawedzki and A.~Kupiainen, {\it {G/h Conformal Field Theory from Gauged WZW
  Model}},  {\em Phys. Lett. B} {\bf 215} (1988) 119--123.

\bibitem{Gawedzki:1991yu}
K.~Gawedzki, {\it {Noncompact WZW conformal field theories}},  in {\em {NATO
  Advanced Study Institute: New Symmetry Principles in Quantum Field Theory}},
  pp.~0247--274, 10, 1991.
\newblock \href{http://arxiv.org/abs/hep-th/9110076}{{\tt hep-th/9110076}}.

\bibitem{Etingof:2019guc}
P.~Etingof and D.~Stryker, {\it {Short Star-Products for Filtered
  Quantizations, I}},  {\em SIGMA} {\bf 16} (2020) 014,
  [\href{http://arxiv.org/abs/1909.13588}{{\tt arXiv:1909.13588}}].

\bibitem{Gaiotto:2023ezy}
D.~Gaiotto, G.~W. Moore, A.~Neitzke, and F.~Yan, {\it {Commuting Line Defects
  At $q^N=1$}},  \href{http://arxiv.org/abs/2307.14429}{{\tt
  arXiv:2307.14429}}.

\bibitem{niu2022local}
W.~Niu, {\it Local operators of 4d $\mathcal{N}=2$ gauge theories from the
  affine grassmannian},  2022.

\bibitem{cautis2018cluster}
S.~Cautis and H.~Williams, {\it Cluster theory of the coherent satake
  category},  2018.

\bibitem{Gaiotto:2014lma}
D.~Gaiotto, {\it {Open Verlinde line operators}},
  \href{http://arxiv.org/abs/1404.0332}{{\tt arXiv:1404.0332}}.

\bibitem{Allegretti:2024svn}
D.~G.~L. Allegretti and P.~Shan, {\it {Skein algebras and quantized Coulomb
  branches}},  \href{http://arxiv.org/abs/2401.06737}{{\tt arXiv:2401.06737}}.

\bibitem{2019}
{\em Representations and Nilpotent Orbits of Lie Algebraic Systems: In Honour
  of the 75th Birthday of Tony Joseph}.
\newblock Springer International Publishing, 2019.

\bibitem{schrader2019ktheoretic}
G.~Schrader and A.~Shapiro, {\it $k$-theoretic coulomb branches of quiver gauge
  theories and cluster varieties},  2019.

\bibitem{Braverman:2022zei}
A.~Braverman, G.~Dhillon, M.~Finkelberg, S.~Raskin, and R.~Travkin, {\it
  {Coulomb branches of noncotangent type (with appendices by Gurbir Dhillon and
  Theo Johnson-Freyd)}},  \href{http://arxiv.org/abs/2201.09475}{{\tt
  arXiv:2201.09475}}.

\bibitem{teleman2023coulomb}
C.~Teleman, {\it Coulomb branches for quaternionic representations},  2023.

\bibitem{Teschner:1997fv}
J.~Teschner, {\it {The Minisuperspace limit of the sl(2,C) / SU(2) WZNW
  model}},  {\em Nucl. Phys. B} {\bf 546} (1999) 369--389,
  [\href{http://arxiv.org/abs/hep-th/9712258}{{\tt hep-th/9712258}}].

\bibitem{Gaiotto:2023kuc}
D.~Gaiotto, J.~Hilburn, J.~Redondo-Yuste, B.~Webster, and Z.~Zhou, {\it
  {Twisted traces on abelian quantum Higgs and Coulomb branches}},
  \href{http://arxiv.org/abs/2308.15198}{{\tt arXiv:2308.15198}}.

\bibitem{Witten:1989rw}
E.~Witten, {\it {Gauge Theories, Vertex Models and Quantum Groups}},  {\em
  Nucl. Phys. B} {\bf 330} (1990) 285--346.

\bibitem{Gaiotto:2012sf}
D.~Gaiotto and J.~Teschner, {\it {Irregular singularities in Liouville theory
  and Argyres-Douglas type gauge theories, I}},  {\em JHEP} {\bf 12} (2012)
  050, [\href{http://arxiv.org/abs/1203.1052}{{\tt arXiv:1203.1052}}].

\bibitem{masuda1990unitary}
T.~Masuda, K.~Mimachi, Y.~Nakagami, M.~Noumi, Y.~Saburi, and K.~Ueno, {\it
  Unitary representations of the quantum group su q (1, 1): Structure of the
  dual space of u q (sl (2))},  {\em letters in mathematical physics} {\bf 19}
  (1990) 187--194.

\bibitem{WorZakr94}
S.~L. Woronowicz and S.~Zakrzewski, {\it Quantum deformations of the {L}orentz
  group. {T}he {H}opf {$^*$}-algebra level},  {\em Compositio Math.} {\bf 90}
  (1994), no.~2 211--243.

\bibitem{WorZakr92}
S.~L. Woronowicz and S.~Zakrzewski, {\it Quantum {L}orentz group having {G}auss
  decomposition property},  {\em Publ. Res. Inst. Math. Sci.} {\bf 28} (1992),
  no.~5 809--824.

\bibitem{Pusz}
W.~Pusz, {\it Irreducible unitary representations of quantum {L}orentz group},
  {\em Comm. Math. Phys.} {\bf 152} (1993), no.~3 591--626.

\bibitem{MR0274663}
W.~R\"{u}hl, {\em The {L}orentz group and harmonic analysis}.
\newblock W. A. Benjamin, Inc., New York, 1970.

\bibitem{Han:2024nkf}
M.~Han, {\it {Representations of a quantum-deformed Lorentz algebra,
  Clebsch-Gordan map, and Fenchel-Nielsen representation of quantum complex
  flat connections at level-$k$}},  \href{http://arxiv.org/abs/2402.08176}{{\tt
  arXiv:2402.08176}}.

\bibitem{2023arXiv230214734J}
D.~{Jordan}, {\it {Langlands duality for skein modules of 3-manifolds}},  {\em
  arXiv e-prints} (Feb., 2023) arXiv:2302.14734,
  [\href{http://arxiv.org/abs/2302.14734}{{\tt arXiv:2302.14734}}].

\bibitem{Collier:2023fwi}
S.~Collier, L.~Eberhardt, and M.~Zhang, {\it {Solving 3d gravity with Virasoro
  TQFT}},  {\em SciPost Phys.} {\bf 15} (2023), no.~4 151,
  [\href{http://arxiv.org/abs/2304.13650}{{\tt arXiv:2304.13650}}].

\bibitem{FaddeevMD}
L.~Faddeev, {\it Modular double of a quantum group},  in {\em Conf\'{e}rence
  {M}osh\'{e} {F}lato 1999, {V}ol. {I} ({D}ijon)}, vol.~21 of {\em Math. Phys.
  Stud.}, pp.~149--156.
\newblock Kluwer Acad. Publ., Dordrecht, 2000.

\bibitem{schraderCluster}
G.~Schrader and A.~Shapiro, {\it A cluster realization of $\mathcal{U}_q
  (\mathfrak{sl}_ n)$ from quantum character varieties},  {\em Inventiones
  mathematicae} {\bf 216} (2019) 799--846.

\bibitem{IpCluster}
I.~C.~H. Ip, {\it Cluster realization of $\mathcal{U}_q(\mathfrak{g})$ and
  factorizations of the universal $r$-matrix},  {\em Selecta Math. (N.S.)} {\bf
  24} (2018), no.~5 4461--4553.

\bibitem{2021arXiv210803453L}
I.~{Losev}, L.~{Mason-Brown}, and D.~{Matvieievskyi}, {\it {Unipotent Ideals
  and Harish-Chandra Bimodules}},  {\em arXiv e-prints} (Aug., 2021)
  arXiv:2108.03453, [\href{http://arxiv.org/abs/2108.03453}{{\tt
  arXiv:2108.03453}}].

\bibitem{Ishtiaque:2017trm}
N.~Ishtiaque, {\it {2D BPS Rings from Sphere Partition Functions}},  {\em JHEP}
  {\bf 04} (2018) 124, [\href{http://arxiv.org/abs/1712.02551}{{\tt
  arXiv:1712.02551}}].

\bibitem{Benini:2012ui}
F.~Benini and S.~Cremonesi, {\it {Partition Functions of ${\mathcal{N}=(2,2)}$
  Gauge Theories on S$^{2}$ and Vortices}},  {\em Commun. Math. Phys.} {\bf
  334} (2015), no.~3 1483--1527, [\href{http://arxiv.org/abs/1206.2356}{{\tt
  arXiv:1206.2356}}].

\bibitem{Doroud:2012xw}
N.~Doroud, J.~Gomis, B.~Le~Floch, and S.~Lee, {\it {Exact Results in D=2
  Supersymmetric Gauge Theories}},  {\em JHEP} {\bf 05} (2013) 093,
  [\href{http://arxiv.org/abs/1206.2606}{{\tt arXiv:1206.2606}}].

\bibitem{Gaiotto:2013bwa}
D.~Gaiotto and P.~Koroteev, {\it {On Three Dimensional Quiver Gauge Theories
  and Integrability}},  {\em JHEP} {\bf 05} (2013) 126,
  [\href{http://arxiv.org/abs/1304.0779}{{\tt arXiv:1304.0779}}].

\end{thebibliography}\endgroup

\end{document}